\documentclass[a4paper,fleqn,usenatbib]{mnras}

\usepackage{newtxtext,newtxmath}

\usepackage[T1]{fontenc}
\usepackage{ae,aecompl}

\usepackage{graphicx}	
\usepackage{amsmath}	

\usepackage{color}
\usepackage{ulem}

\title[Light and colour variations of Miras in the SMC]{Light and colour variations of Mira variables in the Small Magellanic Cloud}

\author[Y. Ita et al.]{Yoshifusa Ita$^{1}$\thanks{E-mail: yita@astr.tohoku.ac.jp}, John W. Menzies$^{2}$, Patricia A. Whitelock$^{2,3}$, Noriyuki Matsunaga$^{4}$,
\newauthor Masaki Takayama$^{1}$, Yoshikazu Nakada$^{4,5}$, Toshihiko Tanab\'{e}$^{5}$, Michael W. Feast$^{2,3}$\thanks{Deceased April 2019, before the paper was finalised.},
\newauthor and Takahiro Nagayama$^{6}$
\\
$^{1}$Astronomical Institute, Graduate School of Science, Tohoku University, 6-3 Aramaki Aoba, Aoba-ku, Sendai, Miyagi 980-8578, Japan\\
$^{2}$South African Astronomical Observatory, PO Box 9, 7935 Observatory, South Africa\\
$^{3}$Department of Astronomy, University of Cape Town, 7701 Rondebosch, South Africa\\
$^{4}$Department of Astronomy, School of Science, The University of Tokyo, 7-3-1 Hongo, Bunkyo-ku, Tokyo 113-0033, Japan\\
$^{5}$Institute of Astronomy, School of Science, The University of Tokyo, 2-21-1 Osawa, Mitaka, 181-0015, Tokyo, Japan\\
$^{6}$Department of Science and Engineering, Kagoshima University, Korimoto, Kagoshima 890-0065, Japan 
}

\date{Accepted XXX. Received YYY; in original form ZZZ}

\pubyear{2015}

\begin{document}
\label{firstpage}
\pagerange{\pageref{firstpage}--\pageref{lastpage}}
\maketitle

\begin{abstract}The goal of this paper is to characterise the light variation properties of Mira variables in the Small Magellanic Cloud. We have investigated a combined optical and near infrared multi-epoch dataset of Mira variables based on our monitoring data obtained over 15 years. Bolometric correction relations are formulated for various near-infrared colours. We find that the same bolometric correction equation holds for both the bolometricly brightest and faintest pulsation phases. Period-bolometric magnitude relations and period-colour relations were derived using time-averaged values. Phase lags between bolometric phase and optical and near-infrared phases were detected from the O-rich (the surface C/O number ratio is below unity) Mira variables, while no significant systematic lags were observed in most of the C-rich (the C/O ratio is over unity) ones. Some Miras show colour phase inversions, e.g., $H-K_{\rm s}$ at its bluest while $J-H$ and $J-K_{\rm s}$ are at their reddest values at about the bolometricly brightest phase. Their occurrence conditions were studied but no clear direct or indirect trigger was found. A large NIR colour change unassociated with stellar pulsation was observed in Miras with long secondary periods, and its possible explanation is described. 
\end{abstract}




\section{Introduction}
Recent large scale photometric surveys toward the Small Magellanic Cloud (SMC) such as the Magellanic Clouds Photometric Survey (optical, \citealt{zaritsky2002}), the Two Micron All Sky Survey (near-infrared, \citealt{skrutskie2006}),
the survey by the Infrared Survey Facility (near-infrared, \citealt{kato2007}), and the Surveying the Agents of a Galaxy's Evolution survey (near- to far-infrared, \citealt{meixner2006}, \citealt{gordon2011}) provide multi-wavelength photometric data, but only for a single epoch. When it comes to variable stars, single-epoch photometric data are not enough to characterise their variability and multi-epoch data are required. Thanks to the surveys aiming to detect gravitational microlensing events (e.g., OGLE, MACHO, EROS), optical multi-epoch data for tens of thousands of variable stars in the SMC are available as a byproduct. While the optical multi-epoch data helped a lot to advance our understanding of stellar variability (see, e.g., \citealt{soszynski2006} and series of their work), commensurate data at longer wavelengths such as in the near-infrared (NIR) were not available. Recently, \citet{ita2018} published their more than 15-year long NIR monitoring data set toward the SMC obtained at the Infrared Survey Facility (IRSF). Other than these data, the Vista Magellanic Clouds survey (\citealt{cioni2011}), and the SAGE-VAR survey (\citealt{riebel2015}) will provide us with wider, deeper data in the NIR.

In this paper, a sample of Mira variables (Miras) in the SMC is chosen from the OGLE survey catalogue (\citealt{soszynski2011}) and its publicly available optical and NIR multi-epoch photometric data and also publicly available single epoch data are compiled to discuss NIR and bolometric variability of Miras. Because the OGLE survey (\citealt{soszynski2011}) and the NIR survey were conducted at roughly the same time, their combined data allow us to calculate time series bolometric magnitudes for an unprecedented number of Miras in the SMC. Relatively well sampled light curves allowed us to investigate possible phase differences between bolometric, optical and NIR brightness variations. Also, the time averaged bolometric magnitudes are calculated and they are correlated with the pulsation periods to determine period-bolometric magnitude relations.

The remainder of this paper is organised as follows. Section 2 describes the data used in this study. Section 3 describes how we determined the pulsation parameters, i.e., periods, phases, amplitudes, mean magnitudes and bolometric magnitudes of Miras in the SMC. Based on the results, we discuss bolometric variability, phase lag, and other pulsation properties in Section 4. Finally, the summary is given in Section 5.

\section{Data}
This section describes the data used in this study and how sample stars were selected.

\subsection{Multi-epoch data}
\subsubsection{OGLE-III data}
The OGLE project released its third phase (OGLE-III; 2001 Jun 12 -- 2009 May 3) survey data in the Magellanic Clouds \citep{udalski2008}. They provided catalogues of several types of variable stars in the Magellanic Clouds and also their multi-epoch photometric data. The catalogues list coordinates, pulsation periods, amplitudes, variability types and subtypes, and mean $V-$ and $I-$band magnitudes. According to \citet{soszynski2009} and \citet{soszynski2013}, they used the Fourier-based Fnpeaks code by Z. Ko{\l}aczkowski to determine pulsation periods; they define the amplitude as the difference between the maximum and minimum values of the third-order Fourier series fit to the observed light curve.

Using the OGLE-III on-line database\footnote{visit the OGLE web page at \url{http://ogle.astrouw.edu.pl/}}, stars with variable star type of ``LPV" (standing for long period variables, \citealt{soszynski2011}) and subtype of ``Mira" in the SMC were chosen. \citet{soszynski2009} used the $I-$band pulsation amplitude to distinguish between variables of the Mira and semi-regular variable subtypes. They defined variables with {\bf peak-to-peak} $I-$band amplitudes greater than 0.8 mag as Miras. In this study, we follow their classification.

\subsubsection{IRSF/SIRIUS NIR variable star survey data}
\citet{ita2018} published NIR ($J, H,$ and $K_{\rm s}$) monitoring data toward a one square degree area in the central part of the SMC centred on 00$^h$55$^m$00.00$^s$ -72$^\circ$50$^\prime$00.00$^{\prime\prime}$ (J2000). This survey area is wholly contained in the OGLE-III survey field. The 10-$\sigma$ detection limits for a moderately populated field and the saturation limits of the survey are 18.3, 17.5, and 16.7 mag and 9.6, 9.6, and 9.0 mag for the $J, H,$ and $K_{\rm s}$ bands, respectively. Their NIR monitoring survey was initiated in July 2001 in the SMC, and is still ongoing. The baseline observation period exceeds 15 years, long enough to see long-term trends in light variation. For more details of the survey, refer to the paper mentioned above. 

We examined the time-series data and found that some of the stars have spurious data. We therefore calculated the change from the mean luminosity and examined the correlation of the change between the bands of $H$ and $K_{\rm s}$, $J$ and $K_{\rm s}$, and $J$ and $H$. Although a clear linear relationship is found, there are scattered points out of the main relationship in $H$ and $K_{\rm s}$, and $J$ and $K_{\rm s}$, where the formal errors do not match the deviations from the main relationship. These points are likely to be the result of measurement errors much larger than the formal errors. These spurious points were removed by the 3-sigma clipping method, for the purpose of the discussion below.

\subsubsection{SAGE-VAR data}
\citet{riebel2015} presented a total of six epochs of photometry at 3.6 and 4.5 $\mu$m obtained by the \textit{Spitzer} space telescope for stars in the 3.7 degrees $\times$ 1.5 degrees area along the bar of the LMC and in the central 1.7 degrees $\times$ 1.7 degrees area of the SMC. Their variable source catalogue lists mean magnitudes and amplitudes in the [3.6] and [4.5] bands\footnote{Throughout this paper, the numbers bracketed by $[~]$ designate \textit{Spitzer} photometry, for example, [3.6] indicates the photometry in the 3.6 $\mu$m band.}. Certainly, six epochs of photometric data for a long period variable is not enough to determine its full variability. Their definition of amplitude is the difference between the brightest and dimmest magnitudes observed for a source, representing a lower limit on the source's full variability.

\subsection{Single epoch data}
The Magellanic Clouds have been the targets of many large scale photometric surveys. In the optical, the Magellanic Clouds Photometric Survey (\citealt{zaritsky2002}) provides photometry in the $U-, B-, V-$, and $I-$bands. A $UBVR$ survey of the Magellanic Clouds (MASSEY: \citealt{massey2002}) provides somewhat shallower, but better data for the brighter stars, a higher precision calibration (particularly at $U$), and complements \citet{zaritsky2002} by including $R-$band data. In the infrared, the 2MASS survey (\citealt{vandyk1999}) provides uniform $J-$, $H-$, and $K_{\rm s}-$ photometry. The $IJK$ Deep Near-Infrared Survey (DENIS: \citealt{cioni2000}) is a complement to the 2MASS survey. The \textit{Spitzer} SAGE-SMC survey catalogue \citep{gordon2011} provides photometry at longer wavelengths, in the \textit{Spitzer} IRAC \citep{fazio2004} and MIPS \citep{rieke2004} wave bands.
 
\subsection{Sample stars}
Sample stars used in this study are selected as follows: 

\subsubsection{Optical sample}
First, the OGLE-III on-line database was queried with a positional criterion\footnote{11.980507 $<=$ R.A. [degree] $<=$ 15.497504 and $-$73.344353 $<=$ Decl. [degree] $<=$ $-$72.321939} to extract known OGLE variables situated within the observed area of the IRSF/SIRIUS NIR variable star survey in the SMC. The query found 95 Miras. All but one of these have counterparts in the IRSF/SIRIUS NIR variable source catalogue. The sole missing Mira (OGLE SMC-LPV-06978) is located at the very edge of the IRSF/SIRIUS survey. Note that the edges of the IRSF/SIRIUS survey area are not rectilinear. Hereafter, we refer to these 94 Miras as the sample stars.

\subsubsection{Cross-identification}
The sample stars were further cross-identified with the existing single epoch photometric catalogues mentioned above except for the MASSEY and DENIS catalogues. The IRSF/SIRIUS position was employed as a working reference and a positional tolerance of 3 arcsec was used for the crossmatch. If more than one star was present within the search radius, the closest one was adopted and others were discarded.

\begin{figure}
\centering
\includegraphics[scale=0.53,angle=0]{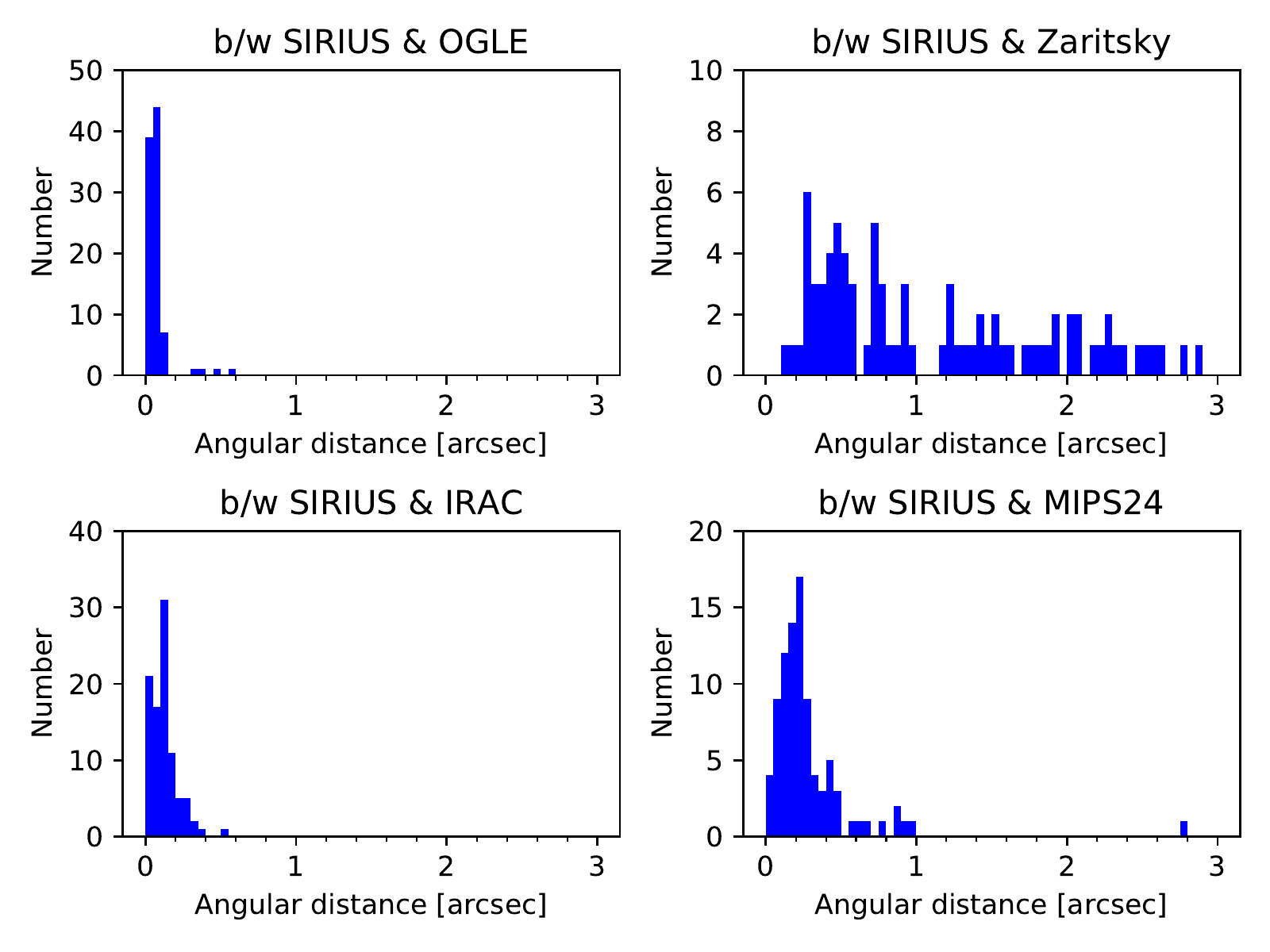} 
\caption{Histograms of the angular separations between the IRSF/SIRIUS positions and the positions in the matched catalogue. The size of the bins is set to 0.05 arcsec.}
\label{adistance}
\end{figure}


Fig.~\ref{adistance} is a histogram of the angular separations between the IRSF/SIRIUS positions and the positions in the matched catalogues. The IRSF/SIRIUS, OGLE and IRAC positions agree very well. On the other hand, there are cases where the angular distance of the matched source is more than an arcsec in the IRSF/SIRIUS - MIPS and the IRSF/SIRIUS - Zaritsky matches. The sole MIPS source with more than one arcsec separation is OGLE SMC-LPV-10929. We checked its MIPS image and found that the star may be suffering from blending. The matching between IRSF/SIRIUS and Zaritsky is rather poor, and we decided not to use the Zaritsky catalogue data in this study. The matching results will be used for discussion in the rest of this paper.

\subsection{Surface chemistry}\label{sec_surface}
The surface chemistry (Oxygen-rich or Carbon-rich) of the sample stars reported in the OGLE-III catalogue was inferred from photometric colours. This type of colour threshold method has been widely used (e.g., \citealt{cioni2003a}, \citealt{cioni2003b}, \citealt{raimondo2005}), but it is a statistical method. While recognising the possibility of contamination of C-rich stars in the O-rich sample or vice versa, we accept the OGLE-III classification and use it in the following discussion. Of the 94 stars discussed here 77 were classified by OGLE-III as C-rich and the remaining 17 as O-rich.

\section{DATA ANALYSES}
\subsection{Light curves and Period determination}
We developed software that performs a period search using the so-called Lomb-Scargle periodogram (\citealt{lomb1976}; \citealt{scargle1982}), adopting the multi-band periodogram approach of \citet{vanderplas2015} so that the same period was found for all wavebands. We used the software to search for significant signals throughout the trial periods between 50 and $T$ days, where $T$ is the total available observation baseline of the time series data. Note that $T$ varies from one star to another, and ranges from about 6,000 to 7,600 days. The trial period is successively incremented by variable step values ($\coloneqq$ resolution, i.e., $P_{\textrm{trial, new}} = P_{\textrm{trial, old}} +$ resolution), which are summarised in Table~\ref{resolution}. We did not try to search for periods longer than $T$ days. This is to ensure that the observation baseline should be long enough to cover at least one cycle for the longest period determined. Possible artefacts arising from setting the period search limit at $T$ days are accounted for and removed appropriately. In addition to the limitation by the size of $T$, we also define a working limit, $T_w$, that is $T$/3. This means that periods shorter than $T_w$ can be repeated at least 3 times within the $T$ days. It should also be noted that we apply the same weights for all the observed data, regardless of their photometric errors.

\begin{table}
\caption[]{Resolution used in period search}
\label{resolution}
\begin{center}
\begin{tabular}{rrr}
\hline
\multicolumn{2}{c}{trial period [day]} & \\
\multicolumn{1}{r}{start($\ge$)} & \multicolumn{1}{r}{end($<$)} & \multicolumn{1}{r}{resolution [day]} \\
\hline
50.0 & 100.0 & 0.01 \\
100.0 & 1000.0 & 0.1 \\
1000.0 & $T$ & 1 \\
\hline
\end{tabular}
\end{center}
\end{table}

Since many stars show multi-periodic or complex light curves (e.g., \citealt{bedding1998}), we looked for up to five significant periods. First, the period with the largest power was found and a third order Fourier fit was made to the data using this period as the fundamental. Then the data set was ``whitened" by removing the fitted curve and the process was repeated for up to a further four periods.

Consequently, the observed light curve can be represented by the following equation:
\begin{equation}
\label{fit}
y(t) = c + \sum_{j=1}^{N_s} \sum_{f=1}^{N_f} \left[ {}a_{j_f} \cos f\omega_{j} t + b_{j_f} \sin f\omega_{j} t \right],
\end{equation}
$\omega = 2\pi/P$, P is the period, $N_s$ is the number of significant periods (up to 5) after removing ones with high false alarm probability (see below), and $N_f$ is the number of terms of the Fourier series ($N_f = 3$ here). 
The constant term $c$ is the fit mean magnitude. 

We checked all the fitting results by eye. There are several sample stars that show more or less similar complex light curves with long-term change that may be attributed to a so called long secondary period (e.g., \citealt{wood2009}, LSP; the word `secondary' implies `not due to pulsation' or `after removing pulsation'), and/or can be due to variable dust obscuration (e.g., \citealt{feast2003}). The primary periods found for these stars are longer than $T_w$ days but are unlikely to be pulsation periods. There are 13 stars for which we set $T_w$ to be 1,000 days regardless of the $T$ of them and preferentially chose periods shorter than 1,000 days. These 13 stars are listed in table~\ref{Tw1000}.

For each frequency, $\omega_{j}$, in equation \ref{fit}, the maximum and minimum magnitudes ($m_{\textrm{max}}$ and $m_{\textrm{min}}$, respectively) were estimated from the fitted light curve with a step of 0.1 day within the observed HJD range. The difference between the two extreme values was defined as the amplitude and it will be designated as ${}^{\lambda}\Delta$ (i.e, ${}^{\lambda}\Delta = m_{\textrm{max}} - m_{\textrm{min}}$), where $\lambda$ indicates one of the photometric bands, $I, J, H,$ and $K_{\rm s}$, or the bolometric ($m_{\textrm{bol}}$, see following) case. The $j$-th initial phase was defined with respect to HJD=2450000.0, and was calculated as $\phi_j = \textrm{arctan2}(b_{j_1}, a_{j_1})/2\pi, (-0.5 \leqq \phi_j < 0.5)$. Practically, the shape of the Mira's light curve can be different from cycle to cycle and also from waveband to waveband. This initial phase definition is employed to see the overall trend. Hereafter, we denote the $j$-th amplitude and initial phase by ${}^\lambda A_j$ and ${}^\lambda \phi_j$, respectively. Recall that we adopted the multi-band periodogram approach, so the $j$-th period is common among the $\lambda$.

To evaluate the significance of the periods, we calculated the false alarm probabilities (FAP; how often a certain signal is observed just by chance) by using bootstrap simulations. First, we randomly shuffled the multi-epoch data keeping the times of observation (i.e., $t_i$) fixed. Then the periodogram of the randomly shuffled data was calculated around the periods of interest. After that, the data were reshuffled and the process repeated. This procedure was repeated 100,000 times ($= N_{\textrm{total}}$) and we counted the number of ``random'' periodograms with $\chi^2$ equal to or smaller than that of the period of interest ($= N_{\textrm{chance}}$). Finally, we calculated the FAP as $\textrm{FAP} = N_{\textrm{chance}}/N_{\textrm{total}}$. We consider periods with FAP smaller than 0.001 to be significant, i.e., real signals. 

Henceforth, any magnitudes at $\lambda$ with an overbar (for example, $\overline{H}$) will mean that their values are fit mean magnitudes. In addition to the fit mean magnitude, the flux mean magnitude was also calculated by integrating the instantaneous flux obtained from the best fit. We define any magnitudes in angle brackets (for example, $\langle H \rangle$) to be flux means. Any magnitudes without an overbar or angle brackets refer to individual observed magnitudes.

The observing data have yearly blank intervals that could be a problem for variable stars with a pulsation period of about a year. However, our exploratory study showed that these yearly gaps hardly affect the period determination in most cases. In very rare cases, the yearly gaps may, unfortunately, coincide with the timings of the peak luminosity. Such cases can affect the determination of pulsation amplitude, but they will be rare.

\begin{table}
\caption[]{Stars whose $T_w$ was set to 1,000 days.}
\label{Tw1000}
\begin{center}
\begin{tabular}{c}
\hline
\multicolumn{1}{c}{Name} \\
\hline
OGLE SMC-LPV-07375 \\ 
OGLE SMC-LPV-07385 \\ 
OGLE SMC-LPV-07771 \\ 
OGLE SMC-LPV-08082 \\ 
OGLE SMC-LPV-08803 \\ 
OGLE SMC-LPV-09683 \\ 
OGLE SMC-LPV-10708 \\ 
OGLE SMC-LPV-11279 \\ 
OGLE SMC-LPV-11806 \\ 
OGLE SMC-LPV-13009 \\ 
OGLE SMC-LPV-13676 \\ 
OGLE SMC-LPV-14778 \\ 
OGLE SMC-LPV-14991 \\ 
\hline
\end{tabular}
\end{center}
\end{table}

\subsubsection{Comparison with OGLE results}
\begin{figure}
\hspace*{-1cm}
\includegraphics[scale=0.4]{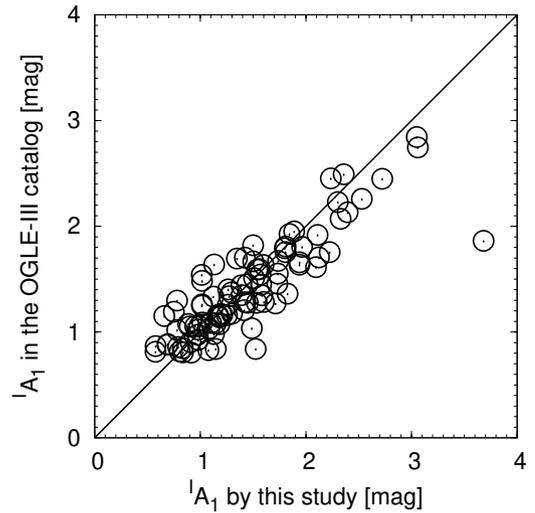} 
\hspace*{-1cm}
\includegraphics[scale=0.4]{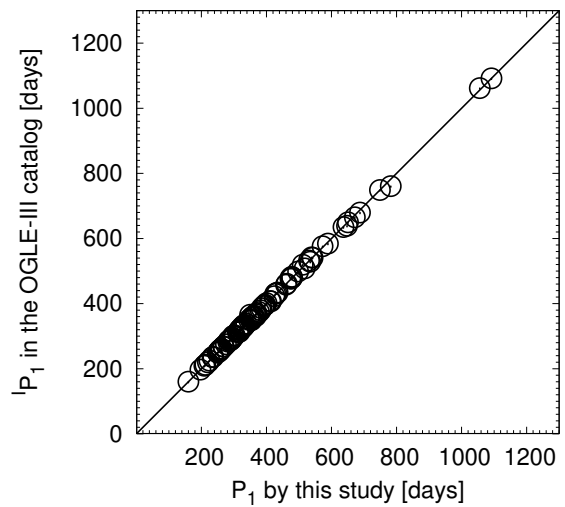} 
\caption{Comparisons of the primary amplitudes (top) and periods (bottom) between those in the OGLE-III catalogue and those calculated in this study. The solid lines show equivalence relations.}
\label{period}
\end{figure}

The periods and amplitudes calculated by our software were compared with those in the OGLE-III catalogue. Here we only compare the primary ones. Note that the primary period in the OGLE-III catalogue is not necessarily associated with the largest amplitude. To require the same definition, we rearranged the order of OGLE-III periods with descending order of their associated amplitudes before comparison. The results of the comparisons are shown in Fig.~\ref{period}.


\begin{figure*}
\centering
\includegraphics[scale=0.50,angle=0]{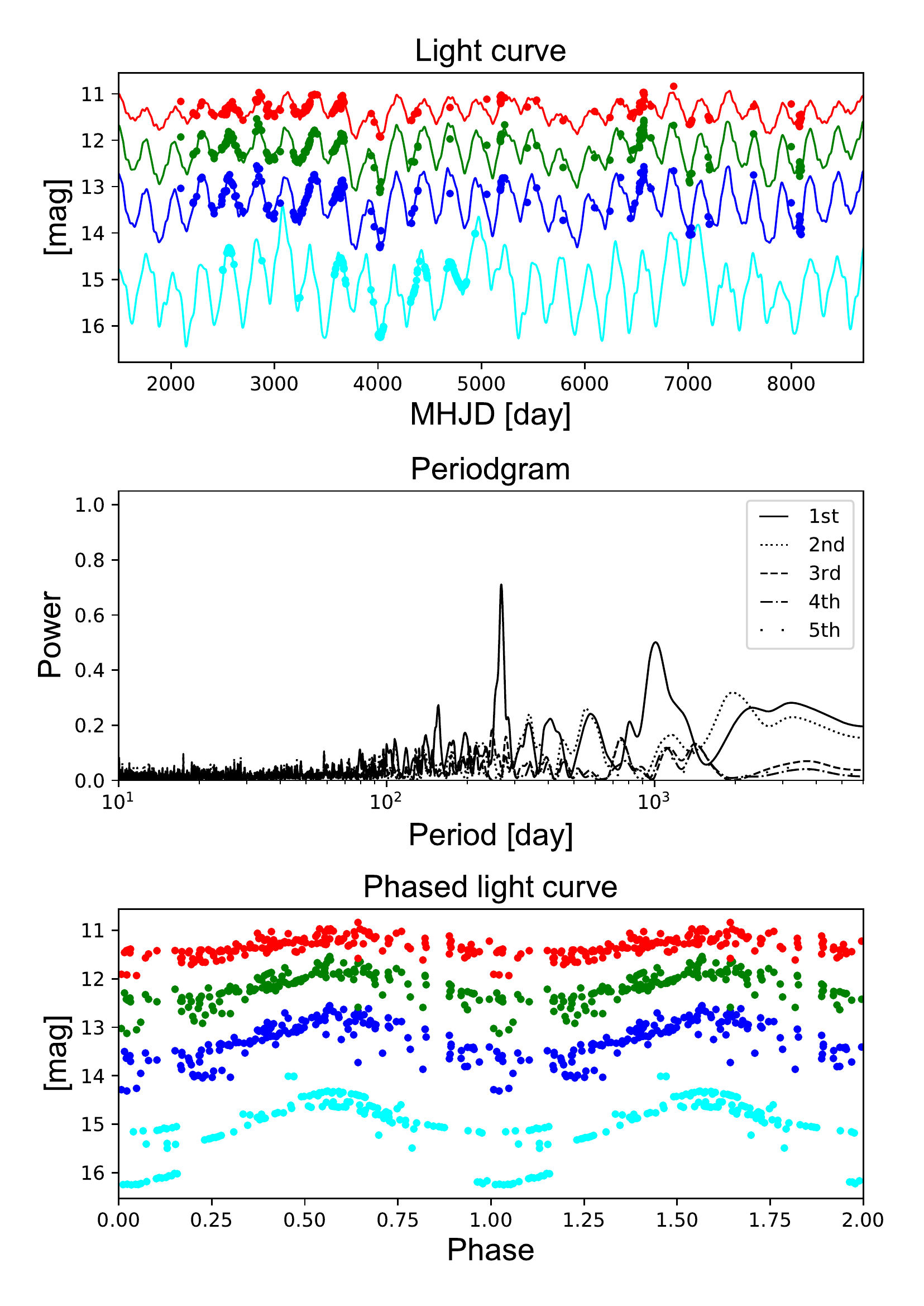} 
\includegraphics[scale=0.50,angle=0]{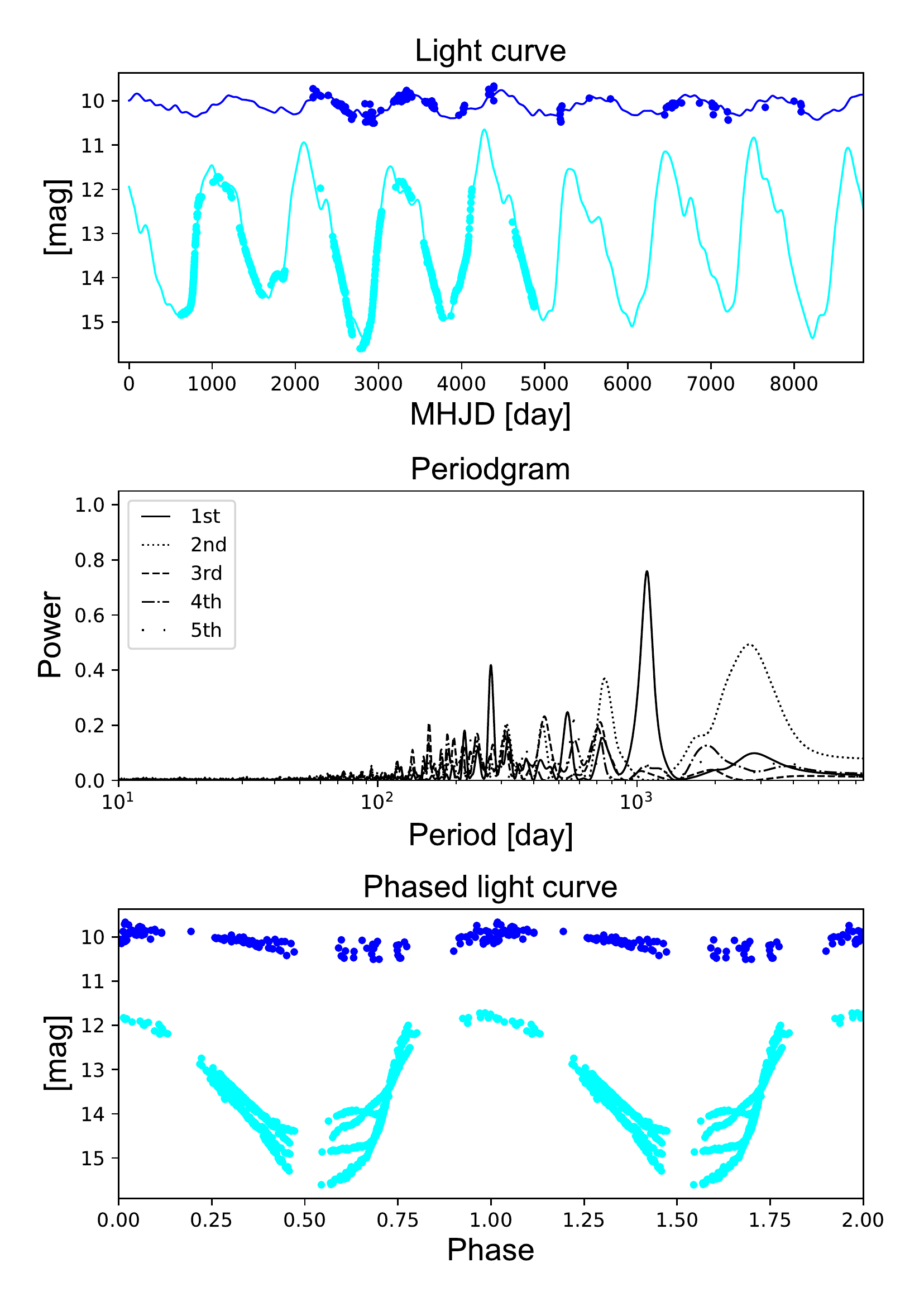} 
\caption{The light curves, multi-band periodograms, and phased (with $P_1$) light curves of OGLE SMC-LPV-08723 (left) and 14462 (right). Cyan, blue, green, and red dots show $I$, $J$, $H$, and $K_{\rm s}$ bands, respectively. Error bars are smaller than the size of the symbols. The OGLE SMC-LPV-14462 image is saturated in both the $H$ and $K_{\rm s}$ bands. The solid lines in the top panels are the best fit results.}
\label{OGLE-SMC-LPV-14462}
\end{figure*}

Pulsation amplitudes are in good agreement between the OGLE-III catalogue values and the ones calculated by our software with a few exceptions. There are two stars with a discrepancy larger than 50\% between the two values, namely, OGLE SMC-LPV-08723 and 14462. The OGLE-III catalogue gives ${}^{I}A_1$ of 0.840 and 1.862 mag, respectively, for the two stars while our software gives 1.524 and 3.682 mag, respectively. In Fig.~\ref{OGLE-SMC-LPV-14462}, we show the light curves, multi-band periodograms, and phased (with $P_1$) light curves for the two stars. The discrepancies in amplitudes probably arise from the details of the fitting process, e.g. the order of removing frequencies in the whitening process. 

We see quite a good agreement in the period comparison. The very long-term (over 15 years) NIR time series data enabled us to confirm the periods in the OGLE-III catalogue.

We interpret the comparison results to suggest that our period and amplitude determination procedures are fair. Throughout this work we will use periods, amplitudes, mean magnitudes and flux mean magnitudes calculated with our software, and do not use the OGLE-III catalogue values unless otherwise explicitly mentioned.

\subsection{Bolometric magnitudes}
Our sample stars are red giant variables in the SMC with optical identifications. Therefore, most, if not all, of their radiant energy distributions attain their peaks at around wavelengths corresponding to $I-, J-, H-, K_{\rm s}-, [3.6]-$, or $[4.5]-$bands. In other words, radiant energies at around these wavelengths should dominate their bolometric luminosities. With this in mind we calculated the time dependent bolometric magnitudes of sample stars as follows:

\subsubsection{Merging multi-epoch data}
First, we merged the OGLE-III multi-epoch data with the NIR multi-epoch data. In practice, the timings of observations are not necessarily the same between the two data sets. Then we linked data points from these two data sets if they were taken $\pm$3 days of each other. In most cases, the smallest difference in timings of observation was smaller than 1 day. Next, we added the time-dependent data obtained at longer wavelengths, namely [3.6] and [4.5] data. Unfortunately, there were only 6 epochs of photometric data in these wavebands. Therefore, we used the following equation to estimate the phase-dependent magnitude in these two wavebands,
\begin{equation}
m_\Lambda (t) = \langle \Lambda \rangle + \frac{{}^{\Lambda}\Delta}{2} \cos \left( 2\pi \left[ \frac{t - t_0}{P_1} - {}^{K_{\rm s}}\phi_{1} \right] \right),
\end{equation}
where $\Lambda$ denotes either [3.6] or [4.5], $t$ is in HJD, and $t_0$ is set to be 2450000.0. Obviously, here we assume that [3.6] and [4.5] light variations are in phase with that at $K_{\rm s}$. Values of $\langle \Lambda \rangle$ and ${}^{\Lambda}\Delta$ were taken from \citet{riebel2015}. Their definition of amplitude is the brightest flux they observed minus the dimmest flux they observed. Because their observations were randomly phased, this should be a lower bound on the true difference between the two extreme (i.e., maximum and minimum) values. 

\subsubsection{Interstellar extinction correction}
The data were corrected for interstellar extinction. \citet{keller2006} derived a mean reddening of $E(B-V)=0.12$ mag including interstellar extinction in our Galaxy and the SMC itself. Using that value and the extinction law of \citet{cardelli1989} with $R_V=3.2$, the photometry of all sample stars at wave bands shorter than or equal to the $K_{\rm s}$ band were corrected for reddening. We assume that the reddening is negligible at wave bands longer than or equal to 3.6 $\mu$m. The NIR photometric data was based on the Las Campanas Observatory (LCO) system \citep{ita2018}. We transformed them using the relations between the LCO and the 2MASS systems given in \citet{cutri2006}. Experiments showed that the choice of the mean reddening has only a small influence on the calculated bolometric magnitude. If we chose a mean reddening of $E(B-V)=0.15$ (or $=0.08$) instead of the employed value of 0.12 mag, the bolometric magnitude would systematically become brighter (fainter) by up to 0.02 (0.03) mag for the vast majority of the sample stars, and by 0.04 (0.05) mag for the bluest sample stars.

\begin{table}
  \caption{The adopted zero magnitude flux densities and their corresponding reference wavelengths.}\label{table:zeromag}
  \begin{center}
    \begin{tabular}{c r r r}
    \hline
    Wavebands & \multicolumn{1}{c}{$f_0$$^1$} & \multicolumn{1}{c}{$\lambda_r$$^2$} & Ref.$^3$ \\
              & \multicolumn{1}{c}{[Jy]} & \multicolumn{1}{c}{[$\mu$m]} &  \\
    \hline
    $I$ & 2459 & 0.8071 & a \\
    $J$ & 1594 & 1.235  & b \\
    $H$ & 1024 & 1.662  & b \\
    $K_{\rm s}$   & 666.8 & 2.159 & b  \\
    $[3.6]$ & 280.9 & 3.550   & c \\
    $[4.5]$ & 179.7 & 4.493   & c \\
    $[5.8]$ & 115.0 & 5.731   & c \\
    $[8.0]$ & 64.1  & 7.872   & c \\
    $[24]$  & 7.14  & 23.68   & d \\
    \hline
    \end{tabular}
  \end{center}
$~^1$ Zero magnitude flux density, $~^2$ Reference wavelength, $~^3$ {\bf References :}
$a$ \citet{cohen2003a}; $b$ \citet{cohen2003b}; $c$ \citet{irac2006}; $d$ \citet{mips2008}
\end{table}

\subsubsection{Calculation method and possible systematics}
The resultant reddening-free magnitudes were converted into Janskies by using the zero magnitude fluxes tabulated in Table~\ref{table:zeromag}. Next, we interpolated between all the available photometric data by using a piecewise cubic Hermite interpolating polynomial to re-sample the spectral energy distribution (SED). Finally, the re-sampled SED was integrated from 0.8071 to 23.68 $\mu$m to calculate the bolometric magnitude of a sample star at each selected variability phase. We used the scale that the zero point of the apparent bolometric magnitude $m_{{\rm bol}} = 0$ corresponds to an irradiance of $2.518 \times 10^{-8}$ [W/m$^2$] (\citealt{mamajek2015}). We calculated bolometric magnitudes only for stars with [8.0] data. For stars without [24] data, we estimated the corresponding flux density from their [8.0] data by extrapolation using the Rayleigh-Jeans law (Flux $\propto \lambda^{-4}$).

Obviously, we ignore the flux contributions at wavelengths outside the range of the integral. Also, we assume that the light variations at longer wavelengths (longer than 4.5 $\mu$m) have only small or negligible effects on the bolometric magnitude variation, because we reluctantly have to use single epoch data for them when calculating time dependent bolometric magnitudes (i.e., we assume that the energy emission at wavelengths longer than 4.5 $\mu$m is constant over time). Of course, the principal reason to make the assumption is the lack of multi-epoch data at longer wavelengths. We admit that the assumption would be inappropriate for very red sources whose wavelengths of peak energy emission are beyond 4.5 $\mu$m. For such extremely red sources, the calculated bolometric fluxes would be underestimated to some extent because the energy emissions from wavelengths longer than 23.68\,{$\mu$}m are not included. However, again, we do not expect many such red sources in our optically selected sample of stars, and actually there is only one such very red star in our sample, namely, OGLE SMC-LPV-12762. In any case, it is fair to say that the calculated bolometric fluxes are lower limits. 

\begin{figure}
\centering
\includegraphics[scale=0.34]{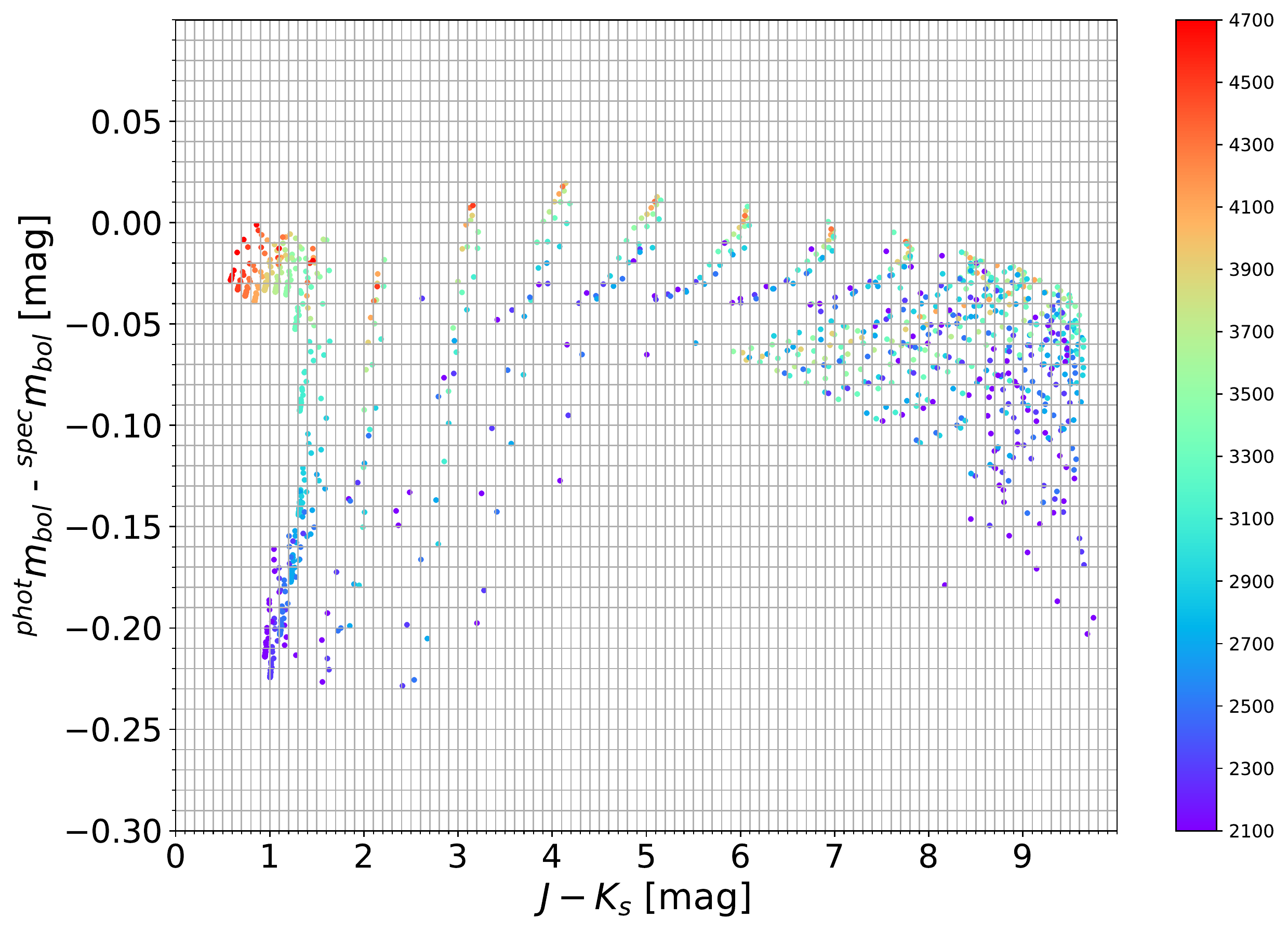} 
\includegraphics[scale=0.34]{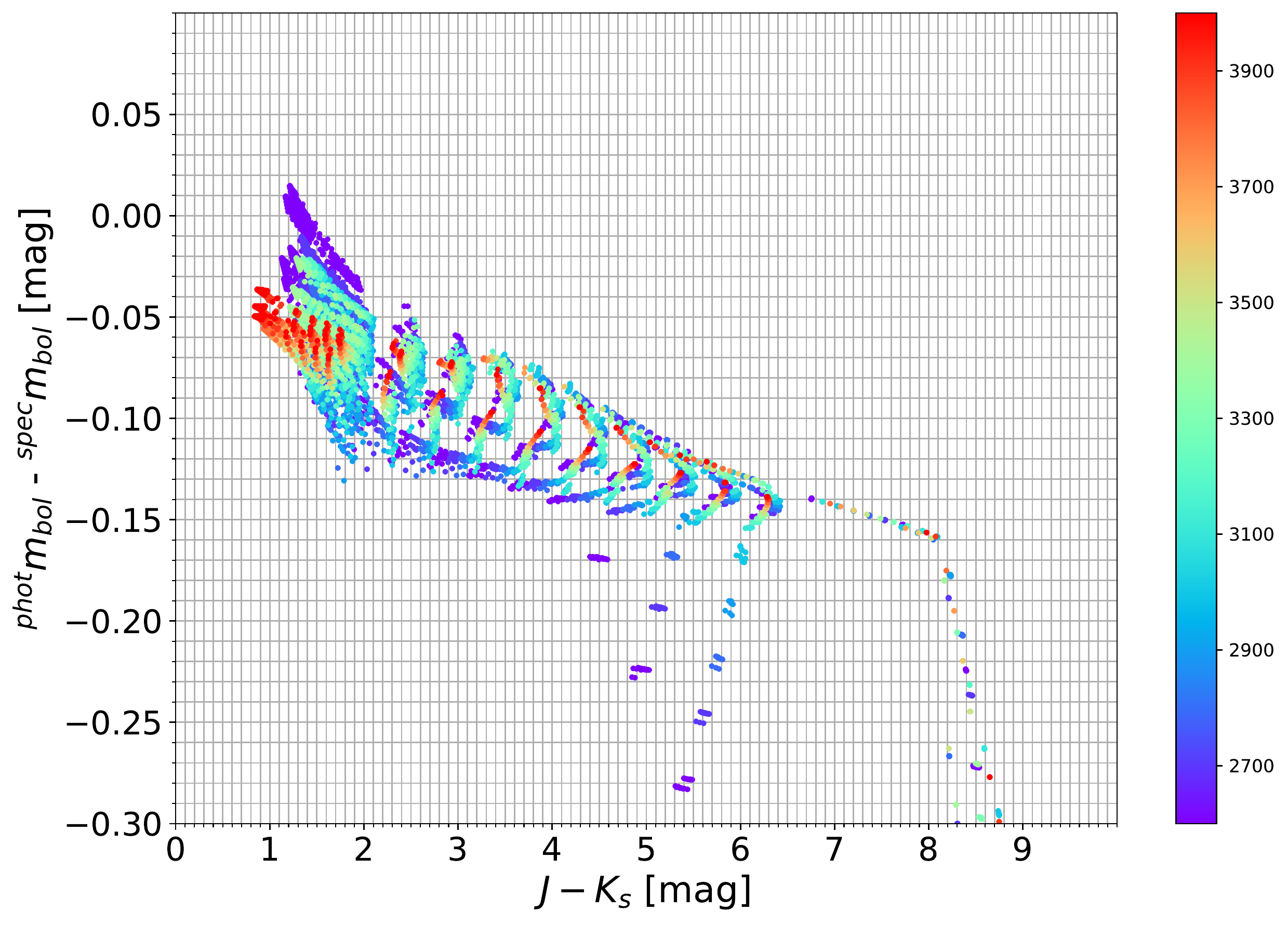} 
\caption{The systematic difference between the ${}^{{\rm phot}}m_{{\rm bol}}$ and the ${}^{{\rm spec}}m_{{\rm bol}}$ along with the $J - K_{\rm s}$ colour for O-rich (upper) and C-rich (lower) GRAMS models stars. The colour bar shows the effective temperature in Kelvin of the model stars (see text).}
\label{systematics1}
\end{figure}

One of the problems with interpolating between data points, particularly for red giants, is that it systematically overestimates the flux between the bands which will be strongly absorbed by molecules in the stellar atmosphere (e.g., \citealt{robertson1981}). One way of allowing for this is fitting a model atmosphere, but here we do not attempt to do so for the sake of simplicity and also of avoiding any uncertainties originating from model choice (among the realistic range of discrete parameter grids) and fitting processes. Instead, we just estimated the extent of the systematic effect by using GRAMS models \footnote{available at \url{https://2dust.stsci.edu/grams_models.cgi}} (\citealt{sargent2011}, \citealt{srinivasan2011}). GRAMS is a grid of oxygen- and carbon-rich circumstellar dust radiative transfer models for asymptotic giant branch and red supergiant stars. First, we integrated GRAMS models spectra from 0.8071 to 23.68 {$\mu$}m to calculate ${}^{{\rm spec}}m_{{\rm bol}}$. Next, we interpolated between the synthetic flux densities that are also given by GRAMS models at the 9 filters/wavelengths listed in Table~\ref{table:zeromag} in the same way as we described above. Then the re-sampled SED was integrated to calculate ${}^{{\rm phot}}m_{{\rm bol}}$. 

Fig.~\ref{systematics1} is a plot of the difference between the ${}^{{\rm phot}}m_{{\rm bol}}$ and ${}^{{\rm spec}}m_{{\rm bol}}$ versus the synthetic $J - K_{\rm s}$ colour. The colour bar indicates the effective temperatures $T_{{\rm eff}}$ of the model stars. For O-rich stars, ${}^{{\rm phot}}m_{{\rm bol}}$ is systematically bright, and the difference is small ($\sim$ 0.03 mag) for relatively hot (say, $T_{{\rm eff}} > 3300$ K) red giants, and is large for ones with low $T_{{\rm eff}}$, regardless of other model parameters. For C-rich stars, ${}^{{\rm phot}}m_{{\rm bol}}$ is systematically bright, and the difference becomes gradually larger with increasing $J-K_{\rm s}$ colour for relatively hot (say, $T_{{\rm eff}} > 3300$ K) red giants, and the difference changes greatly for ones with low $T_{{\rm eff}}$, regardless of other model parameters. In the present context, only the part of the diagram with $J-K_{\rm s} < 1.4$ is relevant for O-rich stars (see e.g., \citealt{cioni2003a}) and the part with $J-K_{\rm s} < 4.0$ for C-rich stars. Refer to \cite{kucinskas2005} and \cite{kucinskas2006} for O-rich stars and \cite{aringer2009} for C-rich stars for a detailed discussion on what makes the characteristic features seen in the Fig.~\ref{systematics1}.

\begin{figure}
\centering
\includegraphics[scale=0.34]{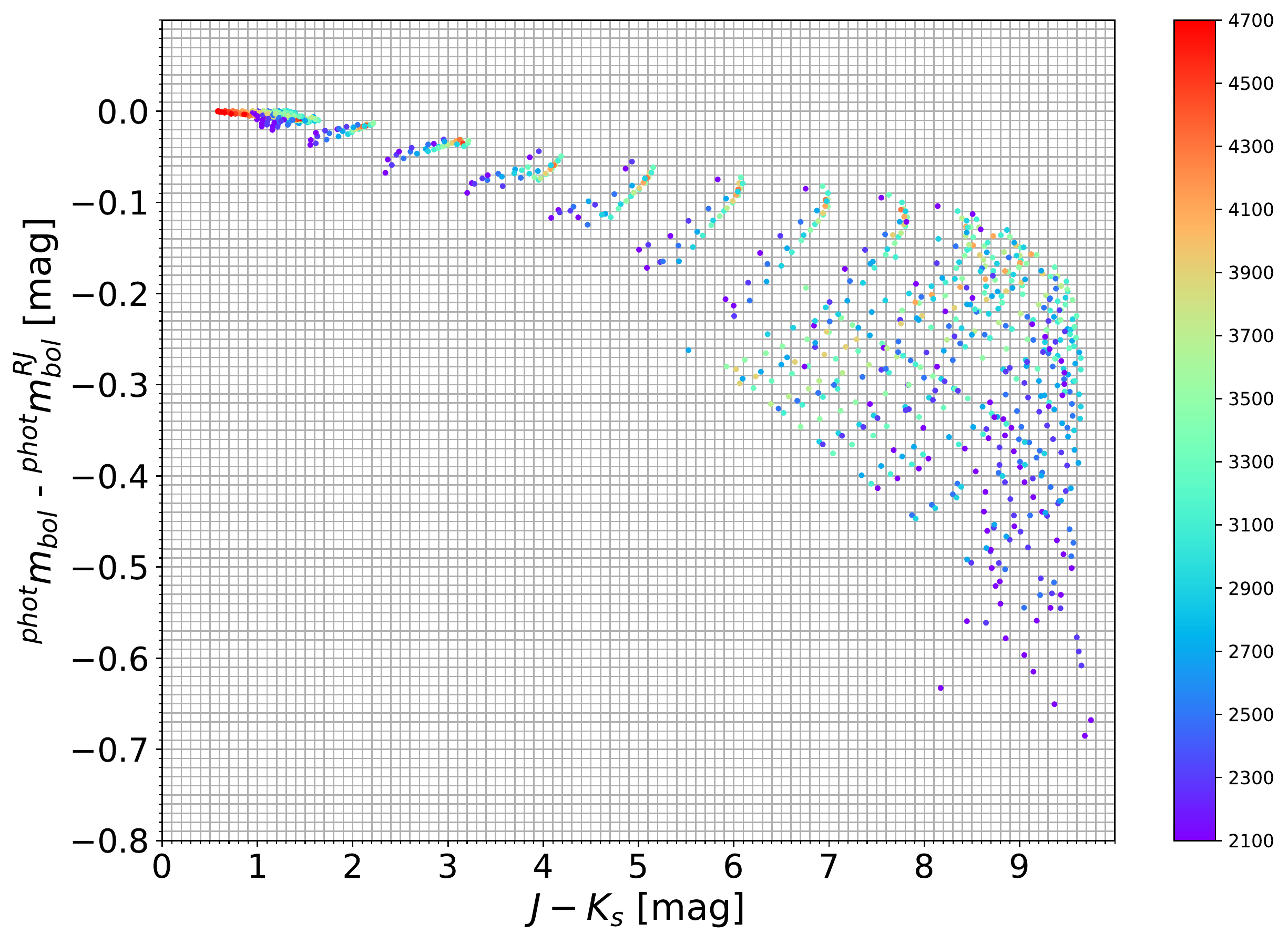} 
\includegraphics[scale=0.34]{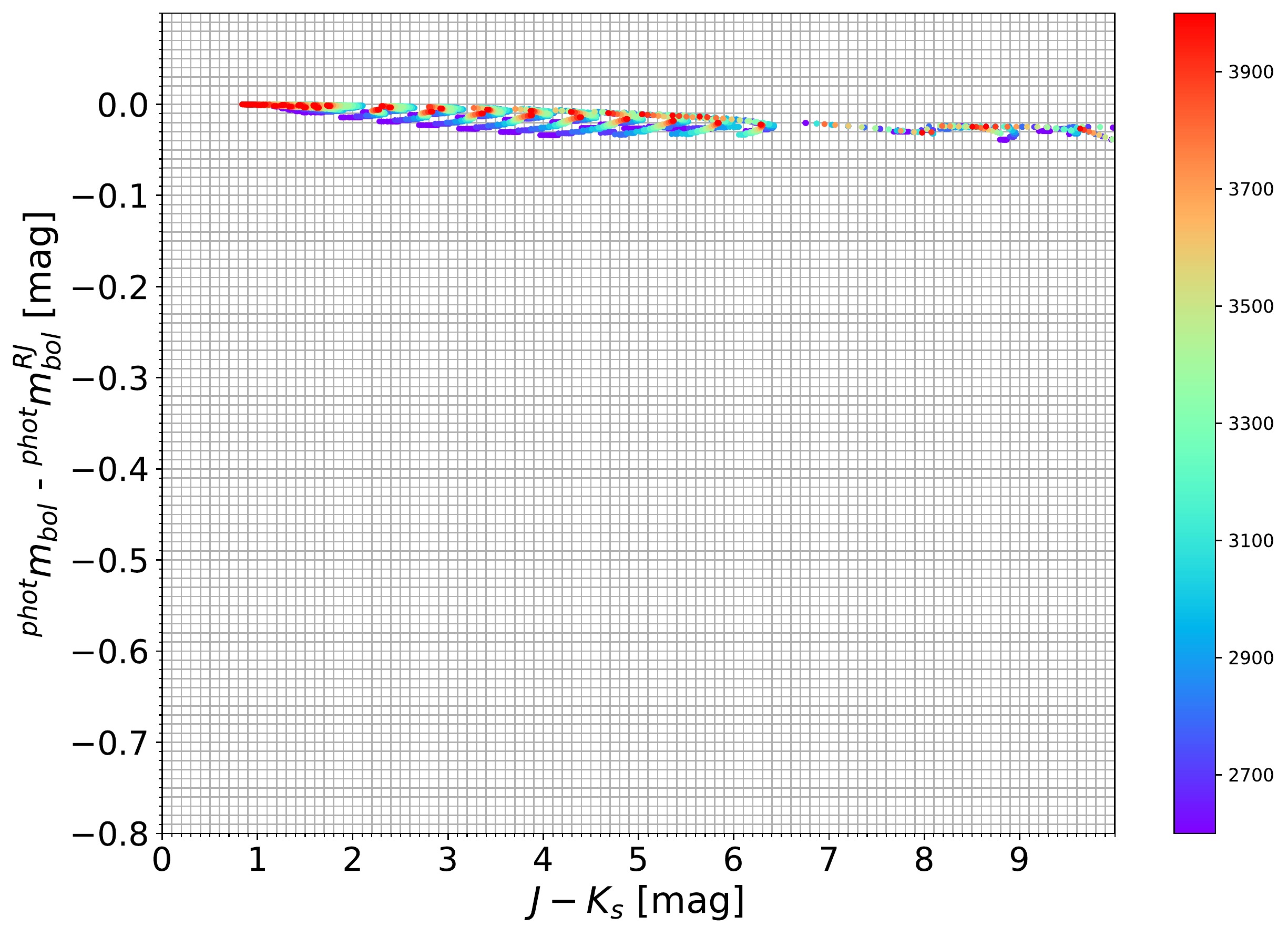} 
\caption{The systematic difference between the ${}^{{\rm phot}}m_{{\rm bol}}$ and the ${}^{{\rm phot}}m_{{\rm bol}}^{{\rm RJ}}$ along with the $J - K_{\rm s}$ colour for O-rich (upper) and C-rich (lower) GRAMS models stars. The colour bar shows the effective temperature in Kelvin of the model stars (see text).}
\label{systematics2}
\end{figure}
 
As was described above, we estimated [24] flux densities from [8.0] data by Rayleigh-Jeans law extrapolation for sample stars without [24] data. The extent of the systematic difference due to this procedure can also be estimated by using GRAMS models. We made another re-sampled SED, whose [24] flux densities were estimated via a Rayleigh-Jeans law from [8.0] data. Then we integrated the SED in the same way as described above to calculate ${}^{{\rm phot}}m_{{\rm bol}}^{{\rm RJ}}$. Fig.~\ref{systematics2} is a plot of the difference between the ${}^{{\rm phot}}m_{{\rm bol}}$ and ${}^{{\rm phot}}m_{{\rm bol}}^{{\rm RJ}}$ versus the synthetic $J - K_{\rm s}$ colour. For O-rich stars, ${}^{{\rm phot}}m_{{\rm bol}}$ is systematically bright, but the difference is small (up to 0.03 mag) for stars with $J-K_{\rm s}$ colours bluer than $1.4$ mag, as anticipated for most optical O-rich Miras (e.g., \citealt{cioni2003a}). For C-rich stars, the difference is quite small, regardless of model parameters. We checked the offsets suggested by Fig.~\ref{systematics2} by using Miras with [24] data and found that the offsets are indeed very small and they are at most 0.04 mag, fairly consistent with the estimation. 

Readers should keep in mind that it is not certain if our sample contains Miras with $T_{{\rm eff}}$ lower than the GRAMS models lower limits, and that the calculated bolometric fluxes are nonetheless suffering from a systematic effect to the extent shown in Figs.~\ref{systematics1} and ~\ref{systematics2}.

\begin{figure}
\hspace*{-0.5cm}
\includegraphics[scale=0.34]{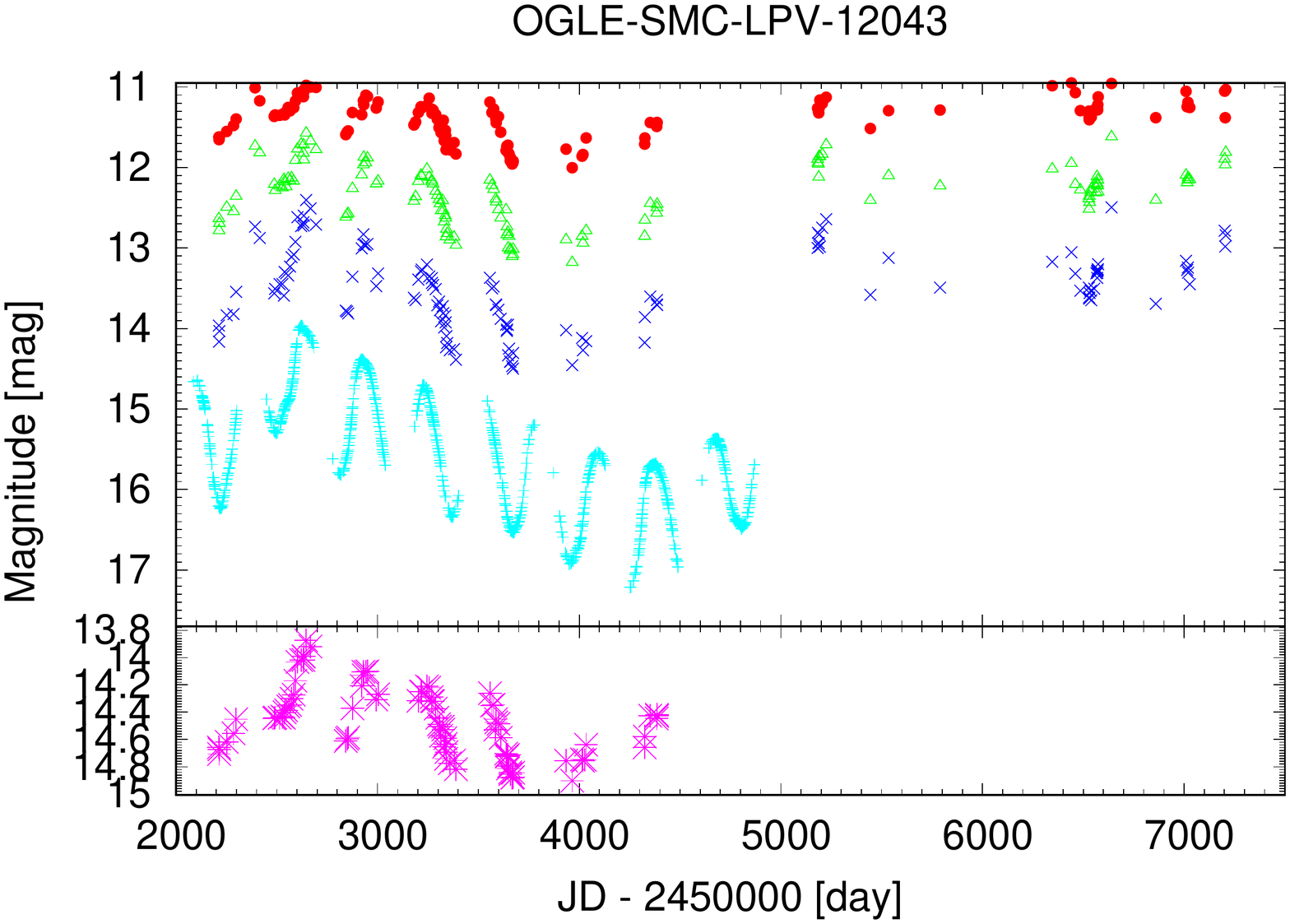}
\hspace*{-0.5cm}
\includegraphics[scale=0.34]{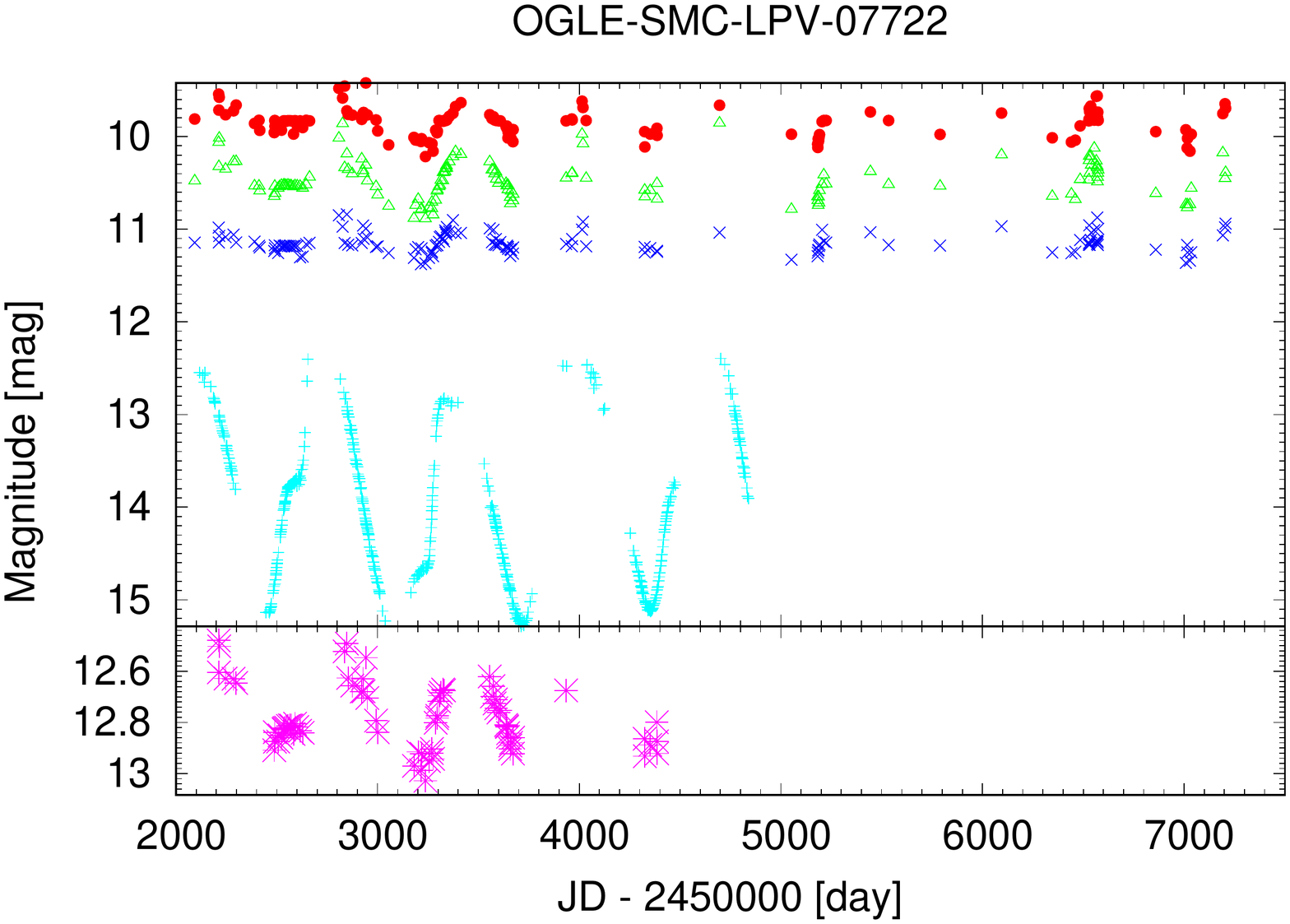}
\caption{Optical and near infrared (upper part ) and bolometric (lower part) light curves for two Mira variables in our sample as examples. {\bf Each upper part:} Cyan, blue, green, red show $I, J, H,$ and $K_{\rm s}$ brightnesses, respectively.}
\label{mbol}
\end{figure}

The time averaged bolometric magnitude and amplitude of bolometric light variation are calculated in the same way as described in the previous section. The same goes for bolometric light variation phases. In Fig.~\ref{mbol}, we show the bolometric light curves of two Miras in our sample.

Some of the key parameters calculated in this section are presented in the table~\ref{table:data} in the appendix.

\section{Discussion}
\subsection{Bolometric variability}
To determine the bolometric luminosities of variable red giants observationally is laborious because their luminosities vary with time. Usually, only limited (e.g., one epoch or a small number of photometric measurements in a few wavebands) data are available for them. In this section, we discuss the properties of bolometric variability and try to relate them to other observables. These kinds of studies would be of benefit in light curve modelling and also in future observations of Miras in more distant galaxies. Furthermore, we discuss the bolometric correction factors and their pulsation phase dependencies by taking advantage of multi-epoch data over a wide range of wavelengths.

\subsubsection{Optical and NIR amplitudes and bolometric light variation amplitudes}
\begin{figure}
\hspace*{-0.5cm}
\vspace*{-1.0cm}
\includegraphics[scale=0.34]{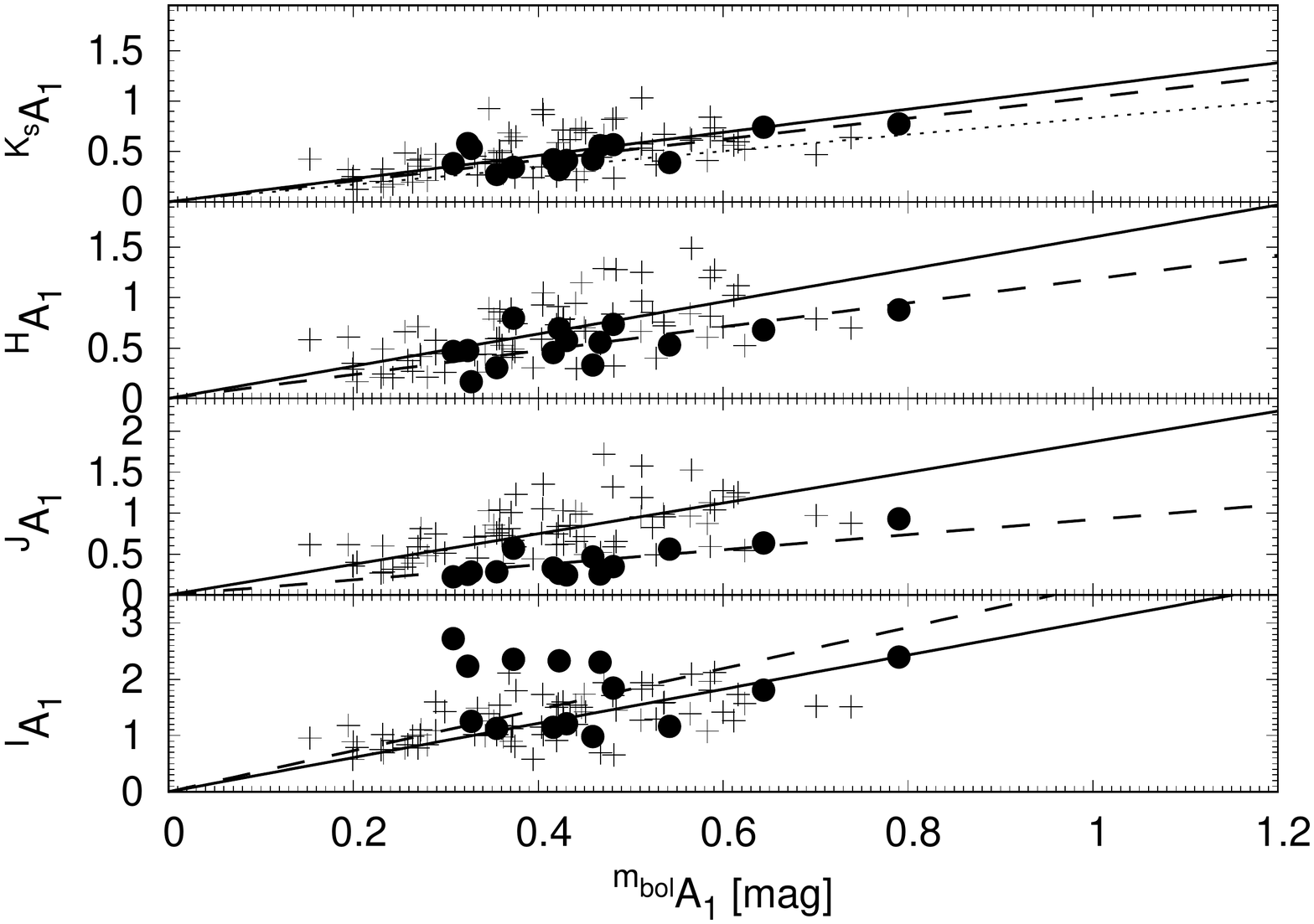} 
\hspace*{-0.5cm}
\vspace*{-1.0cm}
\includegraphics[scale=0.34]{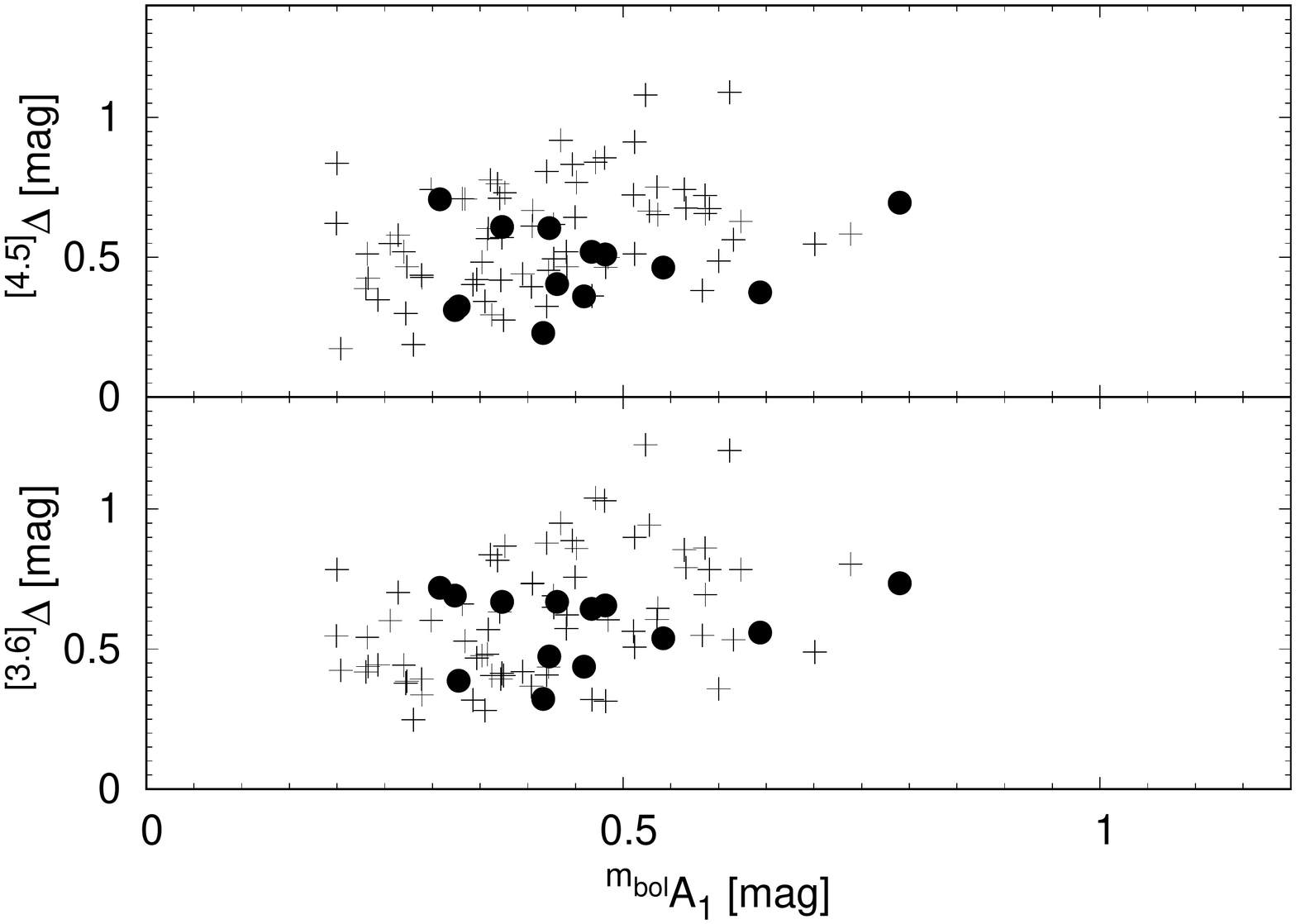}
\hspace*{-0.5cm}
\vspace*{-1.0cm}
\includegraphics[scale=0.34]{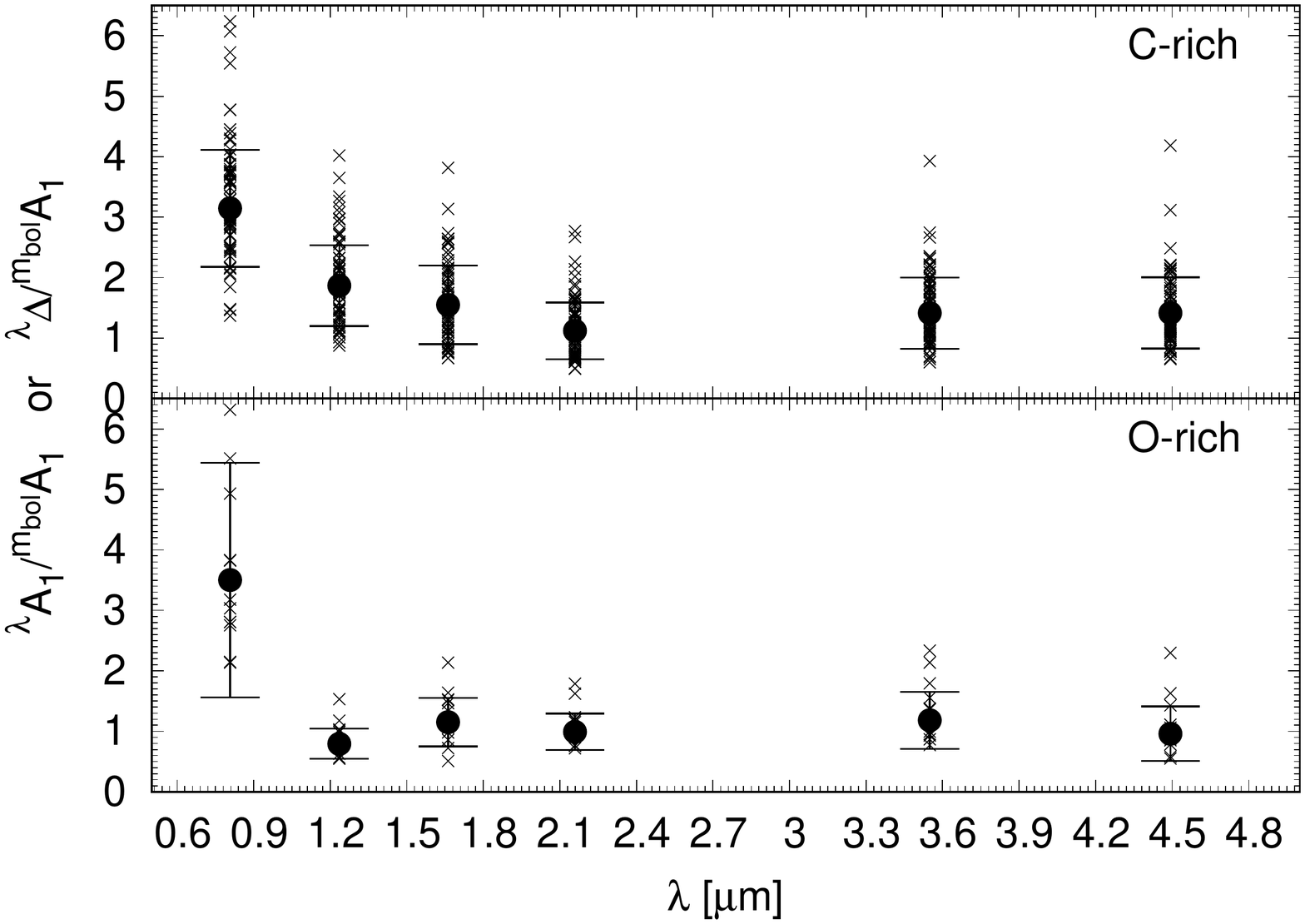}
\caption{Bolometric amplitudes are compared with the optical and NIR amplitudes. For the top and middle panels, dots and crosses show the O-rich and C-rich sample, respectively. Dashed and solid lines show the best fit linear relation of the form ${}^{\lambda}A_1 = a_\lambda \times {}^{m_{{\rm bol}}}A_1.$ for the O-rich and C-rich sample, respectively (see table~\ref{table:ampfit2}). A linear relation taken from \citet{whitelock2000} is shown by a dotted line in the ${}^{m_{{\rm bol}}}A_1$ vs. ${}^{K_{\rm s}}A_1$ plot. The bottom panel is a different representation of the top and the middle panels employing the $\lambda$ versus amplitude ratio plane for O-rich and C-rich samples, respectively. Crosses show the observed data. The black dots correspond to the median of the data, and the error bars show the standard deviation.}
\label{amplitude3}
\end{figure}

\begin{table}
  \caption{${}^{\lambda}A_1$ against ${}^{m_{{\rm bol}}}A_1$ for sample stars of the form ${}^{\lambda}A_1 = a_\lambda \times {}^{m_{{\rm bol}}}A_1.$}
  \label{table:ampfit2}
  \begin{center}
    \begin{tabular}{c r r }
    \hline
    $\lambda$ & $a_\lambda$ & $\sigma_{a_\lambda}$ \\
    \hline
    \multicolumn{3}{c}{{\bf O-rich}}\\ 
    $I$   & 3.654 & 0.435 \\
    $J$   & 0.921 & 0.068 \\
    $H$   & 1.185 & 0.092 \\
    $K_{\rm s}$ & 1.038 & 0.066 \\
    \multicolumn{3}{c}{{\bf C-rich}}\\ 
    $I$   & 3.040 & 0.099 \\
    $J$   & 1.872 & 0.074 \\
    $H$   & 1.600 & 0.068 \\
    $K_{\rm s}$ & 1.150 & 0.050 \\
    \hline
    \end{tabular}
  \end{center}
\end{table}

First, we compare the primary amplitudes of bolometric, optical and NIR light curves, and also the $Spitzer$ amplitudes in Fig.~\ref{amplitude3}. The sole extremely red star in our sample OGLE SMC-LPV-12762 was omitted from the figure and discussions in this section. For typical optical Miras in the SMC these amplitudes are not more than $\sim$1.0 mag, regardless of their surface chemistry. This means that their total luminosity change is not more than a factor of $\sim$ 2.5 (peak-to-valley). The cause of the great contrast between the large optical variation and the small bolometric variation reinforces the idea that the molecules play a critical role in the light variation, especially in the optical (see, e.g., \citealt{smak1964}, \citealt{reid2002} and discussions in section \ref{dis:amplitude}). \citet{whitelock2000} formulated a linear relation between ${}^{m_{{\rm bol}}}A$ and ${}^{K}A$ for O-rich Mira variables in the Milky Way galaxy, namely, ${}^{K}A = 0.826 \times {}^{m_{{\rm bol}}}A + 0.007$. Inspired by their work, we also fitted linear functions to the sample stars. The results of the fits are shown in Fig.~\ref{amplitude3} and their coefficients are given in Table~\ref{table:ampfit2}. Our fit result is consistent with the one in \citet{whitelock2000} within 3 sigma, despite the difference of samples, such that theirs are Galactic and ours are in the SMC. We also did not try to transform their SAAO photometry to our IRSF/SIRIUS system. Considering the overall difference of metallicity between the two samples, effects of metallicity on the amplitude ratios can be small, for at least O-rich Miras. Although we have made a fit that goes through the origin, the O-rich stars seem to have an ${}^{I}A_1$ that seems independent of the ${}^{m_{{\rm bol}}}A_1$. Similarly, in Fig.~\ref{amplitude2}, the ${}^{J}A_1$ for O-rich stars seems independent of the ${}^{I}A_1$. This suggests that the amplitudes in some bands may be dominated by the molecular spectral variations, and that these band variations may saturate (see also section~\ref{dis:amplitude}).

\subsubsection{Bolometric corrections and their phase dependencies}
The bolometric correction is a factor that relates the observed flux density at a certain wavelength with the total luminosity of a star. In fact, bolometric correction is widely used to estimate bolometric luminosities of stars and then compare them with stellar evolutionary models. Because of its usefulness and importance, many authors (e.g., \citealt{whitelock2000}, \citealt{kerschbaum2010}, and references therein) derived bolometric corrections using different sets of data. Here we calculate the bolometric corrections using a new data set and study whether the bolometric correction depends on the pulsation phase.

\begin{figure}
\hspace*{-0.5cm}
\includegraphics[scale=0.34]{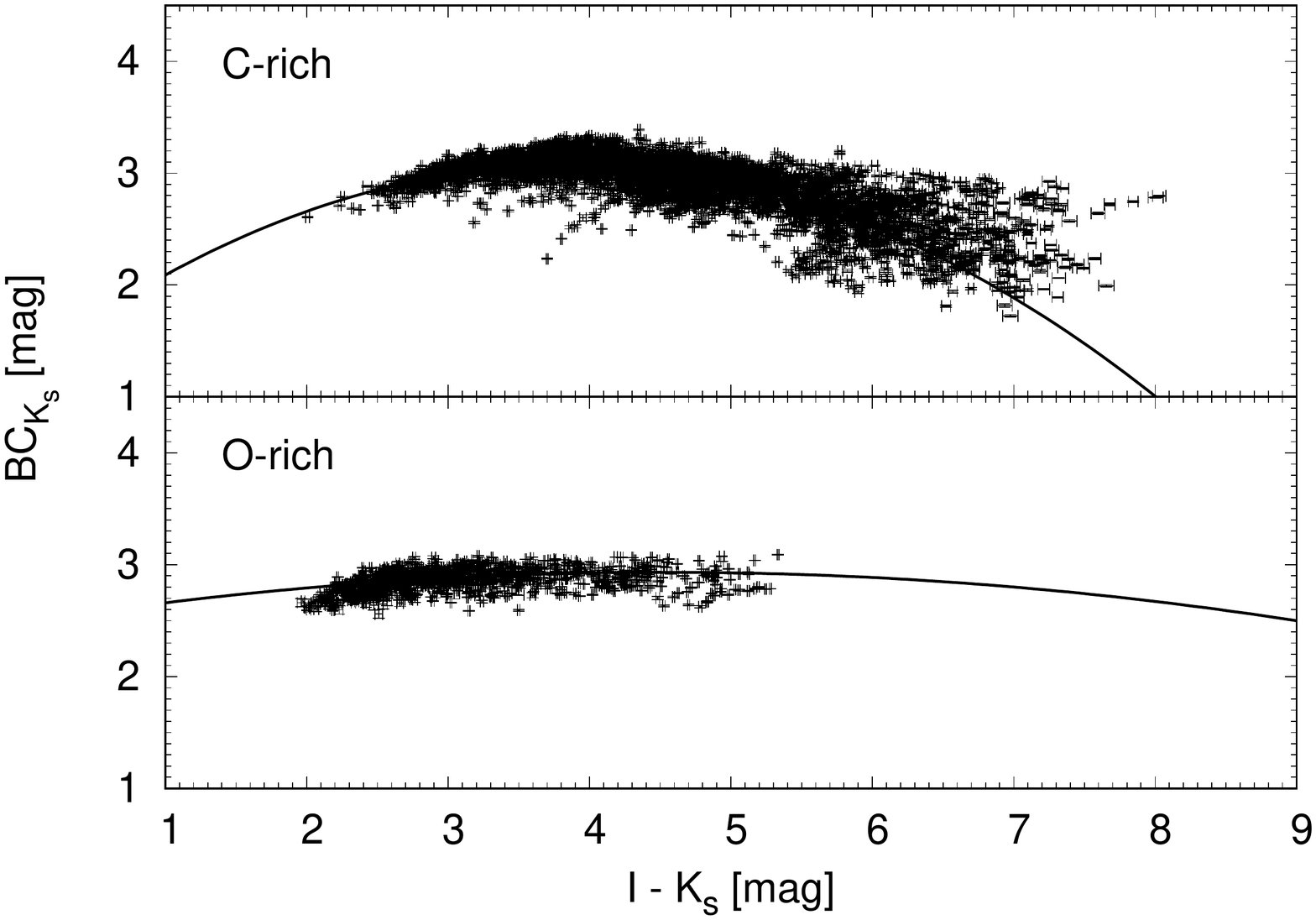} 
\hspace*{-0.5cm}
\includegraphics[scale=0.34]{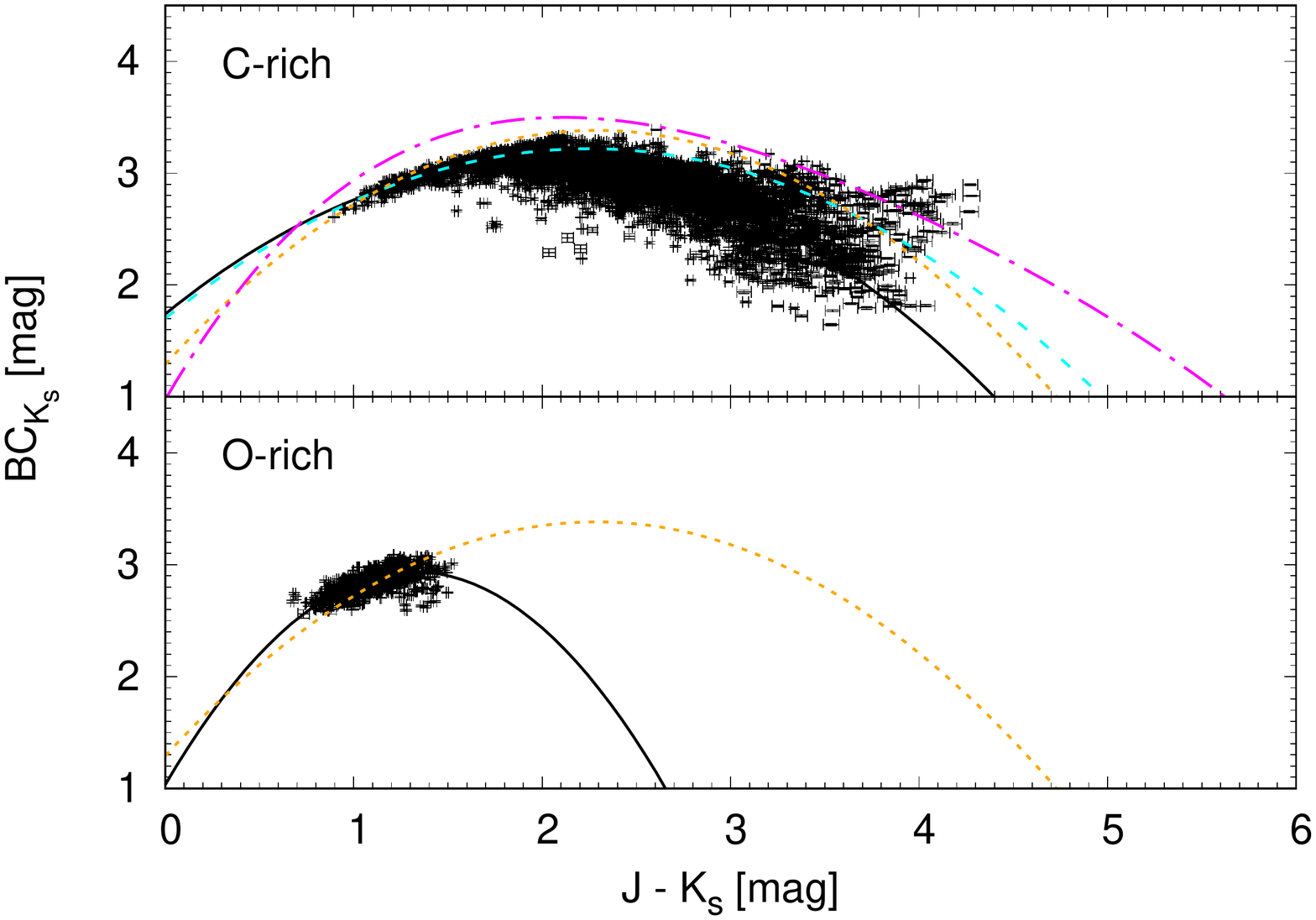} 
\caption{Bolometric corrections at $K_{\rm s}$ as a function of $I-K_{\rm s}$ and $J-K_{\rm s}$. The solid (black) line is the best-fitting least-squares second-order polynomial to the data and the dashed (cyan), dotted (orange), and dot-dashed (magenta) lines are the bolometric corrections derived by \citet{kerschbaum2010}, \citet{riebel2012}, and \citet{whitelock2006}, respectively.}
\label{bolallphase}
\end{figure}

\begin{figure}
\hspace*{-0.5cm}
\includegraphics[scale=0.34]{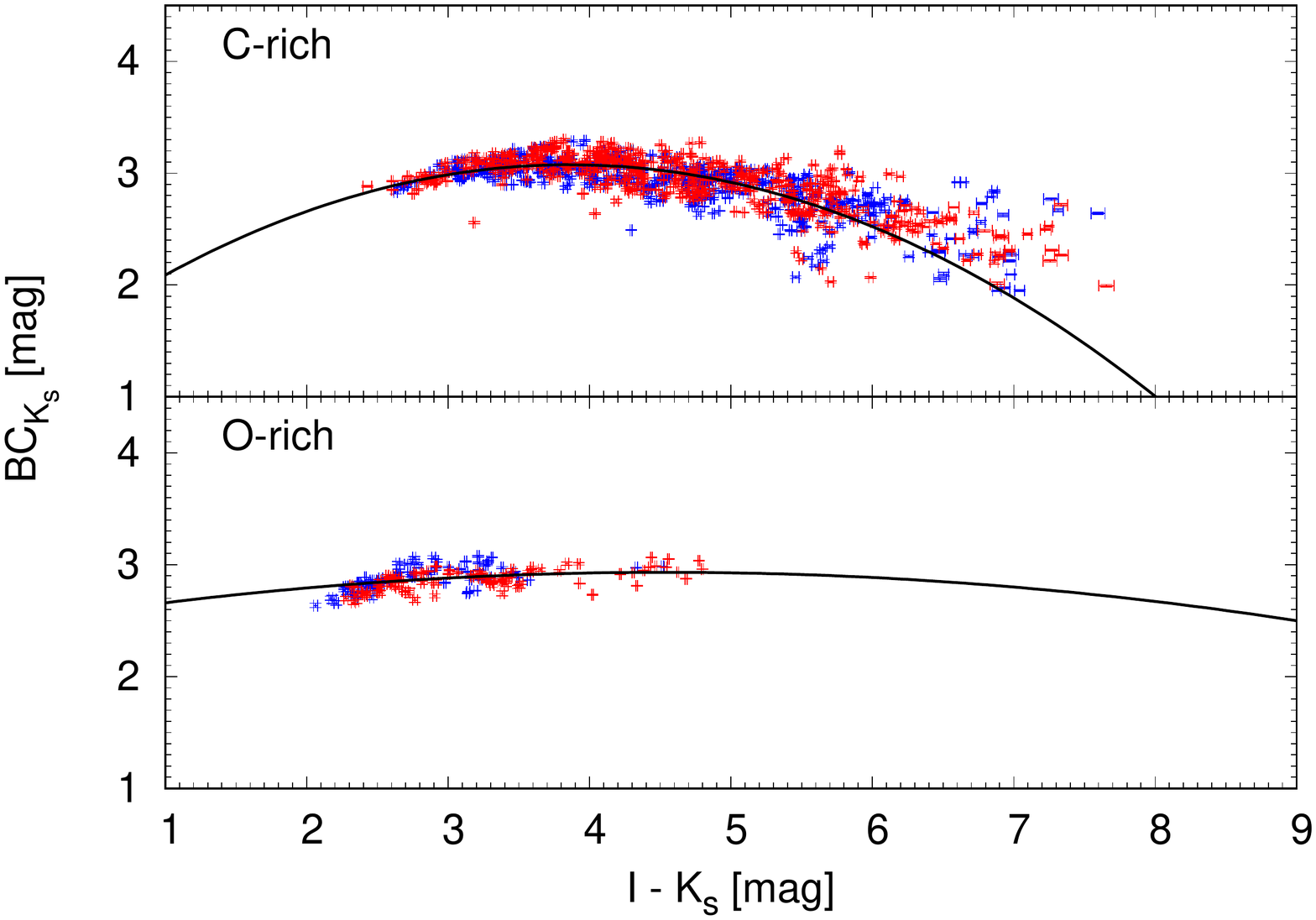} 
\hspace*{-0.5cm}
\includegraphics[scale=0.34]{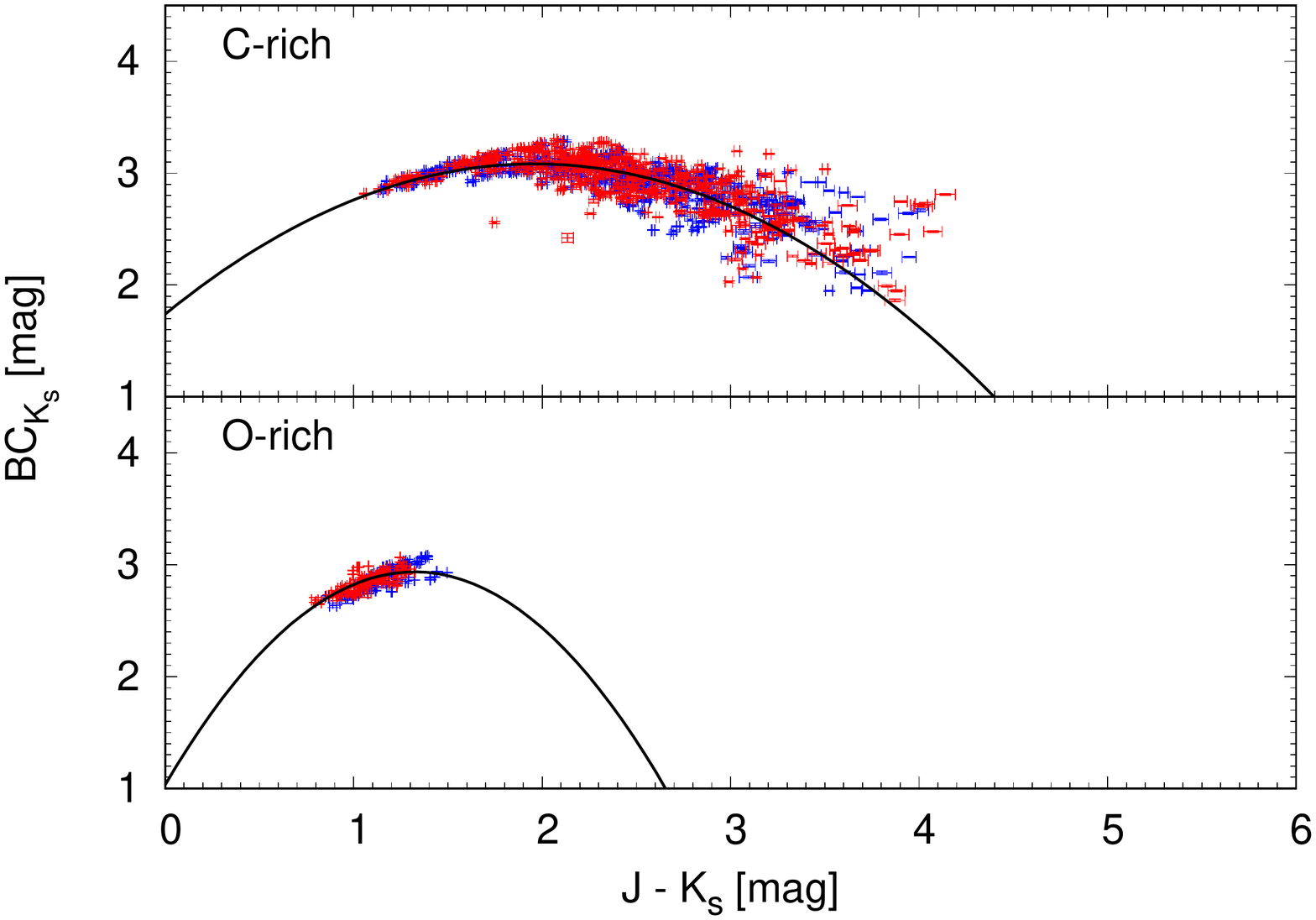} 
\caption{Almost the same as Fig.~\ref{bolallphase}, but shows bolometric corrections calculated at the bolometricly brightest (blue points) and dimmest (red points) pulsation phases of the primary period. The solid (black) line is the relation determined in Fig.~\ref{bolallphase}.}
\label{bolphase}
\end{figure}

Fig.~\ref{bolallphase} shows the bolometric correction to the $K_{\rm s}$ magnitude as a function of $I - K_{\rm s}$ or $J - K_{\rm s}$ colours, calculated by using each individual set of time series data. The time dependent bolometric correction should appear as a sequence or a clump on the diagram for each sample star. The solid (black) lines are the best-fitting least-squares second-order polynomial to the data. Only data with a colour error of less than 0.05 mag were used. Bolometric correction relations given in \citet{kerschbaum2010}, \citet{riebel2012}, and \citet{whitelock2006} are shown by dashed (cyan), dotted (orange), and dot-dashed (magenta) lines, respectively. No corrections were applied to account for the difference in photometric systems used in the literature or of the method used to determine bolometric magnitudes. The sampled stars are different among the four studies (SMC red giants for this work, Galactic C-rich stars for \citealt{whitelock2006} and \citealt{kerschbaum2010}, and LMC red giants for \citealt{riebel2012}), which makes the direct comparison of the bolometric correction difficult. Our bolometric magnitudes are calculated maximumly utilising the $I, J, H, K_{\rm s}, [3.6],$ and $[4.5]$ multi-epoch data, presumably covering the peak of the SED of optical red giants, as well as using the $Spitzer$ single epoch data at longer wavelengths. We provide, in Table~\ref{table:bolcol}, the coefficients of the fits of the form
\begin{equation}
\label{eq:bolcol}
BC_{K_{\rm s}} = m_{{\rm bol}} - K_{\rm s} = c_0 + c_1 \times {\rm colour} + c_2 \times {\rm colour}^2,
\end{equation}
where the colour means the employed colour. Our fit result seems to agree well with that of \citet{kerschbaum2010} at bluer colours. Nevertheless, the relatively large deviations from the fitted relation at redder colours suggest that the underlying model assumption (i.e., second-order polynomial) is not good. Our experiments showed that increasing the order of polynomial did not reduce the scatter, which is probably a manifestation of the intrinsic properties of Miras. 



\begin{table}
  \caption{Coefficients for calculating the bolometric correction at the $K_{\rm s}$ band as a function of colour according to Equation~\ref{eq:bolcol} with the colour range used for the fit.}
  \label{table:bolcol}

  \begin{center}
    \begin{tabular}{crrrrr}
	\hline
 Employed & $c_0$ & $c_1$ & $c_2$ & min & max \\
 colour & \multicolumn{3}{c}{} & \multicolumn{2}{c}{[mag]} \\
	\hline
 \multicolumn{6}{c}{{\bf O-rich}} \\	        
 {$J - H$}         & 2.981 & $-0.268$ & 0.163     & 0.114 & 1.158 \\
 {$H - K_{\rm s}$} & 2.570 & 0.934    & $-0.543$  & 0.008 & 0.984 \\
 {$J - K_{\rm s}$} & 1.032 & 2.881    & $-1.090$  & 0.673 & 1.525 \\
 {$I - K_{\rm s}$} & 2.483 & 0.198    & $-0.022$  & 1.959 & 5.332 \\
	\hline
 \multicolumn{6}{c}{{\bf C-rich}} \\	        
 {$J - H$}         & 1.797 & 2.246 & $-1.008$  & 0.388 & 2.462 \\
 {$H - K_{\rm s}$} & 2.609 & 1.174 & $-0.768$  & 0.141 & 2.532 \\
 {$J - K_{\rm s}$} & 1.740 & 1.372 & $-0.350$  & 0.895 & 4.278 \\
 {$I - K_{\rm s}$} & 1.278 & 0.932 & $-0.121$  & 2.007 & 8.028 \\
 \hline
    \end{tabular}
  \end{center}
\end{table}

The bolometric correction depends on effective temperature, gravity, and chemical composition. Our dataset is not large enough to show their individual impacts but good enough to see if the bolometric correction relations change significantly with the pulsation phase. We calculated bolometric correction relations at around the two extreme pulsation phases associated with the primary period, namely, at $\pm 0.05$ phase with respect to the bolometricly brightest and dimmest phases. Figure~\ref{bolphase} suggests that while the bolometric correction of an individual star does change with the pulsation phase, there seems no significant difference in the bolometric correction relations within the colour range that the observational data cover. 

\subsection{Miscellaneous results related to stellar pulsation}
\subsubsection{Pulsation amplitude}
\label{dis:amplitude}

\begin{table}
	\caption{Main absorbing molecules in each photometric band for O-rich and C-rich red giants.}
  \label{table:molecule}
  \begin{center}
    \begin{tabular}{cll}
	\hline
   Band & O-rich & C-rich \\         
	\hline
   $I$   & VO, TiO & C$_2$, CN \\
   $J$   & VO, TiO, H$_2$O & C$_2$, CN \\
   $H$   & CO, H$_2$O & C$_2$, CN, CO \\
   $K_{\rm s}$ & CO, H$_2$O & C$_2$, CN, CO \\
	\hline
    \end{tabular}
  \end{center}
\end{table}

\begin{figure}
\hspace*{-0.5cm}
\includegraphics[scale=0.34]{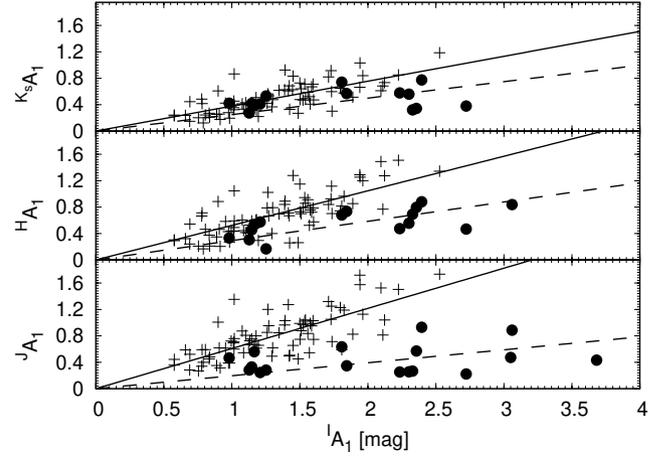} 
\caption{A comparison of ${}^{\lambda}A_1$. Differences in the shapes of the symbols represent surface chemistry, dots for O-rich and pluses for C-rich Miras. The dashed and solid lines are the best fits of a linear relation of the form ${}^{\lambda}A_1 = a_{\lambda} \times {}^{I}A_1$ to the O-rich and C-rich sample stars, respectively. The results of the fits are summarised in Table~\ref{table:ampfit1}.}
\label{amplitude2}
\end{figure}

\begin{table}
  \caption{${}^{\lambda}A_1$ against ${}^{I}A_1$ for sample stars of the form ${}^{\lambda}A_1 = a_{\lambda} \times {}^{I}A_1$.}
  \label{table:ampfit1}
  \begin{center}
    \begin{tabular}{c r r }
    \hline
    $\lambda$ & $a_\lambda$ & $\sigma_{a_\lambda}$ \\
    \hline
	\multicolumn{3}{c}{{\bf O-rich}}\\ 
    $J$   & 0.196 & 0.026 \\
    $H$   & 0.292 & 0.022 \\
    $K_{\rm s}$ & 0.249 & 0.027 \\
	\multicolumn{3}{c}{{\bf C-rich}}\\ 
    $J$   & 0.608 & 0.019 \\
    $H$   & 0.523 & 0.017 \\
    $K_{\rm s}$ & 0.378 & 0.012 \\
    \hline
    \end{tabular}
  \end{center}
\end{table}

It is known that the optical and NIR brightnesses of red giants vary with the same pulsation period, but often with noticeably different phases. Their optical amplitudes are much greater than those of infrared or bolometric brightnesses (e.g., \citealt{reid2002}). These phenomena can be understood as a consequence of time variation in molecular opacity associated with temperature changes. The main absorbing molecules in the optical and NIR wavebands for O-rich and C-rich red giants (\citealt{lancon2000}) are listed in Table~\ref{table:molecule}. The temperature sensitivities of their absorbing effects differ greatly from one to another. Table~\ref{table:molecule} indicates that our multi-band ($I, J, H,$ and $K_{\rm s}$) multi-epoch data are an ideal tool for showing how the chief molecules in each waveband play roles in the observed light variations. 

First, in Fig.~\ref{amplitude2} we compare the optical and NIR primary amplitudes. While the optical and NIR primary amplitudes do not need to be exclusively linearly related, we fitted linear relations between them by least squares for comparison purposes. The dotted and solid lines in the figure are the best fits to the O-rich and C-rich sample stars, respectively. The coefficients of the best fit linear relations are given in Table~\ref{table:ampfit1}. We see a marginal trend that the slopes of the linear relations do not change much with increasing wavelength for O-rich sample stars, while the slopes decrease with increasing wavelength for C-rich stars. Also, Fig.~\ref{amplitude2} indicates that C-rich stars have systematically larger amplitudes in the NIR than O-rich ones if compared at a fixed ${}^{I}A_1$. Indeed, the 1.6 $\mu$m amplitude was used by \citet{huang2018} to separate C- and O-rich Miras in NGC 4258. These results may be understood in terms of the difference in prevailing molecular species between O-rich and C-rich stars. In C-rich stars, the main absorbing molecules at the wavelengths of interest are C$_2$ and CN (and CO). The systematic impacts of the same molecules on various photometric bands lead to the tight correlation of the amplitudes in the optical and NIR bands. In contrast, the main absorbing molecules in the optical are different from those in the NIR for O-rich stars. The TiO molecule, which gives strong absorption over the entire optical region, is a temperature sensitive molecule. On the other hand, the NIR bands are more affected by CO and H$_2$O. This fact makes the $I-$band amplitude very large, even if the NIR amplitude is relatively small (see also Figs.~\ref{OGLE-SMC-LPV-14462}, ~\ref{acrelation}).

\begin{figure}
\hspace*{-1.0cm}
\includegraphics[scale=0.48,angle=0]{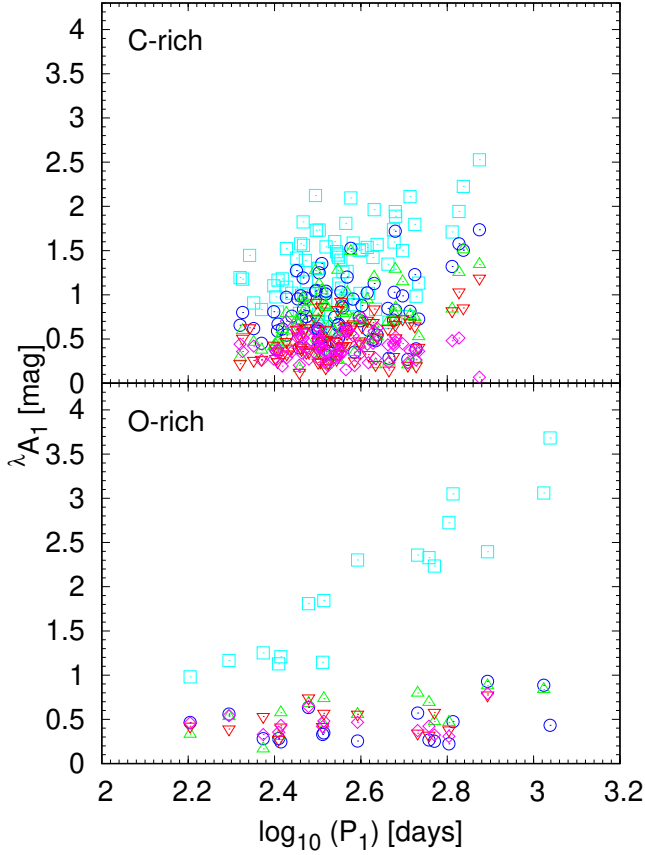} 
\caption{The ${}^{\lambda}A_1$ vs. log$P_1$ relation for the sample stars. Cyan, blue, green, red, and magenta points denote $I, J, H, K_{\rm s},$ and $m_{{\rm bol}}$ data, respectively.}
\label{aprelation}
\end{figure}

\begin{figure}
\hspace*{-0.5cm}
\includegraphics[scale=0.48,angle=0]{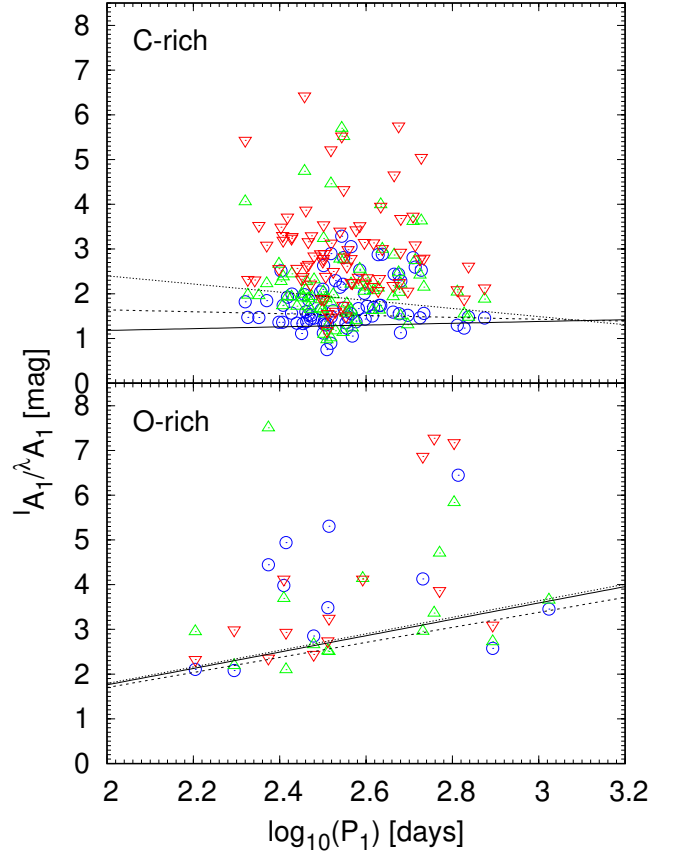} 
\caption{Almost the same as Fig.~\ref{aprelation}, but this figure employs amplitude ratios for the vertical axis. The solid, dashed, and dotted lines are the best fit results given by \citet{yuan2018} for $\lambda$ equals $J, H,$ and $K_{\rm s}$, respectively.}
\label{aprelation_yuan}
\end{figure}

\begin{figure}
\hspace*{-1.0cm}
\includegraphics[scale=0.48,angle=0]{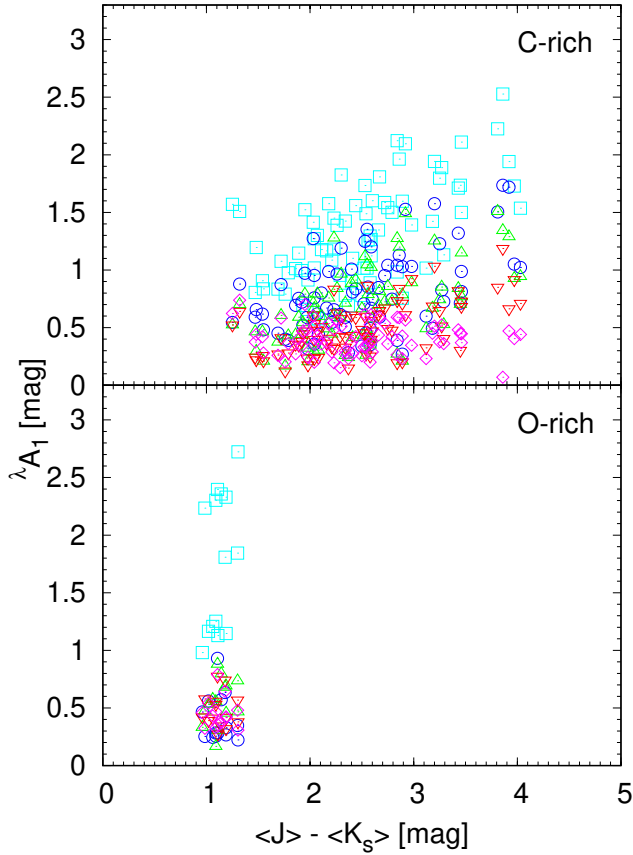} 
\caption{The ${}^{\lambda}A_1$ vs. $\langle J \rangle - \langle K_{\rm s} \rangle$ relation for Miras. Meanings of the symbols are the same as in Fig.~\ref{aprelation}}
\label{acrelation}
\end{figure}

Next, we explore possible correlations of optical, NIR, and bolometric amplitudes with period or colour. Fig.~\ref{aprelation} is the period-amplitude relation of the sample stars. There is a weak trend whereby their amplitudes increase with increasing period. \citet{yuan2018} give $I$-to-$JHK_{\rm s}$ amplitude ratio vs. period relations for Miras in the LMC that were derived with three epoch NIR data. A comparison of their fits and our data is made in Fig.~\ref{aprelation_yuan}. Our results are not consistent with those of \citet{yuan2018}. It is not clear if this is due to the difference in the samples, such that theirs are in the LMC and ours are in the SMC, but our light-curves are much better defined as our temporal coverage is much greater.

The NIR colour-amplitude relation of Miras is shown in Fig.~\ref{acrelation}. We chose the $\langle J \rangle - \langle K_{\rm s} \rangle$ colour for the abscissa. There is a trend such that the optical and NIR amplitudes grow larger as $\langle J \rangle - \langle K_{\rm s} \rangle$ colours become redder. Surely for O-rich stars no trend can be seen - there is insufficient $\langle J \rangle - \langle K_{\rm s} \rangle$ range. The bolometric amplitudes of Miras stay almost constant (${}^{m_{{\rm bol}}}A_1 \sim 0.4$ mag) over the observed colour range, although the bolometric amplitudes for red (say, $\langle J \rangle - \langle K_{\rm s} \rangle > \sim 3.0$ mag) C-rich Miras could have been underestimated thanks to the lack of multi-epoch data at longer wavelengths. \citet{matsunaga2005} compared the colour-amplitude ($\Delta I$) relations of Miras in the Bulge, the LMC, and the SMC, and pointed out that O-rich and C-rich Miras occupy different regions on the colour-amplitude diagram. A contrasting behaviour of ${}^{I}A_1$ between the O-rich and C-rich Miras is also seen in Fig.~\ref{acrelation} such that ${}^{I}A_1$ seems to be independent of colour for O-rich stars. This probably reflects the temperature-dependent absorption of responsible molecules, such that TiO in O-rich Miras is quite temperature sensitive, but C$_2$ and CN in C-rich Miras are not particularly sensitive to the temperature variation (e.g., \citealt{loidl2001}).

\subsubsection{Period-magnitude relation}
\begin{figure}
\hspace*{-1.0cm}
\includegraphics[scale=0.34]{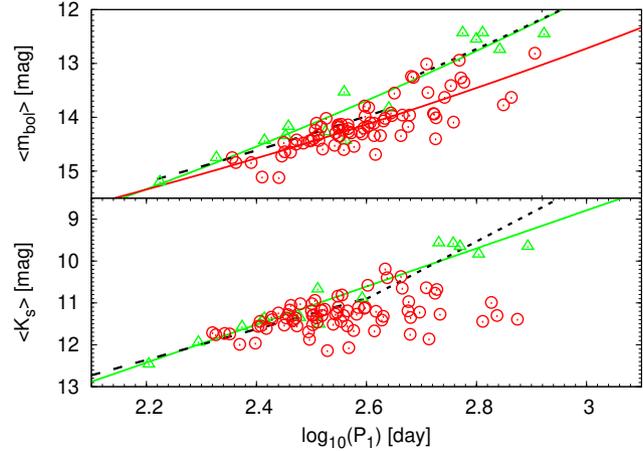} 
\caption{The log$P_1$-$\langle K_{\rm s} \rangle$ and log$P_1$-$\langle m_{{\rm bol}} \rangle$ relations for the sample stars. Circles and triangles denote C-rich and O-rich surface chemistry, respectively. See text for the meanings of the lines.}
\label{plrelation}
\end{figure}

\begin{table}
  \caption{Coefficients of $P_1$-$\langle K_{\rm s} \rangle$ and $P_1$-$\langle m_{{\rm bol}} \rangle$ relations for Miras of the form $\langle m \rangle = a \times \left( \log_{10}(P_1 {\rm [day]}) - 2.3 \right) + b$, where $\langle m \rangle$ is either $\langle K_{\rm s} \rangle$ or $\langle m_{{\rm bol}} \rangle$. The RMSE is the standard deviation of residuals.}
  \label{table:pl}
  \begin{center}
    \begin{tabular}{c r r r}
	\hline
     & $\langle K_{\rm s} \rangle$ & \multicolumn{2}{c}{$\langle m_{{\rm bol}} \rangle$} \\
     & O-rich  & O-rich & C-rich  \\
	\hline
 $a$ & -4.543$\pm$0.379 & -4.462$\pm$0.399 & -3.280$\pm$0.282 \\ 
 $b$ & 11.972$\pm$0.123 & 14.814$\pm$0.129 & 14.958$\pm$0.077 \\ 
RMSE & 0.29             & 0.30             & 0.28 \\
	\hline
    \end{tabular}
  \end{center}
\end{table}

The period-magnitude relation of Miras was discovered by \citet{glass1981} based on observations of Miras in the Large Magellanic Cloud. Since the relation is a useful tool in deriving distances to Miras and their parent populations such as clusters and galaxies, many astronomers have observed Mira variables in the Magellanic Clouds to calibrate and establish the relation. However, it is known that the period-magnitude relation breaks down for Miras suffering from significant circumstellar extinction. \citet{whitelock2012} pointed out that this occurs at very long periods for O-rich Miras, but over quite a large range of periods for C-rich stars (see also figs.~2 and 3 in \citealt{ita2011}). Circumstellar reddening of C-rich Miras has also been found in Local Group dwarf spheroidals (e.g. \citealt{menzies2010}, \citealt{whitelock2009}). An alternative solution is to use a period-luminosity relation that instead employs the bolometric magnitude (\citealt{feast1989}). By definition, the period-bolometric magnitude relation is independent of circumstellar extinction, and above all, the bolometric magnitude can be easily compared to model calculations especially for Miras in the Magellanic Clouds where their distances are reasonably well known.

Fig.~\ref{plrelation} shows the period-$\langle K_{\rm s} \rangle$ and period-$\langle m_{{\rm bol}} \rangle$ relations of the sample stars. Several lines are drawn. Solid green and red lines are the best fit of a linear relation to the O-rich and C-rich samples, respectively, and the results of the fits are summarised in Table~\ref{table:pl}. The black dashed and dotted lines are relations derived by \citet{ita2011} for O-rich Miras in the LMC with periods shorter or longer than about 400 days, respectively. Here we shifted their relations by 0.49 mag to compensate for the distance modulus offset between the LMC and the SMC. C-rich stars barely fall on a linear relation on the period-$\langle K_{\rm s} \rangle$ magnitude plane due probably to the circumstellar extinction. In this context, it seems likely that our sample does not contain O-rich Miras suffering from severe circumstellar extinction. Therefore, we fitted linear relations by least squares to derive a period-$\langle K_{\rm s} \rangle$ relation for O-rich Miras only. The calculated slope of the period-$\langle m_{{\rm bol}} \rangle$ relation for O-rich Miras in the SMC agrees within the errors with that of O-rich Miras with a period longer than 400 days in the LMC. Although we employed flux mean magnitudes, the observed dispersion about the adopted relations is large, compared to the LMC relations (e.g., \citealt{ita2011}), which are based on single-epoch photometry. The complex three dimensional structure of the SMC (e.g., \citealt{subramanian2012}), the intrinsic dispersion and the artificial broadening by the data analysis must all contribute to the scatter.

\subsubsection{Period-colour relations}
\begin{figure}
\hspace*{-1.0cm}
\includegraphics[scale=0.34]{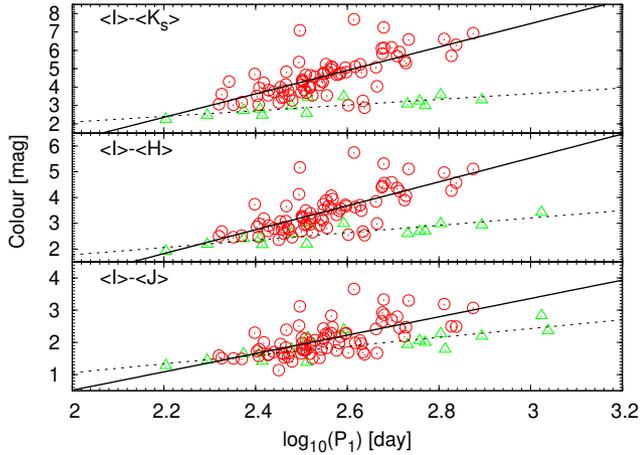} 
\caption{Period-colour relations for the sample stars, employing optical and NIR combined colours. Circles and triangles denote C-rich and O-rich surface chemistry, respectively. The solid and dotted lines are the least squares linear fits to the C-rich and O-rich samples, respectively.}
\label{periodcolor}
\end{figure}

\begin{figure}
\hspace*{-1.0cm}
\includegraphics[scale=0.34]{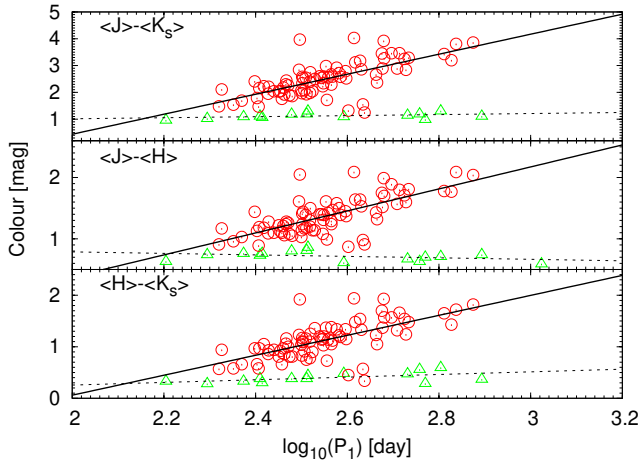} 
\caption{Almost the same as in Fig.~\ref{periodcolor}, but employing NIR colours.}
\label{periodcolor2}
\end{figure}

\begin{table*}
  \caption{Coefficients of period-colour relations for Miras of the form Colour $= a \times \log_{10}({\rm Period [day]}) + b$. The RMSE is the standard deviation of residuals.}
  \label{table:pcr}
  \begin{center}
    \begin{tabular}{c r r r r r r}
	\hline
	 & $\langle I \rangle - \langle J \rangle$  & $\langle I \rangle - \langle H \rangle$ & $\langle I \rangle - \langle K_{\rm s} \rangle$ & $\langle H \rangle - \langle K_{\rm s} \rangle$ & $\langle J \rangle - \langle H \rangle$ & $\langle J \rangle - \langle K_{\rm s} \rangle$ \\
	\hline
	\multicolumn{7}{c}{{\bf O-rich}} \\
	$a$ &  1.369$\pm$0.254 &  1.434$\pm$0.271 &   1.543$\pm$0.375 &  0.257$\pm$0.113 & $-0.121\pm$0.086 &  0.196$\pm$0.131 \\ 
	$b$ & -1.678$\pm$0.670 & -1.096$\pm$0.703 &  -0.989$\pm$0.960 & -0.258$\pm$0.291 &  1.029$\pm$0.223 &  0.623$\pm$0.336 \\ 
  RMSE  &  0.255                    & 0.239                     & 0.283                      & 0.086                     & 0.075                     & 0.099 \\
	\hline
    \multicolumn{7}{c}{{\bf C-rich}} \\
	$a$ &  2.850$\pm$0.393 &  4.644$\pm$0.551 &   6.375$\pm$0.764  &  1.938$\pm$0.265 &  1.794$\pm$0.214 &  3.732$\pm$0.465 \\ 
	$b$ & -5.185$\pm$1.003 & -8.394$\pm$1.408 & -11.669$\pm$1.951 & -3.817$\pm$0.677 & -3.209$\pm$0.548 & -7.026$\pm$1.188 \\ 
  RMSE  &  0.404                    & 0.567                      & 0.803                      & 0.273                     & 0.221                     & 0.479 \\
  	\hline
	\end{tabular}
  \end{center}
\end{table*}

Some types of variable star show tight period-colour relations. Such a relation would tell us the intrinsic colour of the variable star if one gets its pulsation period. The knowledge of intrinsic colour of a star is important not only for stellar astrophysics but also for estimating the amount of interstellar extinction to the star and also of the circumstellar extinction around the star, if any. Mira variables discussed in this work are one such type, but they often suffer from circumstellar extinction due to mass-loss in addition to interstellar extinction. Attempts to understand the circumstellar extinction law have been made (e.g., \citealt{ita2011}), but it is still difficult to tell the amount of circumstellar extinction. In this paper we limit our study to relating the observed colours (corrected for the interstellar extinction) of Miras with their periods. We found relatively tight period-colour relations for the sample stars, as in Figs.~\ref{periodcolor} and \ref{periodcolor2}, using flux-mean colours.

The difference in surface chemistry has a significant effect on the relations. Both optical-NIR and NIR-only colours of the O-rich Miras show a very weak dependency on their periods. It is reasonable to suppose that our O-rich Mira samples do not suffer much from circumstellar extinction (see also Fig.~\ref{plrelation}). Then, their colours should closely represent their intrinsic colours. Even if there is some amount of dust around the O-rich Mira stars, O-rich circumstellar dust is transparent in the NIR (see, e.g., \citealt{woitke2006}), and their NIR colours should not be affected much. C-rich Miras have very steep period-colour relations for all colours employed. In the previous section, we saw that the colours of C-rich Miras suffer substantially from circumstellar extinction. This is because their circumstellar dust is opaque in both the optical and NIR, in contrast to the O-rich situation. Based on the fact that C-rich Miras fall on tight period-colour relations, it is likely that their mass loss rate and pulsation period are related such that the mass loss rate is larger for longer periods. In fact, both \citet{whitelock1994} and \citet{wood2007} pointed out that there is a very tight correlation between the mass loss rate and pulsation period, especially for Miras with pulsation periods longer than 500 days. Our observed period-colour relations for C-rich Miras indicate that even short period ones have a tight correlation between their mass loss rate and period. We have fitted linear relations by least squares to derive the period-colour relations of Miras and the results are summarised in Table~\ref{table:pcr}.

Three supposed C-rich stars appear to lie close to the O-rich lines in all diagrams in Figs.~\ref{periodcolor} and \ref{periodcolor2}, particularly noticeably in Fig.~\ref{periodcolor2}. These stars may have been misclassified, a possibility raised in section~\ref{sec_surface} earlier. Two of the stars, OGLE SMC-LPV-07597 and 07488, have been classified as SC stars (\citealt{tle1985}), and the former as a possible symbiotic star (\citealt{munari2002}), which may have affected the colour-based spectral types adopted by OGLE. The third star, OGLE SMC-LPV-09546, has generally been classed as type C, but is otherwise not unusual.

\subsection{Phase lags in the light curves}

\begin{figure*}
\hspace*{-1.0cm}
\includegraphics[scale=0.63]{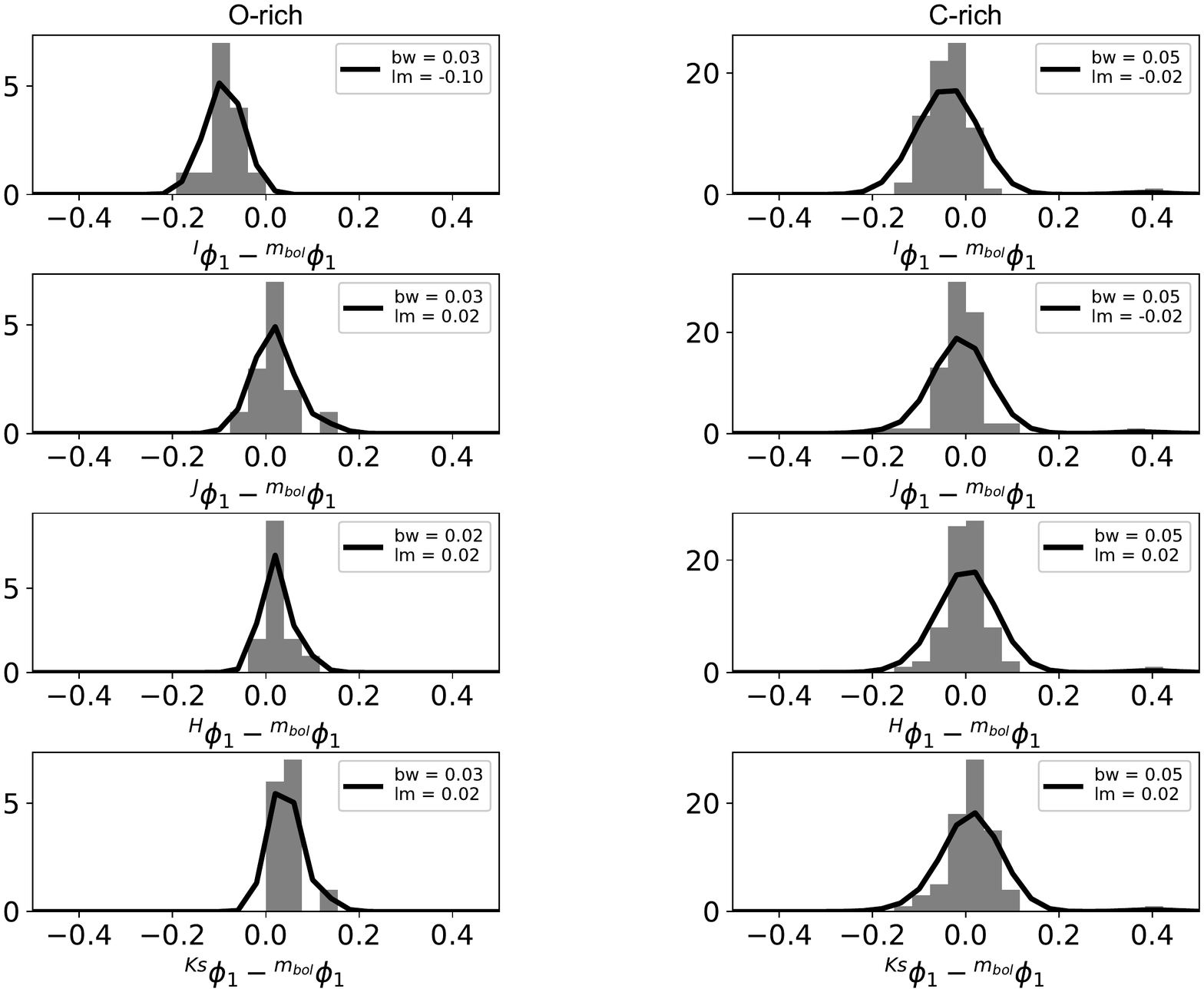} 
\caption{Histograms of phase lag against bolometric phase for O-rich (left column) and C-rich (right column) Miras. The black solid lines indicate the distribution profile estimated by kernel density estimation with a Gaussian kernel. The bandwidth of the kernel (bw) and the location of the maximum peak (lm) for the model with the best score are indicated.}
\label{phaselag1}
\end{figure*}

\citet{smith2006} studied phase lags\footnote{It should be noted that the definition of the initial phase is different between theirs and ours, such that they consider the phase of maxima and ours is anchored to the HJD of 2450000.0 (see section 3.1). As long as one accepts that the shape of Mira's light curve can be approximated by a sinusoid and it does not change much with wavebands, the difference in the initial phase definition should not matter very much.} in the optical-infrared light curves of Galactic red giants. Their final sample consists of 21 sources, of which 16 are Miras and 5 are semi-regular variables. Among the 16 Miras, 2 are S-type (carbon to oxygen number ratio is about 1), 13 are O-rich, and 1 is C-rich. Here we use a larger sample to investigate the phase lags among Miras, trying to provide observational results crucial for constraining dynamical atmosphere models (e.g., \citealt{bessell1996} for O-rich stars, \citealt{hofner2003} for C-rich stars). 

The phase lag can be evaluated by using the light curve fitting results for each star. In the following discussion, the phase means the one that is associated with the primary period, i.e., ${}^{\lambda}\phi_1$. Fig.~\ref{phaselag1} shows histograms of the phase difference between the optical or NIR phases and the bolometric phase for the sample stars. By definition, the phase is {\bf advanced} if the difference is negative. The width of the phase bins is chosen to be 0.04 for both O- and C-rich samples. We used the kernel density estimation (KDE) method to estimate the shape of the original phase lag probability distribution from the observed histogram. The idea of the KDE is that the shape of the original probability distribution can be expressed as a  superposition of the kernel functions. We employed a Gaussian kernel and used the cross-validation approach to find the optimal bandwidth $bw$ of the kernel. The $lm$ denotes the position of the maximum probability density of the original probability distribution estimated in this way. The estimated original probability distribution with the best score is shown in the figure with a thick solid line. We can see a general trend of phase lag for O-rich Miras. Their optical ($I-$band) phase {\bf precedes} the bolometric phase by $\sim$10\% of the primary pulsation period, while their NIR phase {\bf follows} the bolometric phase. The phase difference between $I-$ and $K_{\rm s}-$band is significant for O-rich Miras, as large as about 12\%. This result for O-rich Miras is consistent with that of \citet{pettit1933}. On the other hand, we do not see such a significant phase lag among the C-rich Miras. The scatter about phase synchronisation is comparable for both the O- and C-rich samples, but it is possible that the relatively large scatter about phase synchronisation is merely an expression of the fact that the light curves of some Miras in our sample are complicated and the cadence of their bolometric data is not good enough for this kind of analysis.

\subsection{NIR colour variation}
\subsubsection{Phase inversion}

\begin{figure}
\hspace*{-0.5cm}
\includegraphics[scale=0.33]{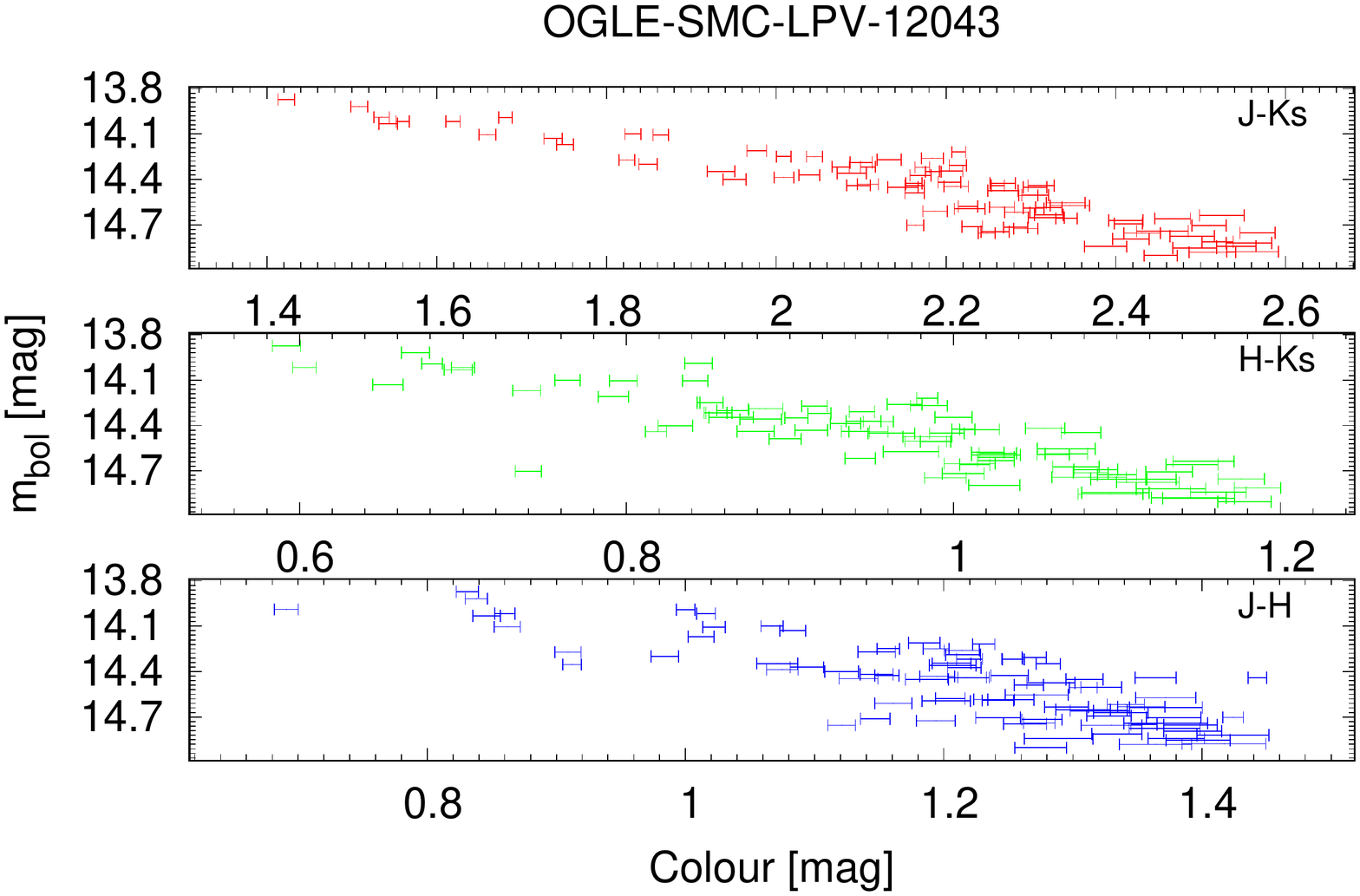}  
\hspace*{-0.5cm}
\includegraphics[scale=0.33]{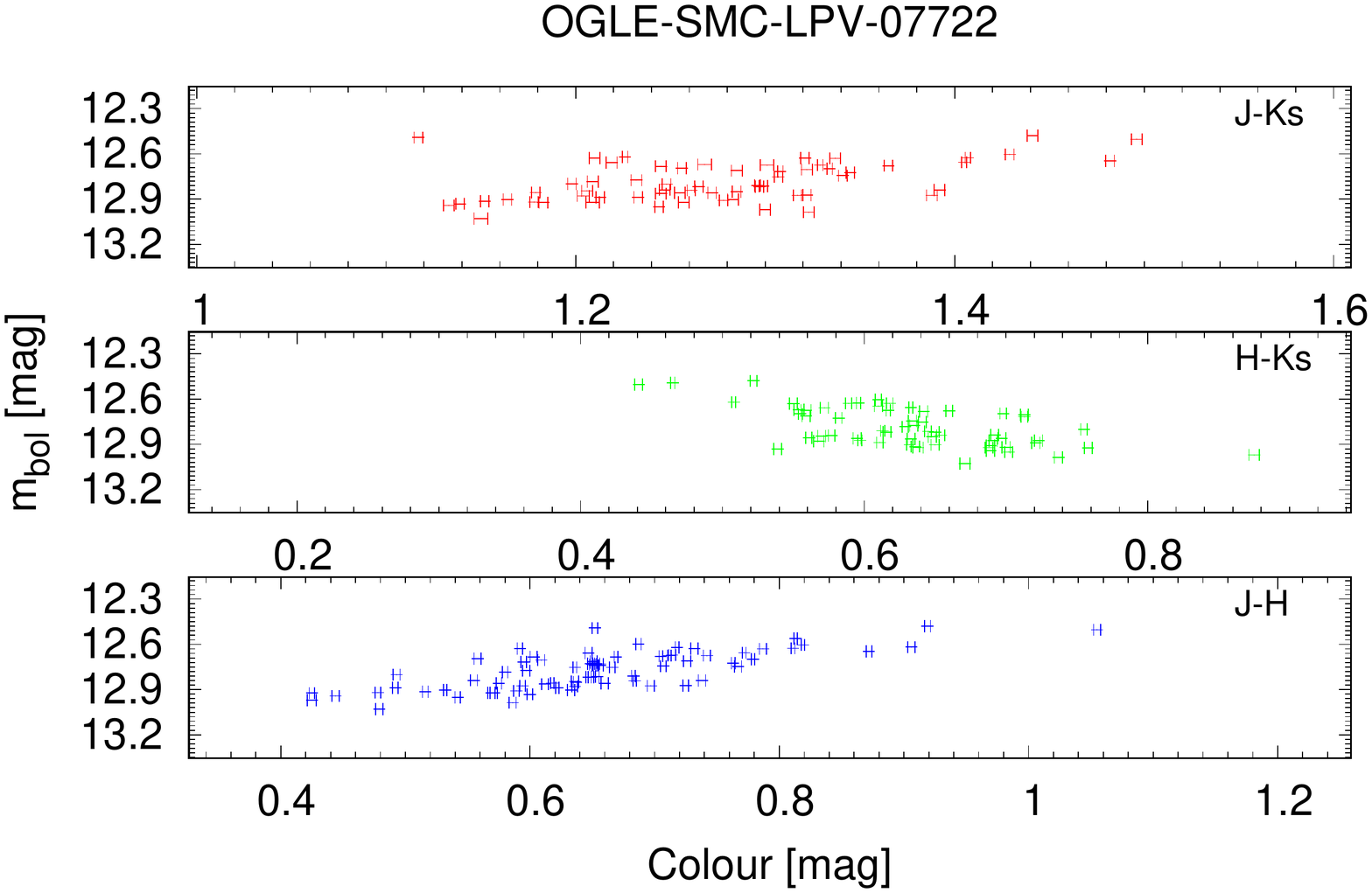} 
\caption{The correlations between the colours and bolometric light for two stars in Fig.~\ref{mbol} as an example.}
\label{colorchange2}
\end{figure}

The phase lags are also seen in the temporal variation of the NIR colours of some Miras. Refer to Fig.~\ref{colorchange2}: for the first star, OGLE SMC-LPV-12043, the bolometric brightest phase ${}^{m_{\rm bol}}\phi_{1}^{\rm max}$ corresponds to the bluest colour for all NIR colours. However, this is not the case for the second star, OGLE SMC-LPV-07722. In that case, the ${}^{m_{\rm bol}}\phi_{1}^{\rm max}$ corresponds to the bluest $H-K_{\rm s}$ colour, but to the reddest $J-H$ and $J-K_{\rm s}$ colours. If we assume that the $^{m_{\rm bol}}\phi_{1}^{\rm max}$ corresponds to the phase of maximum temperature, the results mean that the $J-H$ and $J-K_{\rm s}$ colours are not good indicators of stellar temperature for at least a subgroup of Miras. It is possible that the molecules and/or negative hydrogen ions in their atmospheres play a key role in producing such a phase inversion. \citet{frogel1981} pointed out that H$_2$O molecule formation in the extended atmosphere is critical to producing the colour phase inversion seen in some O-rich Miras.

\begin{figure}
\hspace*{-0.5cm}
\vspace*{-0.5cm}
\includegraphics[scale=0.31]{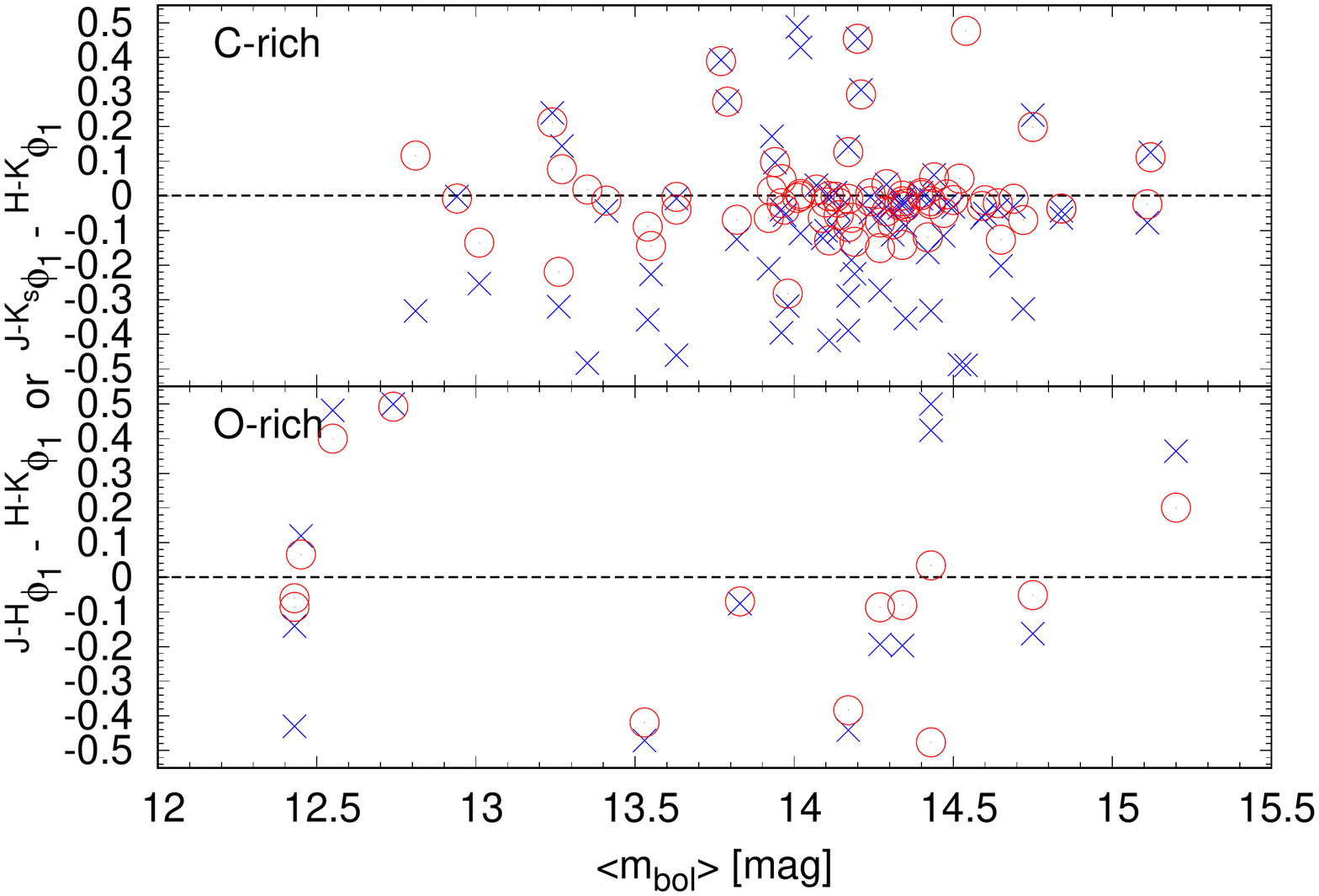}  
\hspace*{-0.5cm}
\vspace*{-1.0cm}
\includegraphics[scale=0.31]{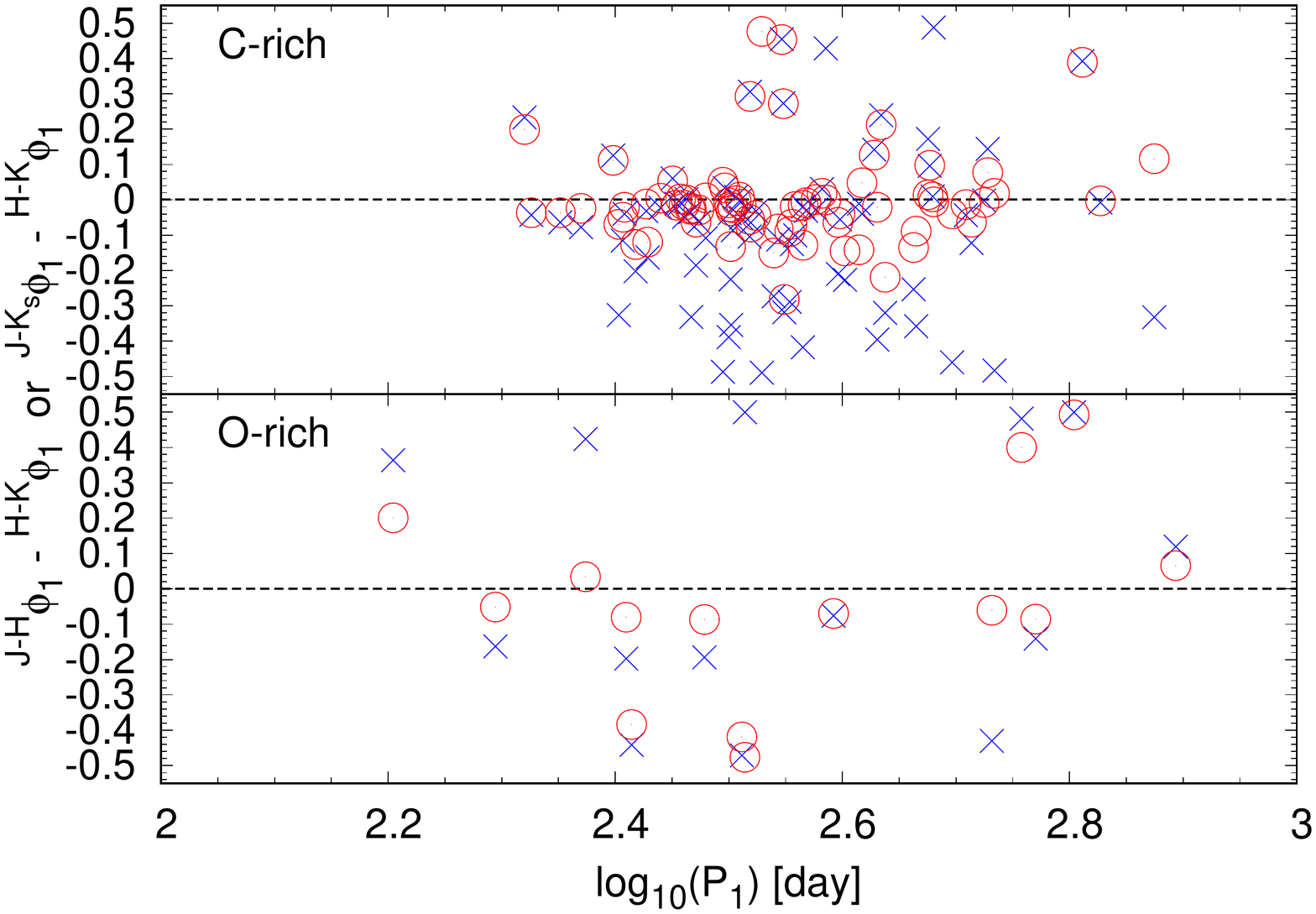} 
\hspace*{-0.5cm}
\vspace*{-1.0cm}
\includegraphics[scale=0.31]{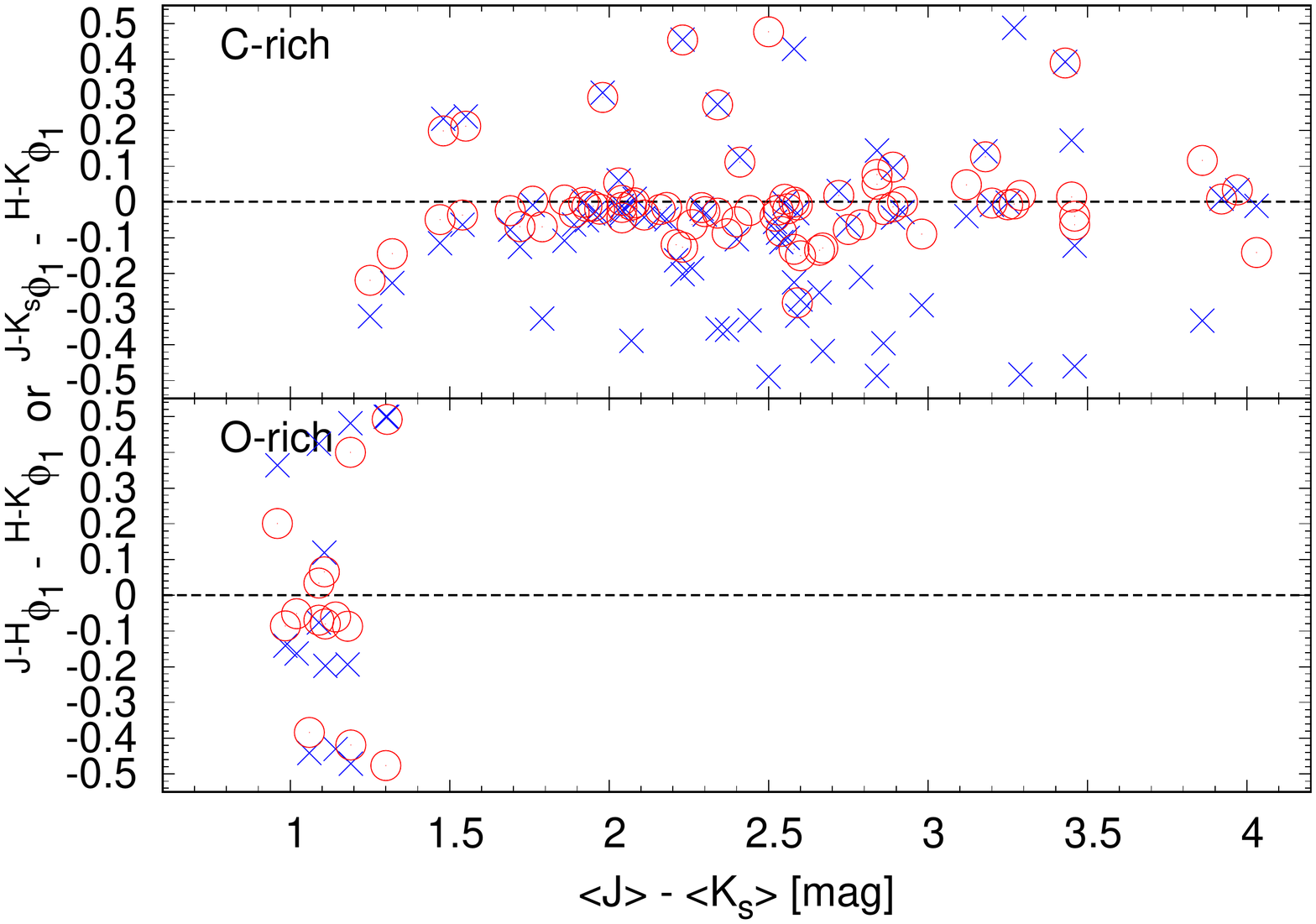} 
\hspace*{-0.5cm}
\vspace*{-1.0cm}
\includegraphics[scale=0.31]{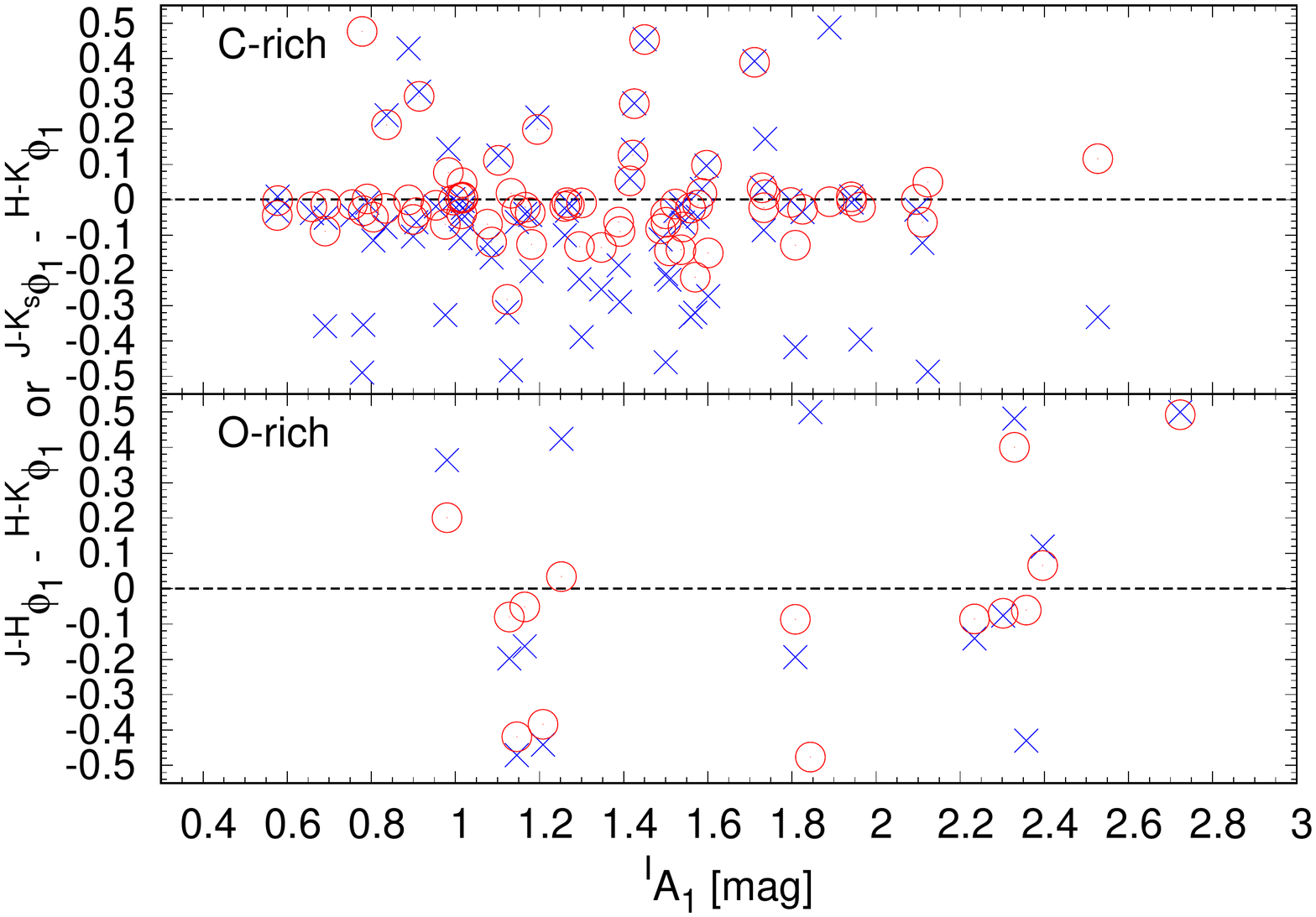} 
\caption{The ${}^{J-H}\phi_{1} - {}^{H-K_{\rm s}}\phi_{1}$ (blue crosses) or ${}^{J-K_{\rm s}}\phi_{1} - {}^{H-K_{\rm s}}\phi_{1}$ (red circles) is plotted against the $\langle m_{\rm bol} \rangle$, $P_1$, $\langle J \rangle - \langle K_{\rm s} \rangle$, and ${}^{I}A_1$, from top to bottom, respectively.}
\label{colorchange}
\end{figure}

To see the colour phase inversion quantitatively, we calculate the phase of the NIR colour time variation corresponding to the primary period (${}^{\rm colour}\phi_{1}$). Then ${}^{J-H}\phi_{1} - {}^{H-K_{\rm s}}\phi_{1}$ or ${}^{J-K_{\rm s}}\phi_{1} - {}^{H-K_{\rm s}}\phi_{1}$ is tested for any relationship with any other observables, such as pulsation period, amplitude or stellar colours to identify what type of Miras could show a colour phase inversion. Fig.~\ref{colorchange} shows the results. The figures suggest that the occurrence of the colour phase inversion does not depend much on the surface chemistry, luminosity, pulsation period, colour or amplitude. 

\begin{figure}
\hspace*{-0.5cm}
\vspace*{-1.0cm}
\includegraphics[scale=0.34]{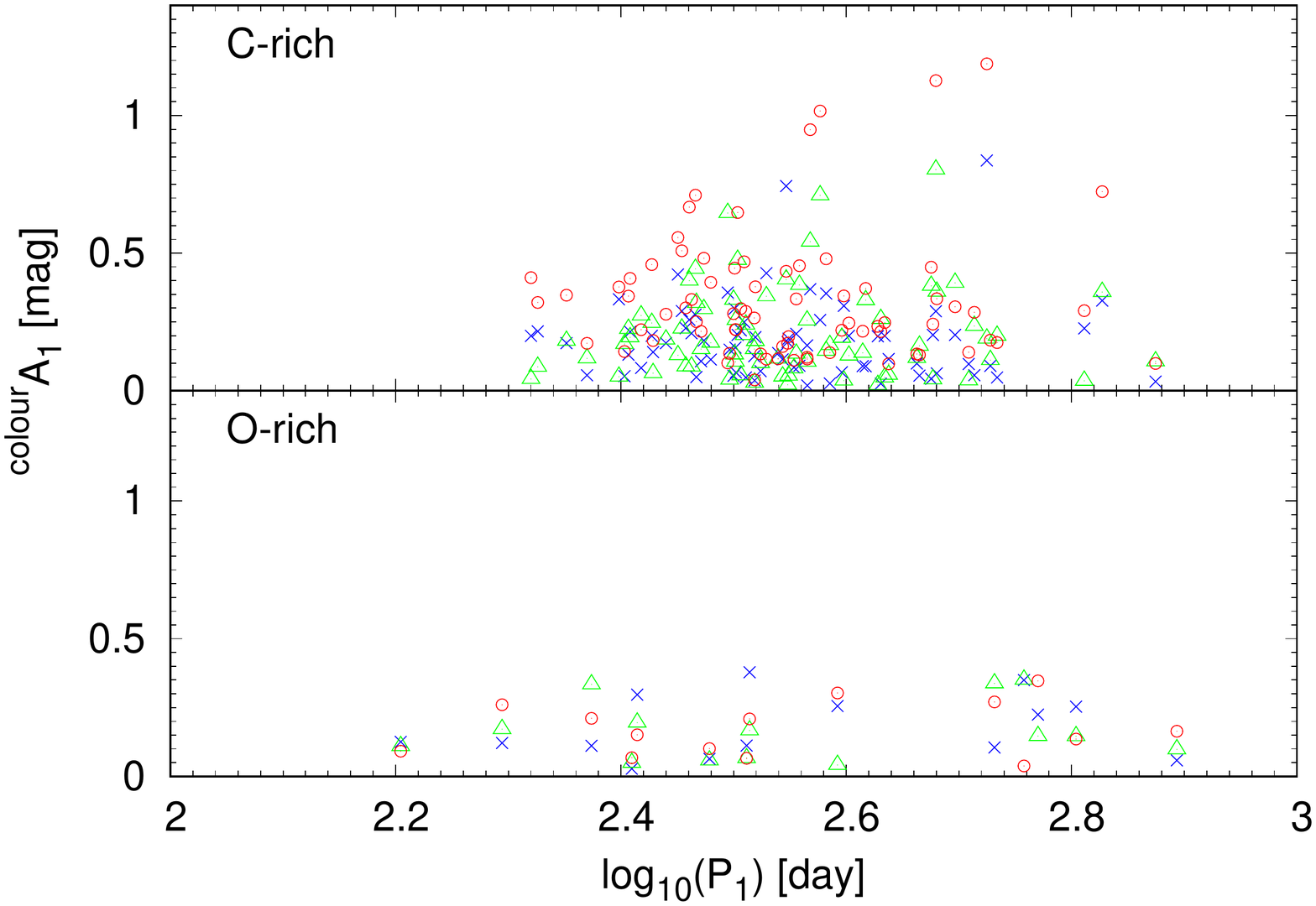} 
\hspace*{-0.5cm}
\vspace*{-1.0cm}
\includegraphics[scale=0.34]{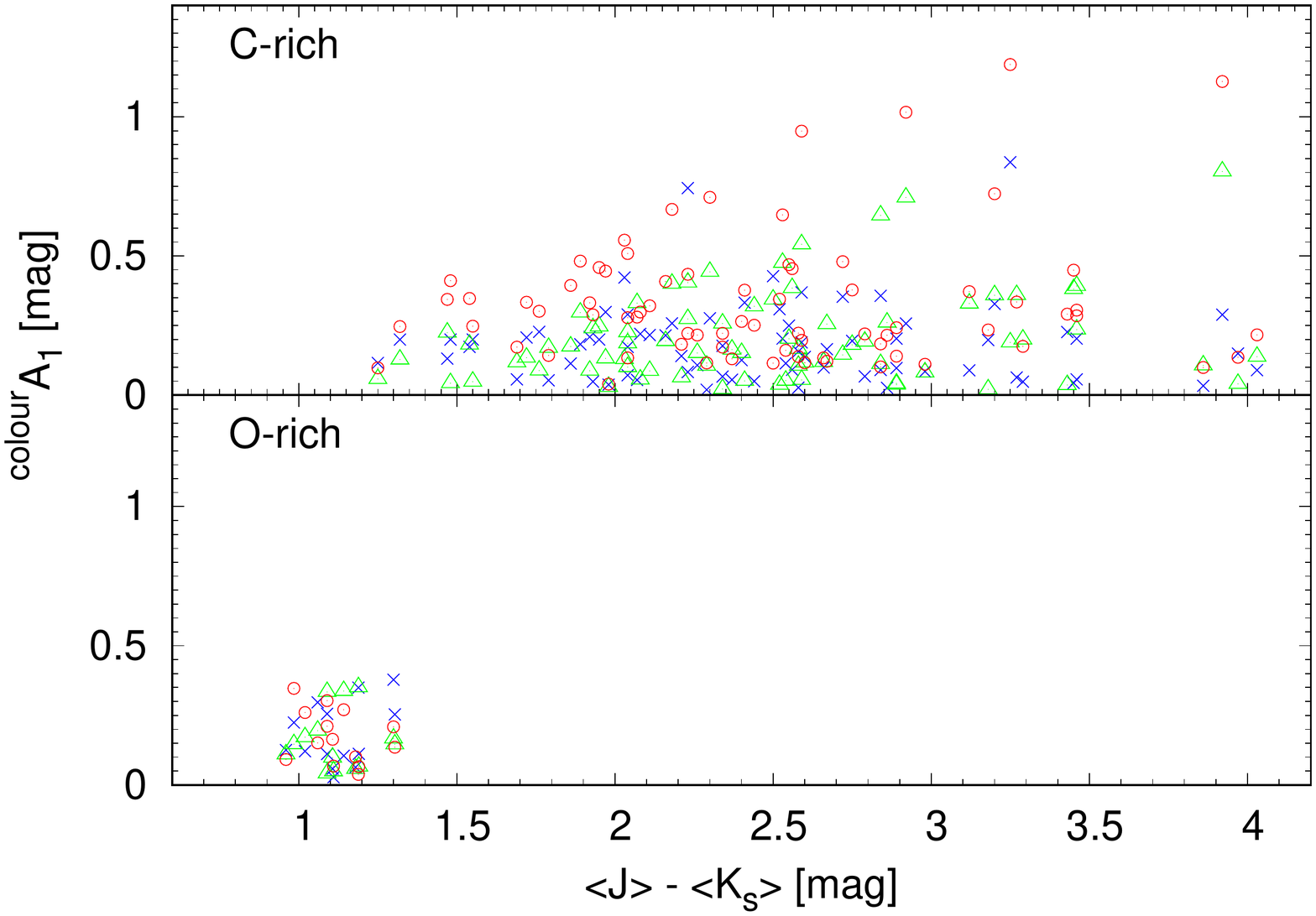} 
\caption{Colour amplitudes against log period and $\langle J \rangle - \langle K_{\rm s} \rangle$ colour for the sample stars. The shapes/colours of the symbols show the employed photometric colour, such that blue crosses, green triangles, and red circles correspond to $J-H$, $H-K_{\rm s}$, and $J-K_{\rm s}$, respectively.}
\label{colorchange22}
\end{figure}

It is also of interest to see whether the amplitudes of colour variation of the sample stars depend on stellar properties. Fig.~\ref{colorchange22} indicates that the colours of Miras can vary by up to about 0.5 mag regardless of their surface chemistries, periods or colours. 

\begin{figure}
\hspace*{-0.5cm}
\includegraphics[scale=0.34]{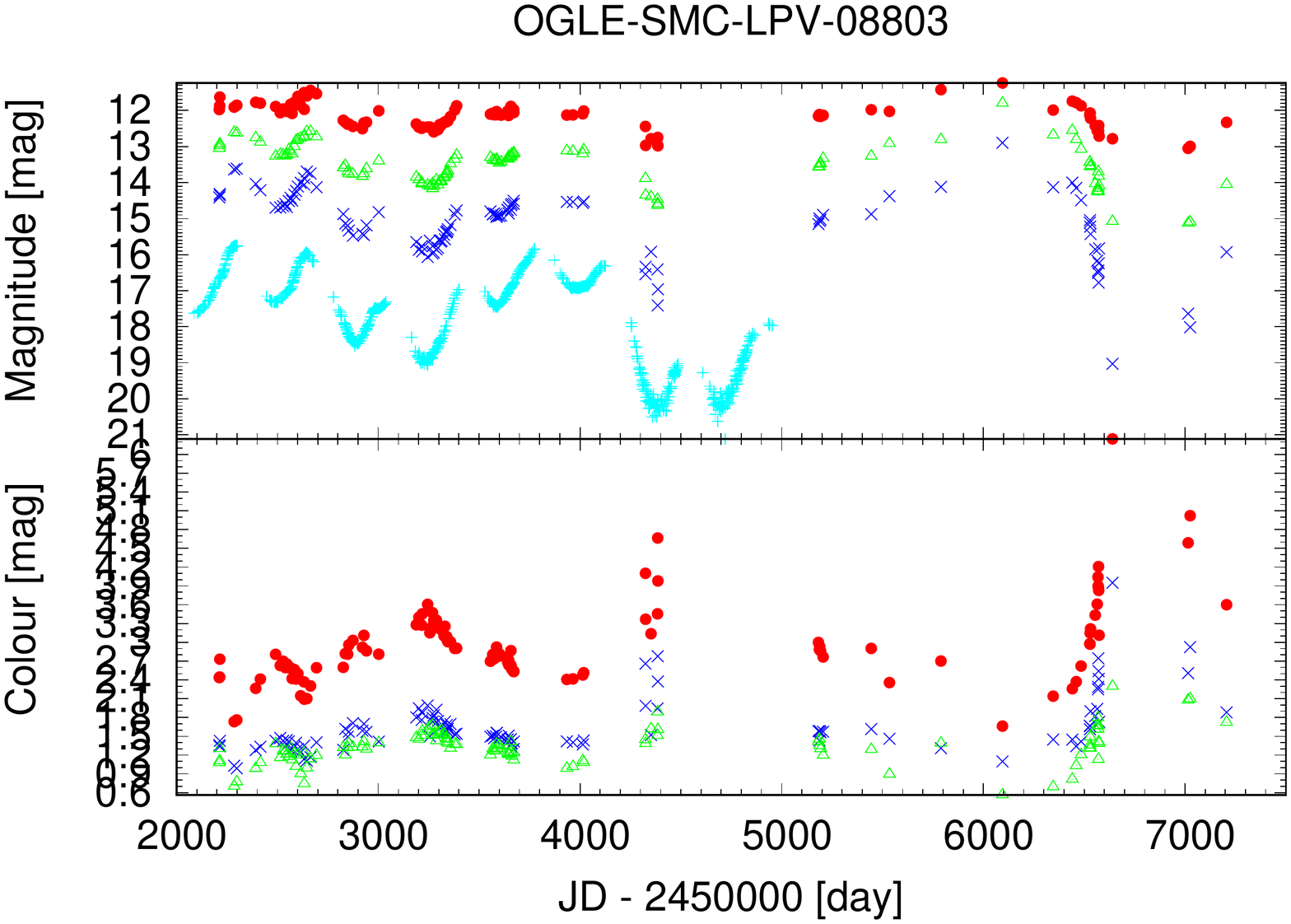} 
\vspace*{-1.0cm}
\hspace*{-0.5cm}
\includegraphics[scale=0.34]{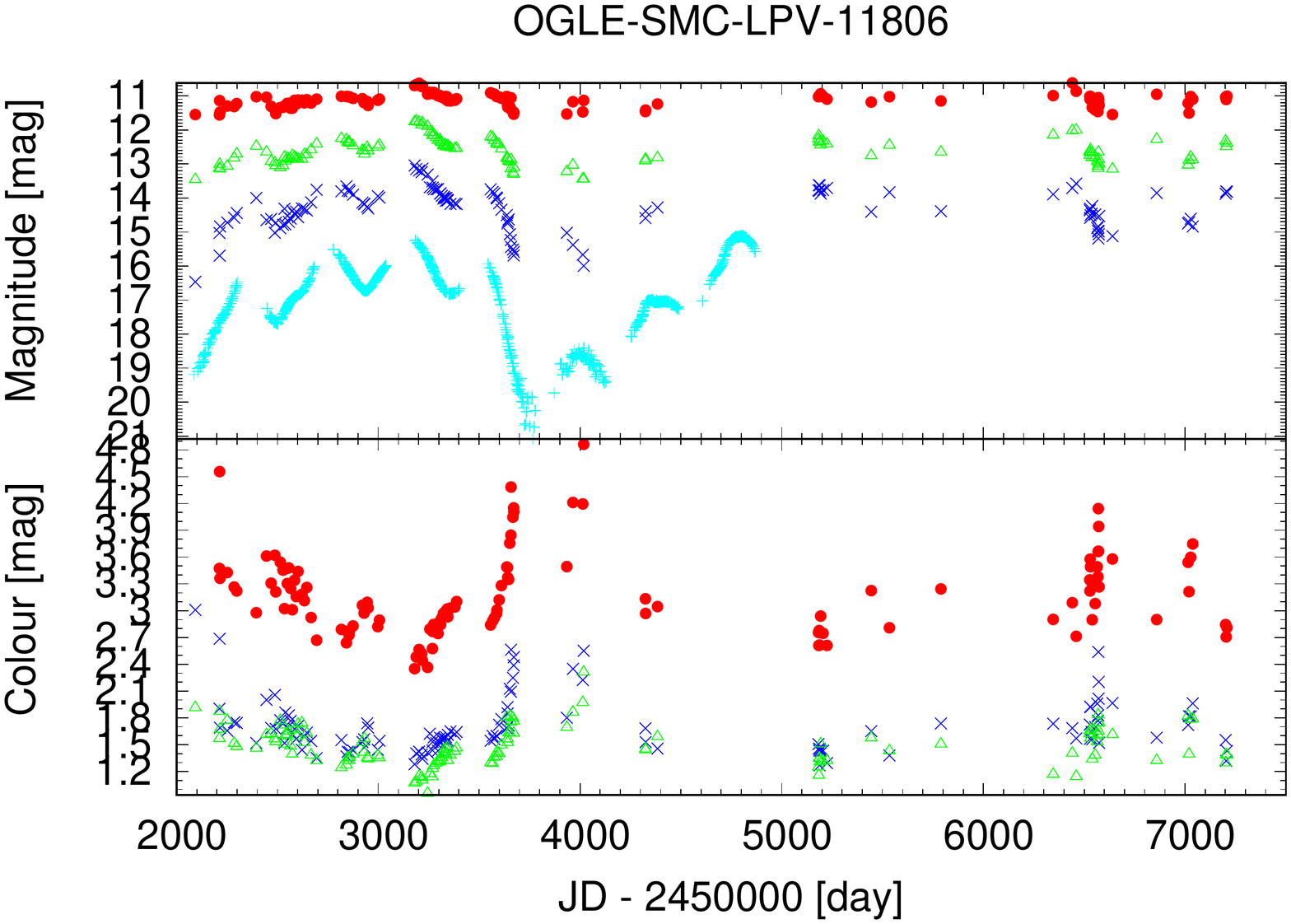} 
\caption{Two examples of Miras that exhibit occasional dimming events
during which their NIR magnitudes and colours vary independently of the pulsation period. The OGLE names of the two stars are indicated at the tops of the panels. The time variation of brightnesses (upper part) and near-infrared colours (lower part) are shown. {\bf Upper part of each panel:} Cyan, blue, green, and red show $I, J, H,$ and $K_{\rm s}$ brightness variations, respectively. {\bf Lower part of each panel:} Blue, green, and red show $J-H, H-K_{\rm s},$ and $J-K_{\rm s}$ colour variations, respectively.}
\label{largecolorchange}
\end{figure}

\begin{figure}
\centering
\includegraphics[scale=0.34]{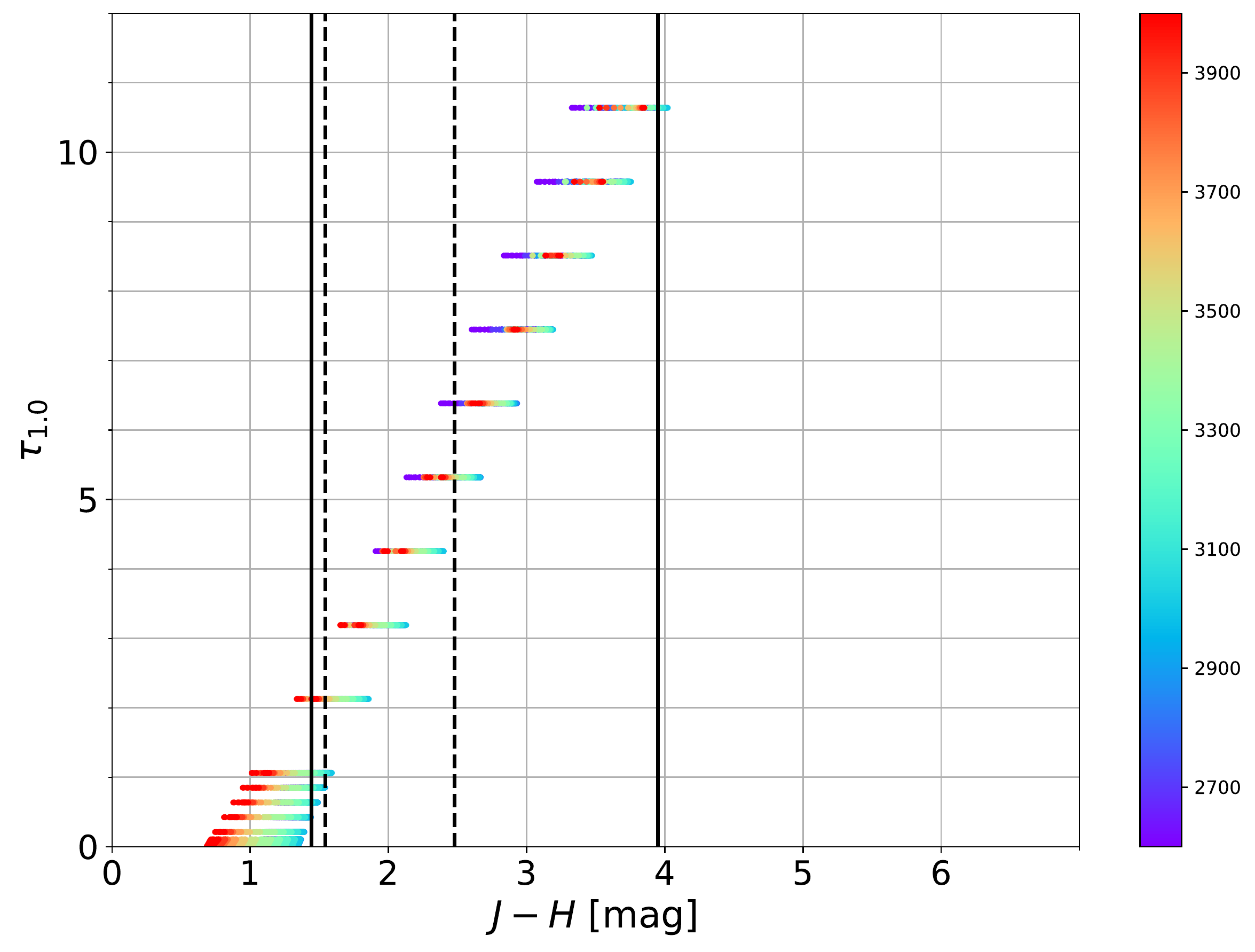} 
\includegraphics[scale=0.34]{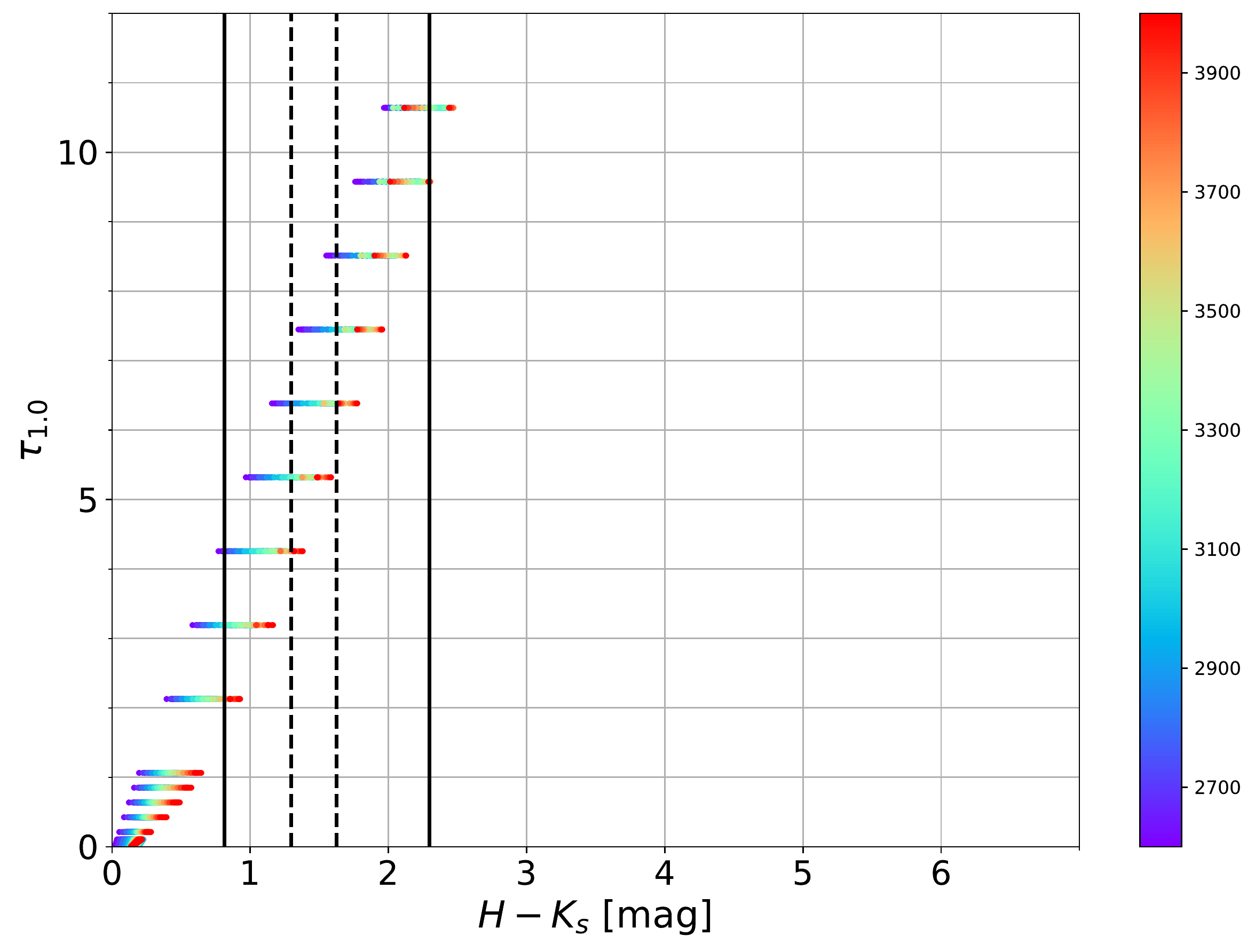} 
\includegraphics[scale=0.34]{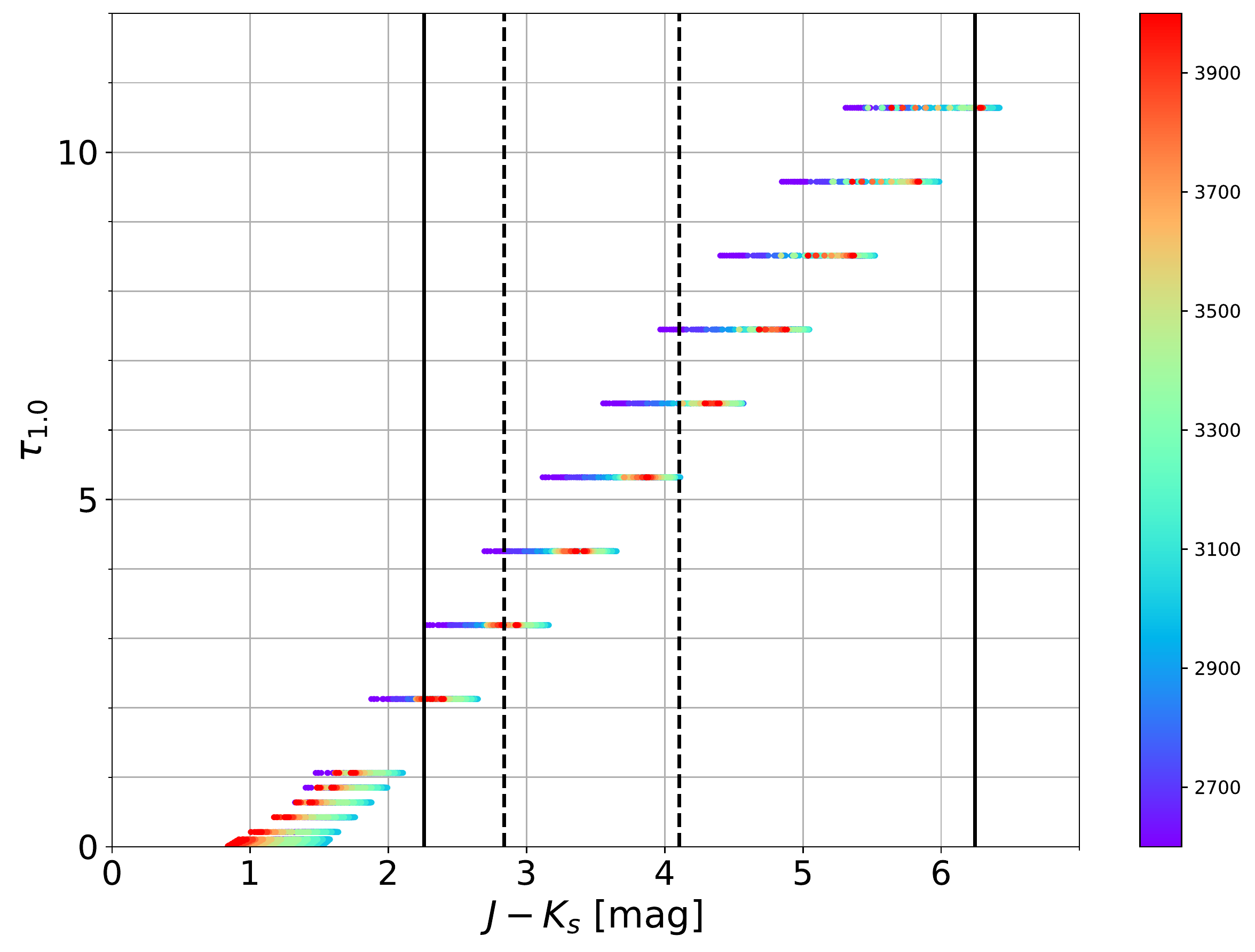} 
\caption{GRAMS models NIR colours are plotted against the model parameter, $\tau_{1.0}$. The solid and dashed lines show the colour changes at the events under consideration (see text) for OGLE SMC-LPV-08803 and 11806, respectively. The colour bar shows the effective temperature in Kelvin of the model photosphere for C-rich stars.}
\label{tauchange}
\end{figure}

\subsubsection{Large colour variation NOT due to stellar pulsation}
Among the stars listed in Table~\ref{Tw1000}, we found two Miras which exhibit occasional dimming events during which their NIR magnitudes and colours vary independently of the pulsation period. Their light and colour curves are shown in Fig.~\ref{largecolorchange}. Both are classified as carbon stars. Here we focus on the dimming events occurring at around HJDs 2456500 and 2453600, for OGLE SMC-LPV-08803 and 11806, respectively. In order to find out the possible causes of their large NIR colour change, we use GRAMS models for C-rich stars and surrounding dust shells again. In Fig.~\ref{tauchange}, GRAMS models NIR colours are plotted against $\tau_{1.0}$, the dust shell optical depth at 1 $\mu$m. The solid and dashed lines in the figure show the colour changes at the events under consideration for OGLE SMC-LPV-08003 and 11806, respectively. 

From Fig.~\ref{tauchange}, it can be seen that the colour change of OGLE SMC-LPV-11806 at the event could be explained by a change of effective temperature of its photosphere from 2600 to 4000K (GRAMS models limits). However, the colour change of OGLE SMC-LPV-08803 is far larger than expected from such a change of effective temperature in which case a large effective temperature change might be a necessary but not sufficient condition to account for the large colour change. 

The large colour change of OGLE SMC-LPV-08803 can be explained if $\tau_{1.0}$ changes from $\sim$2.1 to $\sim$10.6, a factor of approximately 5. If we attribute this change in $\tau_{1.0}$ solely to the circumstellar dust, it means that the cross section for extinction of the dust particle and/or the amount of dust has changed accordingly, since the optical depth is a product of the cross section and the column density of dust particles. The former means that the dust properties (size or species) have changed, and the latter that the dust mass loss rate has changed significantly during the dimming event considered here. This sudden change in the optical depth seems non-periodic or random as does the amount of reddening. These factors remind us of R Coronae Borealis type variables that show sudden and unpredictable drops in brightness due to obscuration by puffs of dust (e.g., \citealt{feast1997}), and are completely consistent with the explanation of asymmetric dust obscuration for the long term trends (or long secondary periods) of C-rich Miras in the LMC and the Galaxy proposed by \citet{feast2003} and \citet{whitelock2006}. It should be also noted that these cool C-rich stars which undergo dimming events are sometimes put in the DY Per class of variable stars (e.g., \citealt{tisserand2009}).

\section{Summary}
We combined optical to NIR multi-wavelength multi-epoch data to study bolometric light and colour changes of Mira variables in the SMC. We derived reliable bolometric corrections for various combinations of colours as well as period-bolometric and period-colour relations. Phase lags between bolometric and optical-to-NIR brightnesses were quantified, revealing that O-rich Miras have a large phase difference between optical and NIR brightness, by up to about 16\% of the primary pulsation period. A subgroup of the Miras was found to show phase inversion in colour variation. Large NIR colour changes independent of stellar pulsation were observed in some Miras and the long term trends are interpreted as being due to variable dust obscuration.

\section*{Acknowledgments}
We thank the anonymous reviewers for their careful reading and their insightful comments and suggestions for improving this paper. This work is partly supported by the Grant-in-Aid for Scientific Research (C) No.18K03690 and (A) No.16H02158 and Grant-in-Aid for Encouragement of Young Scientists (B) No.21740142 from the Ministry of Education, Culture, Sports, Science and Technology of Japan. This work is also partly supported by the Brain Circulation Program (R2301) by Japan Society for the Promotion of Science (JSPS). This paper uses observations made at the South African Astronomical Observatory. PAW and JWM acknowledge support from the South African National Research Foundation (NRF). We appreciate the support for our collaboration, in 2015--2017, provided by the Joint Research Project under agreement between the JSPS and NRF.

\section*{Data Availability}
The data underlying this article are partly available in the article and the rest will be shared on reasonable request to the corresponding author.







\appendix
\section{Data table}
\label{sec:appendix}
\begin{table*}
  \caption{Key values calculated in this paper. For those that could not be calculated are left blank.}
  \label{table:data}
  \begin{center}
    \begin{tabular}{c r r r r r r r r r r r}
	\hline
	 OGLE name & $P_1$ & ${}^{I}A_1$ & ${}^{J}A_1$ & ${}^{H}A_1$ & ${}^{K_{\rm s}}A_1$ & ${}^{m_{\rm bol}}A_1$ & $\langle I \rangle$ & $\langle J \rangle$ & $\langle H \rangle$ & $\langle K_{\rm s} \rangle$ & $\langle m_{\rm bol} \rangle$ \\
	 & [days] & \multicolumn{10}{c}{[mag]} \\
	\hline
  	\hline
OGLE-SMC-LPV-07392 &  197.0 & 1.17 & 0.56 & 0.53 & 0.39 & 0.54 & 14.40 & 12.96 & 12.22 & 11.94 & 14.75 \\ 
OGLE-SMC-LPV-07488 &  434.2 & 1.57 & 0.54 & 0.52 & 0.52 & 0.62 & 13.28 & 11.65 & 10.74 & 10.40 & 13.26 \\ 
OGLE-SMC-LPV-07597 &  400.3 & 1.51 & 0.88 & 0.70 & 0.64 & 0.74 & 13.59 & 11.90 & 11.03 & 10.58 & 13.55 \\ 
OGLE-SMC-LPV-07665 &  384.9 & 0.89 & 0.35 & 0.35 & 0.25 & 0.20 & 15.87 & 13.69 & 12.30 & 11.11 & 14.02 \\ 
OGLE-SMC-LPV-07722 &  636.8 & 2.72 & 0.22 & 0.47 & 0.38 & 0.31 & 13.41 & 11.14 & 10.43 &  9.84 & 12.74 \\ 
OGLE-SMC-LPV-07771 &  357.8 & 1.39 & 1.03 & 0.89 & 0.92 & 0.35 & 16.69 & 14.42 & 12.81 & 11.44 & 14.17 \\ 
OGLE-SMC-LPV-07793 &  353.2 & 1.43 & 0.51 & 0.26 & 0.33 & 0.30 & 15.61 & 13.18 & 11.87 & 10.84 & 13.79 \\ 
OGLE-SMC-LPV-08082 &  316.9 & 0.66 & 0.59 & 0.32 & 0.24 & 0.48 & 15.13 & 12.99 & 11.85 & 11.02 & 14.14 \\ 
OGLE-SMC-LPV-08202 &  219.9 & 1.45 &      &      & 0.63 &      & 16.03 &       &       & 11.73 &       \\ 
OGLE-SMC-LPV-08215 &  255.1 & 0.81 & 0.59 & 0.46 & 0.24 & 0.37 & 14.49 & 12.91 & 12.02 & 11.44 & 14.47 \\ 
OGLE-SMC-LPV-08356 &  323.1 & 1.02 & 1.35 & 1.05 & 0.87 & 0.41 & 16.02 & 13.91 & 12.52 & 11.36 & 14.24 \\ 
OGLE-SMC-LPV-08384 & 1056.0 & 3.06 & 0.89 & 0.84 &      &      & 13.21 & 10.38 &  9.79 &       &       \\ 
OGLE-SMC-LPV-08390 &  396.1 & 1.01 & 0.71 & 0.43 & 0.45 & 0.33 & 15.92 & 13.63 & 12.26 & 11.11 & 13.97 \\ 
OGLE-SMC-LPV-08425 &  329.9 & 0.90 & 1.01 & 0.88 & 0.61 & 0.37 & 15.23 & 13.58 & 12.28 & 11.18 & 14.14 \\ 
OGLE-SMC-LPV-08583 &  572.5 & 2.33 & 0.27 & 0.69 & 0.32 & 0.42 & 12.82 & 10.77 & 10.14 &  9.58 & 12.55 \\ 
OGLE-SMC-LPV-08590 &  290.2 & 1.00 & 0.72 & 0.44 & 0.38 & 0.34 & 14.51 & 12.98 & 11.92 & 11.06 & 14.11 \\ 
OGLE-SMC-LPV-08708 &  671.6 & 1.94 & 1.58 & 1.25 & 1.03 & 0.51 & 16.69 & 14.19 & 12.42 & 10.99 & 13.63 \\ 
OGLE-SMC-LPV-08723 &  267.5 & 1.52 & 0.97 & 0.79 & 0.47 & 0.70 & 14.90 & 13.29 & 12.20 & 11.34 & 14.50 \\ 
OGLE-SMC-LPV-08781 &  330.1 & 0.91 & 0.32 & 0.20 & 0.18 & 0.24 & 15.15 & 13.16 & 11.96 & 11.18 & 14.21 \\ 
OGLE-SMC-LPV-08794 &  361.7 & 1.26 & 0.85 & 0.79 & 0.42 & 0.35 & 16.10 & 13.86 & 12.44 & 11.30 & 14.10 \\ 
OGLE-SMC-LPV-08803 &  369.8 & 1.26 & 1.20 & 1.02 & 0.56 & 0.61 & 17.13 & 14.66 & 13.19 & 12.07 & 14.69 \\ 
OGLE-SMC-LPV-08804 &  256.8 & 1.13 & 0.28 & 0.31 & 0.27 & 0.35 & 14.23 & 12.54 & 11.81 & 11.43 & 14.34 \\ 
OGLE-SMC-LPV-08975 &  318.8 & 1.73 & 1.25 & 1.12 & 0.60 & 0.62 & 15.58 & 13.82 & 12.38 & 11.29 & 14.35 \\ 
OGLE-SMC-LPV-09155 &  293.1 & 1.56 & 0.84 & 0.91 & 0.59 & 0.42 & 16.23 & 13.84 & 12.56 & 11.40 & 14.43 \\ 
OGLE-SMC-LPV-09169 &  224.7 & 0.91 & 0.62 & 0.46 & 0.26 & 0.42 & 14.78 & 13.28 & 12.32 & 11.74 & 14.84 \\ 
OGLE-SMC-LPV-09274 &  459.9 & 1.35 & 0.85 & 0.70 & 0.62 & 0.43 & 15.45 & 13.03 & 11.60 & 10.37 & 13.01 \\ 
OGLE-SMC-LPV-09477 &  255.9 & 1.17 & 0.67 & 0.49 & 0.36 & 0.37 & 15.50 & 13.71 & 12.52 & 11.55 & 14.64 \\ 
OGLE-SMC-LPV-09518 &  782.4 & 2.40 & 0.93 & 0.88 & 0.77 & 0.79 & 12.95 & 10.76 & 10.02 &  9.65 & 12.45 \\ 
OGLE-SMC-LPV-09546 &  430.6 & 0.84 & 0.48 & 0.21 & 0.21 & 0.28 & 13.44 & 11.74 & 10.76 & 10.19 & 13.24 \\ 
OGLE-SMC-LPV-09683 &  349.6 & 1.49 & 0.45 & 0.26 & 0.27 & 0.33 & 15.59 & 13.81 & 12.41 & 11.27 & 14.31 \\ 
OGLE-SMC-LPV-09727 &  353.9 & 1.12 & 0.67 & 0.41 & 0.40 & 0.37 & 15.82 & 13.69 & 12.30 & 11.10 & 13.98 \\ 
OGLE-SMC-LPV-09751 &  330.6 & 1.54 & 1.04 & 0.86 & 0.49 & 0.36 & 15.94 & 14.05 & 12.55 & 11.30 & 14.27 \\ 
OGLE-SMC-LPV-09801 &  534.5 & 0.98 & 0.39 & 0.27 & 0.19 & 0.26 & 15.99 & 13.52 & 11.92 & 10.68 & 13.27 \\ 
OGLE-SMC-LPV-10077 &  295.9 & 1.39 & 0.97 & 0.84 & 0.63 & 0.56 & 15.17 & 13.62 & 12.36 & 11.36 & 14.18 \\ 
OGLE-SMC-LPV-10208 &  359.4 & 1.08 & 0.87 & 0.61 & 0.41 & 0.58 & 14.36 & 12.53 & 11.54 & 10.81 & 13.82 \\ 
OGLE-SMC-LPV-10266 &  297.6 & 1.15 & 0.75 & 0.59 & 0.35 & 0.40 & 15.11 & 13.44 & 12.36 & 11.55 & 14.59 \\ 
OGLE-SMC-LPV-10436 &  473.6 & 1.74 & 0.71 & 0.70 & 0.30 & 0.45 & 17.29 & 14.64 & 12.89 & 11.19 & 13.93 \\ 
OGLE-SMC-LPV-10448 &  292.5 & 1.82 & 1.19 & 0.97 & 0.58 & 0.51 & 15.57 & 13.73 & 12.49 & 11.43 & 14.33 \\ 
OGLE-SMC-LPV-10708 &  320.9 & 0.58 & 0.44 & 0.30 & 0.24 & 0.39 & 14.79 & 13.02 & 11.85 & 10.94 & 14.13 \\ 
OGLE-SMC-LPV-10790 &  236.5 & 1.25 & 0.28 & 0.17 & 0.53 & 0.33 & 14.30 & 12.66 & 11.90 & 11.57 & 14.43 \\ 
OGLE-SMC-LPV-10866 &  301.8 & 1.01 & 0.69 & 0.44 & 0.36 & 0.27 & 14.29 & 12.88 & 11.83 & 11.02 & 14.02 \\ 
OGLE-SMC-LPV-10929 &  326.7 & 1.84 & 0.35 & 0.74 & 0.57 & 0.48 & 14.96 & 12.81 & 11.95 & 11.51 & 14.43 \\ 
OGLE-SMC-LPV-11066 &  324.2 & 0.69 & 0.52 & 0.54 & 0.45 & 0.47 & 15.16 & 13.39 & 12.21 & 11.46 & 14.60 \\ 
OGLE-SMC-LPV-11205 &  346.4 & 1.60 & 0.75 & 0.58 & 0.47 & 0.29 & 15.63 & 14.11 & 12.71 & 11.51 & 14.27 \\ 
OGLE-SMC-LPV-11246 &  427.2 & 1.96 & 1.13 & 1.20 & 0.84 & 0.59 & 16.14 & 14.18 & 12.65 & 11.32 & 13.96 \\ 
OGLE-SMC-LPV-11279 &  541.8 & 1.13 & 0.73 & 0.53 & 0.41 & 0.36 & 17.86 & 14.56 & 12.75 & 11.27 & 13.35 \\ 
OGLE-SMC-LPV-11557 &  301.0 & 1.81 & 0.63 & 0.68 & 0.74 & 0.64 & 14.32 & 12.53 & 11.73 & 11.35 & 14.27 \\ 
OGLE-SMC-LPV-11639 &  391.1 & 2.30 & 0.26 & 0.56 & 0.56 & 0.47 & 14.36 & 11.98 & 11.37 & 10.89 & 13.83 \\ 
OGLE-SMC-LPV-11691 &  497.3 & 1.50 & 0.99 & 1.15 & 0.73 & 0.45 & 17.38 & 14.68 & 12.79 & 11.22 & 13.63 \\ 
OGLE-SMC-LPV-11739 &  650.7 & 3.05 & 0.47 &      &      &      & 11.74 &  9.96 &       &       &       \\ 
OGLE-SMC-LPV-11800 &  478.1 & 1.94 & 1.72 & 1.29 & 0.66 & 0.47 & 19.00 & 15.67 & 13.68 & 11.75 & 14.40 \\ 
OGLE-SMC-LPV-11806 &  414.3 & 1.02 & 0.60 & 0.49 & 0.32 & 0.23 & 16.31 & 14.32 & 12.68 & 11.20 & 13.96 \\ 
OGLE-SMC-LPV-11843 &  411.7 & 1.54 & 1.02 & 0.94 & 0.71 & 0.44 & 19.36 & 15.70 & 13.61 & 11.67 & 14.34 \\ 
OGLE-SMC-LPV-11899 &  316.2 & 1.30 & 0.62 & 0.79 & 0.48 & 0.43 & 14.92 & 13.26 & 12.23 & 11.19 & 14.17 \\ 
OGLE-SMC-LPV-11947 &  382.1 & 1.59 & 0.95 & 0.72 & 0.67 & 0.54 & 15.97 & 13.99 & 12.49 & 11.27 & 14.07 \\ 
	\multicolumn{12}{r}{Continued on next page.}
	\end{tabular}
  \end{center}
\end{table*}

\begin{table*}
  \contcaption{}
  \label{table:data}
  \begin{center}
    \begin{tabular}{c r r r r r r r r r r r}
	\hline
	 OGLE name & $P_1$ & ${}^{I}A_1$ & ${}^{J}A_1$ & ${}^{H}A_1$ & ${}^{K_{\rm s}}A_1$ & ${}^{m_{\rm bol}}A_1$ & $\langle I \rangle$ & $\langle J \rangle$ & $\langle H \rangle$ & $\langle K_{\rm s} \rangle$ & $\langle m_{\rm bol} \rangle$ \\
	 & [days] & \multicolumn{10}{c}{[mag]} \\
	\hline
  	\hline
OGLE-SMC-LPV-12043 &  284.5 & 1.27 & 0.95 & 0.67 & 0.53 & 0.51 & 15.35 & 13.40 & 12.28 & 11.36 & 14.43 \\ 
OGLE-SMC-LPV-12187 &  424.6 & 1.42 & 0.49 & 0.67 & 0.69 & 0.45 & 16.65 & 14.59 & 12.95 & 11.41 & 14.17 \\ 
OGLE-SMC-LPV-12249 &  367.6 & 0.95 & 0.61 & 0.58 & 0.42 & 0.15 & 16.31 & 13.48 & 12.24 & 11.19 & 14.34 \\ 
OGLE-SMC-LPV-12337 &  647.6 & 1.71 & 1.32 & 0.84 & 0.82 & 0.48 & 18.06 & 14.87 & 13.09 & 11.44 & 13.77 \\ 
OGLE-SMC-LPV-12427 &  367.4 & 1.81 & 0.59 & 0.81 & 0.65 & 0.59 & 16.51 & 14.22 & 12.93 & 11.55 & 14.11 \\ 
OGLE-SMC-LPV-12568 &  479.0 & 1.89 & 0.83 & 0.85 & 0.51 & 0.52 & 17.48 & 14.62 & 12.92 & 11.35 & 14.01 \\ 
OGLE-SMC-LPV-12603 &  517.3 & 2.11 & 0.81 & 0.77 & 0.68 & 0.37 & 17.77 & 15.32 & 13.52 & 11.86 & 14.09 \\ 
OGLE-SMC-LPV-12637 &  538.9 & 2.36 & 0.57 & 0.80 & 0.34 & 0.37 & 12.64 & 10.71 & 10.04 &  9.57 & 12.43 \\ 
OGLE-SMC-LPV-12690 &  352.1 & 1.45 & 0.66 & 1.28 & 0.83 & 0.48 & 15.50 & 13.46 & 12.38 & 11.23 & 14.20 \\ 
OGLE-SMC-LPV-12762 &  749.1 & 2.53 & 1.74 & 1.34 & 1.19 & 0.07 & 18.32 & 15.25 & 13.21 & 11.39 & 12.81 \\ 
OGLE-SMC-LPV-12931 &  317.8 & 0.78 & 0.59 & 0.71 & 0.42 & 0.27 & 15.84 & 13.86 & 12.62 & 11.52 & 14.35 \\ 
OGLE-SMC-LPV-12942 &  234.3 & 0.83 & 0.45 & 0.37 & 0.27 & 0.26 & 15.17 & 13.68 & 12.65 & 11.99 & 15.11 \\ 
OGLE-SMC-LPV-12972 &  286.7 & 0.79 & 0.40 & 0.17 & 0.12 & 0.20 & 14.71 & 13.04 & 11.94 & 11.28 & 14.40 \\ 
OGLE-SMC-LPV-12997 &  209.0 & 1.20 & 0.66 & 0.29 & 0.22 & 0.44 & 14.78 & 13.19 & 12.28 & 11.71 & 14.75 \\ 
OGLE-SMC-LPV-13009 &  334.2 & 0.58 & 0.36 & 0.29 & 0.23 & 0.20 & 14.70 & 13.19 & 12.05 & 11.15 & 14.29 \\ 
OGLE-SMC-LPV-13099 &  268.2 & 1.09 & 0.54 & 0.44 & 0.33 & 0.29 & 15.24 & 13.61 & 12.33 & 11.40 & 14.42 \\ 
OGLE-SMC-LPV-13139 &  253.0 & 0.98 & 0.39 & 0.43 & 0.28 & 0.36 & 15.08 & 13.35 & 12.22 & 11.56 & 14.72 \\ 
OGLE-SMC-LPV-13676 &  530.6 & 1.80 & 1.23 & 0.74 & 0.65 & 0.38 & 16.21 & 14.02 & 12.31 & 10.77 & 12.94 \\ 
OGLE-SMC-LPV-13762 &  160.1 & 0.98 & 0.47 & 0.33 & 0.42 & 0.46 & 14.70 & 13.42 & 12.79 & 12.46 & 15.20 \\ 
OGLE-SMC-LPV-13766 &  394.5 & 1.50 & 0.75 & 0.73 & 0.48 & 0.42 & 16.19 & 13.92 & 12.47 & 11.13 & 13.92 \\ 
OGLE-SMC-LPV-13824 &  312.4 & 2.12 & 1.04 & 1.27 & 0.73 & 0.59 & 17.05 & 14.53 & 12.92 & 11.69 & 14.52 \\ 
OGLE-SMC-LPV-13861 &  462.4 & 0.69 & 0.28 & 0.24 & 0.15 & 0.23 & 14.69 & 13.02 & 11.70 & 10.65 & 13.54 \\ 
OGLE-SMC-LPV-13907 &  282.2 & 1.42 & 1.27 & 0.71 & 0.61 & 0.60 & 14.70 & 13.56 & 12.34 & 11.53 & 14.44 \\ 
OGLE-SMC-LPV-14040 &  313.7 & 1.73 & 1.05 & 0.93 & 0.92 & 0.40 & 18.96 & 15.84 & 13.79 & 11.87 & 14.29 \\ 
OGLE-SMC-LPV-14055 &  324.7 & 1.15 & 0.33 & 0.46 & 0.42 & 0.42 & 13.24 & 11.86 & 11.05 & 10.67 & 13.53 \\ 
OGLE-SMC-LPV-14170 &  275.4 & 1.02 & 0.76 & 0.59 & 0.40 & 0.36 & 14.98 & 13.28 & 12.18 & 11.24 & 14.48 \\ 
OGLE-SMC-LPV-14205 &  261.8 & 1.18 & 0.61 & 0.61 & 0.32 & 0.19 & 15.84 & 13.84 & 12.66 & 11.61 & 14.65 \\ 
OGLE-SMC-LPV-14462 & 1092.0 & 3.68 & 0.43 &      &      &      & 12.46 & 10.10 &       &       &       \\ 
OGLE-SMC-LPV-14581 &  317.4 & 1.29 & 0.49 & 0.40 & 0.37 & 0.53 & 15.70 & 13.88 & 12.47 & 11.30 & 14.19 \\ 
OGLE-SMC-LPV-14643 &  288.8 & 1.58 & 0.99 & 0.76 & 0.41 & 0.54 & 14.99 & 13.35 & 12.14 & 11.17 & 14.24 \\ 
OGLE-SMC-LPV-14688 &  211.9 & 1.18 & 0.80 & 0.60 & 0.51 & 0.36 & 15.38 & 13.87 & 12.70 & 11.76 & 14.84 \\ 
OGLE-SMC-LPV-14778 &  475.2 & 1.60 & 1.03 & 0.79 & 0.71 & 0.43 & 16.79 & 13.86 & 12.34 & 10.97 & 13.94 \\ 
OGLE-SMC-LPV-14787 &  259.7 & 1.21 & 0.24 & 0.58 & 0.41 & 0.43 & 13.84 & 12.43 & 11.67 & 11.37 & 14.17 \\ 
OGLE-SMC-LPV-14860 &  589.1 & 2.23 & 0.25 & 0.47 & 0.58 & 0.32 & 12.64 & 10.64 &  9.94 &  9.65 & 12.43 \\ 
OGLE-SMC-LPV-14991 &  338.1 & 0.78 & 0.34 & 0.66 & 0.48 & 0.26 & 16.71 & 14.64 & 13.50 & 12.14 & 14.54 \\ 
	\hline
	\hline
	\end{tabular}
  \end{center}
\end{table*}




\bsp	
\label{lastpage}
\end{document}